%% file: switch_networks_paper.tex
\documentclass[twocolumn,amsmath,amsfonts,amssymb,floatfix]{revtex4-1}

\usepackage{graphicx,epsfig,color,times,bbm}
\usepackage{psfrag}
\usepackage{braket}
\usepackage{float}
\usepackage[section]{placeins}
\usepackage{enumerate}
\AtBeginDocument{\usepackage{booktabs}}
\graphicspath{{figures/}}

\newcommand{\me}{\mathrm{e}}

\newcommand{\diag}{\mathrm{diag}}

\begin{document}

\title{Switch networks for photonic fusion-based quantum computing}

\author{Sara Bartolucci}
\author{Patrick Birchall}
\author{Damien Bonneau} 
\author{Hugo Cable} 
\email[]{Lead author, hugo@psiquantum.com}
\author{Mercedes Gimeno-Segovia}
\author{Konrad Kieling}
\author{Naomi Nickerson}
\author{Terry Rudolph}
\author{Chris Sparrow}

\affiliation{PsiQuantum, Palo Alto}
\date\today

\begin{abstract}
Fusion-based quantum computing (FBQC) offers a powerful approach to building a fault-tolerant universal quantum computer using photonic components --- single-photon sources, linear-optical circuits, single-photon detectors, and optical switching with feedforward control.  Both individual optical switches and sophisticated switch networks are needed where it is necessary to perform operations conditionally, using feedforward of previous photon-detection outcomes, within the lifetime of remaining photons.  Most visibly, feedforward switching is required for fault-tolerant operations at the level of logical qubits, which are needed in turn for useful quantum algorithms.  However, switch networks are also required for multiplexing (``muxing'') stages that are needed for generating specific small entangled resource states, where it is used to boost the probabilities for allocating quantum states to fusion gates and other operations --- a task which dominates the footprint of photonic FBQC.  Despite their importance, limited attention has been paid to exploring possible designs of switch networks in this setting.  Here we present a wide range of new techniques and schemes which enable major improvements in terms of muxing efficiency and reductions in hardware requirements.  Since the use of photonic switching heavily impacts qubit losses and errors, our schemes are constructed with low switch depth.  They also exploit specific features of linear-optical circuits which are commonly used to generate entanglement in proposed quantum computing and quantum network schemes. 
\end{abstract}

\maketitle
\section{\label{sec:Intro} Introduction}
\input{sections/intro}

\section{\label{sec:SwNetReview} Review of common switch-network schemes}
\input{sections/review}

\section{\label{sec:SwitchingTheory} Power of a single layer of switching}
\input{sections/gmzis}

\section{\label{sec:NewSpatialMUXes} Switch networks for efficient spatial multiplexing}
\input{sections/spatial_muxes}

\section{\label{sec:NewTemporalMUXes} Switch networks for efficient temporal multiplexing}
\input{sections/temporal_muxes}

\section{\label{sec:Conclusions} Conclusions}
\input{sections/conclusions}

\acknowledgements{The authors would like to thank Chia-Ming Chang, Eric Dudley, Gary Gibson, Nikhil Kumar, Gabriel Mendoza, Mihai Vidrighin for discussions about the limitations of hardware components, and, our other colleagues at PsiQuantum and Jacob Bulmer for other useful discussions.}

\bibliography{switch_networks_paper}{}

\appendix
\input{sections/appendices}

\end{document}

%% file: sections/intro.tex
{\bf Fusion-based quantum computing (FBQC) provides a compelling paradigm for universal fault-tolerant quantum computing using photonic qubits \cite{Bartolucci21}.}  FBQC lends itself to a modular hardware implementation of a quantum computer using resource-state generators (RSGs) of small, entangled states, and fusion devices which perform entangling measurements between them.  An entire quantum computer can be constructed using networked modules, each comprised of an RSG and associated fusion devices.  The approach of interleaving, introduced recently in \cite{Bombin21}, provides a technique for implementing fault-tolerant quantum computation using these modules in the context of photonic FBQC (meeting some common objectives as Ref.~\cite{Asavanant19} for example for other quantum-computing paradigms).  It is based on a sophisticated method of temporal muxing which enables each module to host thousands of physical qubits using low-loss fiber delay, and to implement the measurements required for logical operations.

Integrated quantum photonics provides a compelling approach to building a quantum computer using chip-based devices \cite{Silverstone16}.  These chips can incorporate passive linear optics, probabilistic single-photon generation using parametric nonlinear sources, single-photon detectors, fast reconfigurable (active) phase shifters, and electrical control for feedforward.  They can also be optically and electrically interconnected to enable powerful subsystems which implement the various stages of entanglement generation needed to create resource states.  Sophisticated processing of classical information can be performed using standard microelectronics for digital logic and electrical buses.

{\bf The physical characteristics of switch networks have major implications for overall hardware footprint, as well as for optical loss and qubit errors.}  The hardware footprint for a full-scale photonic quantum computer is inevitably dominated by the initial steps of resource-state generation \cite{Rudolph17}. Although there are many ways to generate a resource state, the initial steps commonly involve muxing operations on single-photon states generated by nonlinear parametric sources \cite{Migdall02} and by circuits for creating Bell \cite{Zhang08} or GHZ states \cite{Varnava08}.  Muxing uses a switch network to relocate photonic quantum states in target spatio-temporal bins from non-deterministic inputs, and is necessary as the sources and circuits are intrinsically probabilistic, heralding useful output states when they succeed.  Scaling computation in FBQC is achieved by increasing the number of resource states rather than their size, and the generation of each state is allowed to fail with a finite probability, resulting in erasure (the known loss of photons encoding qubits). Because switch networks are used for muxing and routing resource states, their practical constraints have a critical impact on the overall footprint of the machine. In the interleaving approach they have a large impact on the size of the module.  The central goal of this paper is to introduce techniques and schemes for improving the performance of muxing, and switch networks useful for photonic FBQC more generally, with direct gains in turn for full-scale quantum computing hardware.

Despite the importance of switch networks both for photonic FBQC and for photonic quantum computing more generally, limited attention has gone into the design of new switch network schemes, resulting in exploration of a relatively small number of distinct switching architectures.  In fact, every switch network presents different requirements for the performance of optical components, the speeds at which they operate, active power consumption, the complexity of control electronics for routing logic, and optical/electrical connectivity as part of a larger networked system.  An overriding requirement for photonic quantum computing is to minimize the depth of physical operations, so as to minimize loss and errors on photons that transit through them.

Although the principal motivation for muxing is to boost the probabilities for obtaining quantum states, it is often overlooked that the switch networks used for muxing can also naturally enable hardware redundancy for circuits upstream and downstream.  Furthermore, when optical switches can operate faster than other hardware components, it is possible to design switch networks to route around bottlenecks caused by the slower components, to make the quantum-computer architecture more efficient overall.

{\bf High-speed and low-loss switching networks enable universal photonic quantum computing.}  The requirement for active switching and high-performance switch networks comes from multiple considerations.  As noted above, feedforward and adaptive operation are fundamental requirements for universal quantum computation in all known approaches based on photonic hardware (whether working in the FBQC framework, the circuit model \cite{Knill01}, measurement-based quantum computation paradigms \cite{Nielsen04, Browne05, Kieling07, Gimeno_Segovia15, Li15, Rohde15B, Morley_Short17}, or in the continuous-variable approach using Gottesman-Kitaev-Preskill (GKP) qubits \cite{Menicucci14}).  Boson sampling provides a well-known example of a photon-based model of computation which operates passively and does not incorporate feedforward \cite{Aaronson13}.  However despite being believed to defy efficient classical simulation, boson sampling is not equivalent to universal quantum computation.

Strategies exist that allow some feedforward-based switching to be removed within a quantum-computing architecture, but each of them introduces some alternative cost.  For example, generation of single photons --- and even strings of entangled photons --- can in principle be done in a deterministic ``on-demand'' manner using some quantum dot sources \cite{Lindner09}.  However these sources suffer from implementation issues especially regarding noise and photon indistinguishability (as explained in Ref.~\cite{Rudolph17}).  On a different point, percolation-based frameworks such as first proposed in Ref.~\cite{Kieling07} address the randomness of nondeterministic quantum gates while avoiding large amounts of switching.  Recent work has also shown how to eliminate fast switching from some parts of architectures based on GKP qubits \cite{Tzitrin21}.  Despite these developments, low-loss and high efficiency switch networks remain a critical requirement in any useful photonic quantum computer.

\begin{figure*}
\centering
\includegraphics[width=0.8\textwidth]{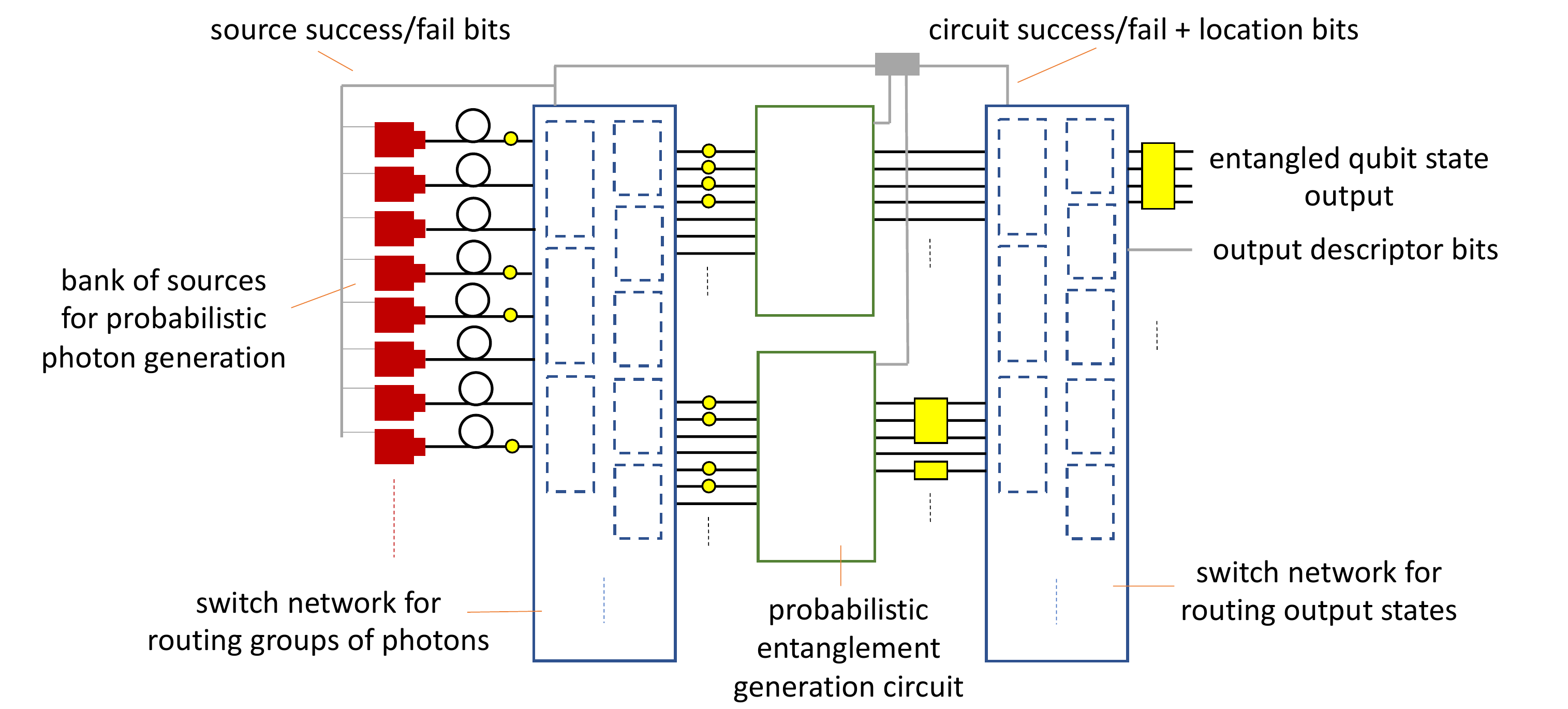}
\caption{\label{fig:intro_spatial_muxes}  {\it Spatial mux schemes.} A large number of strategies can be employed to minimize resource requirements for generating small states of entangled qubits, such as illustrated with mux stages after single-photon and entanglement generation.  Some options explored in Sec.~\ref{sec:spatial_muxes} include exploiting the fact that linear-optical circuits for generating entanglement can use photons in large numbers of input patterns, using alternative circuits defined on large numbers of modes \cite{PsiEntGen} and using a common switch network to feed multiple circuits at the same time.}   
\end{figure*}

\begin{figure*}
\centering
\includegraphics[width=0.9\textwidth]{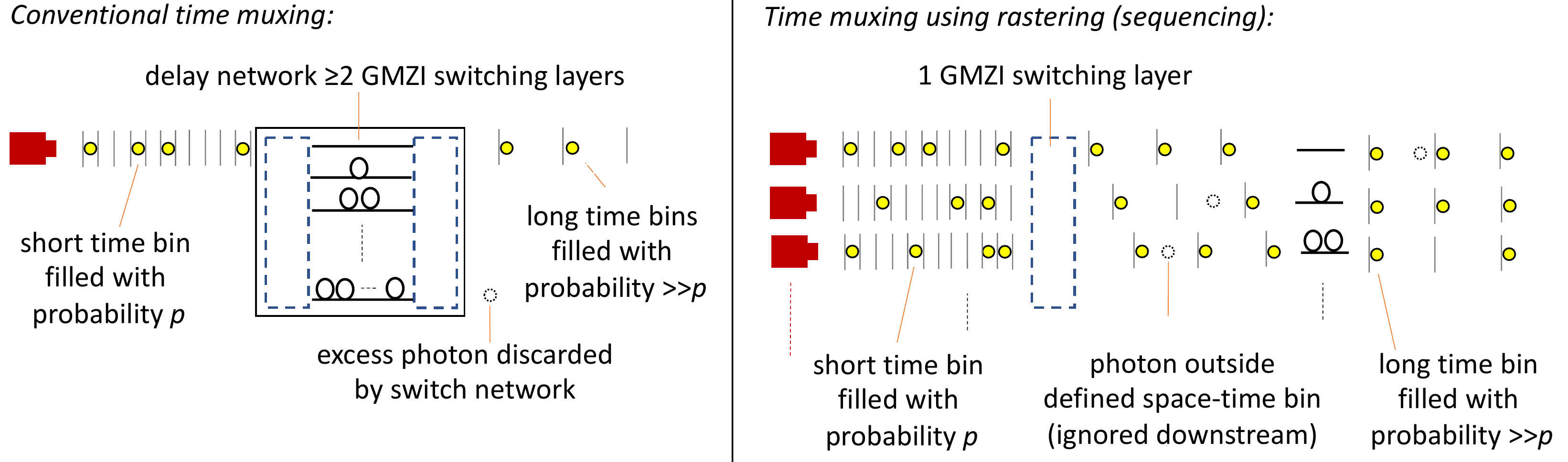}
\caption{\label{fig:intro_temporal_muxes}  {\it Temporal mux schemes.}  Commonly the type of time muxing found in the literature is based on delay networks (such as binary delay networks) as illustrated on the left.  However, many other forms of time muxing are possible which can offer significant improvement.  The example of time mux illustrated on the right uses a single GMZI whereas standard delay networks necessarily have switch depth $\geq2$.  The alternative scheme uses sequencing of operations to fill output space-time bins with high-probability; photons which cannot be used are directed to space-time bins which are unused downstream of the mux, and detections of the photons in those space-time bins can be ignored by electronic gating.}   
\end{figure*}

{\bf Overview of results.} In this paper, we present schemes for spatial and temporal muxing based on networks of (Generalized) Mach-Zehnder interferometers (GMZIs), delay networks and electronic controls which implement routing logic.  The key motivations for these schemes are simplifying hardware requirements for common tasks, minimizing switch depth, and high efficiency --- which is to say ability to successfully route most or all possible inputs (which we quantify in terms of yield as explained in Sec.~\ref{sec:Metrics}). Our key results are:
\begin{enumerate}
	
\item {\it Efficient spatial and temporal ``$N$-to-$M$'' (i.e. multiple input to multiple output) switching schemes with only two active phase-shifter layers}: We present a number of practical schemes that enable the supply of groups of photons to multiple circuits simultaneously while only using two layers of active phase shifters without requiring all-to-all connectivity. 

\item {\it Exhaustive derivation of configurations of networks using a single layer of switching to maximize number of useful routing operations:} A comprehensive analysis of GMZI permutation operations is given in Sec.~\ref{sec:Commuting}, and we show how qubit operations can be integrated into muxing stages without extra switches (see Sec.~\ref{sec:Multimultiplexing}).
	
\item {\it Proof that GMZIs can switch non-classical input and/or output light in more ways than any standard ``mode-permutation'' switch of classical light:} We show how a GMZI can implement operations other than permutations (see Sec.~\ref{sec:AltGMZI}), a hitherto unappreciated functionality. One application is incorporating qubit time-encoders into the single-photon muxing commonly used with a Bell-state generator (see Sec.~\ref{sec:rastermux}).
	
\item {\it  Efficient strategies for muxing groups of outputs (illustrated in Fig.~\ref{fig:intro_spatial_muxes}):} In Sec.~\ref{sec:spatial_muxes} we explore ideas to exploit the large number of input patterns that can be used by typical entanglement-generation circuits --- a na{\"i}ve $m\times N/m$-to-$1$ strategy, i.e. using $m$ copies of a multiple input to single output mux, uses only one such pattern.  We find that: it is as efficient to generate Bell states in random modes, using only blocking switches to dump excess photons, as it is to use initial muxing of single photons; remarkably, a ballistic strategy without any initial switching increases the number of required sources only by a factor $\approx\!3\times$; and, a single (double) layer of MZIs can rearrange entangled qubits in random modes to pre-assigned mode bundles with $>70\%$ ($100\%$) efficiency (see Sec.~\ref{sec:RandomInput}). 
On the other hand, we demonstrate how low-depth networks of small switches (i.e. MZIs and three-mode GMZIs) can be added before standard $m\times$ $N/m$-to-$1$ muxes to make them optimally efficient (see Sec.~\ref{sec:hugmux}).  We also show how single-photon muxes that supply groups of photons to several entanglement-generation circuits simultaneously can achieve high yield across a very broad range of values for input probability $p$, and for one example we show a minimum improvement of mux yield of $2.5\times$ and considerably greater across a range of $p$ (see Sec.~\ref{sec:bnmux}). 
	
\item {\it Greatly enhancing the capabilities of spatial switch networks by the judicious use of temporal delays and sequencing (illustrated in Fig.~\ref{fig:intro_temporal_muxes}):} In Sec.~\ref{sec:temporal_muxes} we explore novel forms of time muxing including: a general ``rastering'' technique which is useful when the number of components on a hardware module (e.g. a bank of sources or a switching chip) does not match what is required for efficient muxing, e.g. so one single-photon mux can be used in place of four (or six) muxes to prepare groups of photons (see Sec.~\ref{sec:rastermux}); a scheme using a pair of $N$-mode GMZIs to enable arbitrary reordering of inputs --- a na{\"i}ve cross-bar topology \cite{Lee19} would use $N^2$ active devices versus $2N$ here (see Sec.~\ref{sec:rastering_permutation_networks}); and, using delay networks based on de Bruijn sequences \cite{deBruijn46} to align random distributions of photons across multiple modes and time bins (see Sec.~\ref{sec:de_Bruijn}).
	
\end{enumerate}

We note that our work comes in the context of a great amount of theoretical and experimental work on muxing over recent years, especially for muxed single-photon sources \cite{Meyer-Scott2020}.   Sec.~\ref{sec:SwNetReview} provides an overview of common muxing concepts and background for our new results in the later sections.  

The techniques we introduce have applications to other photonic quantum technologies and mostly obviously quantum communication. Long-distance communication of quantum information through quantum networks relies on the use of quantum repeaters, one example implementation of such devices being all-optical quantum repeaters \cite{Azuma15, mihir-repeater, repeaters-briegel}. All-photonic quantum repeaters are based on the generation of photonic graph states, usually called repeater graphs states \cite{hilaire2021resource,chan2018optimized}, using the same linear-optical operations needed for FBQC or solid-state emitters \cite{ss-emitters-buterakos,ss-emitters-russo,ss-emitters-sophia}. Therefore many of the techniques and schemes presented in this work are of interest in this context also.

%% file: sections/review.tex
This section reviews switch network schemes found in the literature, highlighting their key properties with a particular focus on their realisation using chip-based platforms for integrated quantum photonics.  Standard components used in these platforms include waveguides, directional couplers, passive and active (fast) phase shifters, crossings, single-photon detectors and heralded single-photon sources (HSPSs) --- see for example Ref.~\cite{Silverstone16} for a review. Switch networks can be categorised according to their primary function as follows. $N$-to-$1$ ($M$) {\it muxes} map one (or multiple $M$) input(s) to designated output ports. The inputs are commonly assumed to be probabilistic and of the same type, although more complicated assumptions apply in some problems. For example, an $N$-to-$4$ photon mux extracts groups of four photons from $N$ HSPSs. Sometimes it is necessary to carefully distinguish the number of output (input) ports from the number of principal target outputs (inputs).  Most commonly, the excess ports must be populated with the vacuum state, and the switch network is required to access specific distributions (``patterns'') of the  outputs (inputs) across the ports.  Note that the description $N\times M$ is used elsewhere to denote the number of input and output ports of a switch network (rather than the number of entities actually being inputted and outputted) but we do not use this terminology to avoid ambiguity.  We refer to switch networks as {\it permutation networks} when their primary purpose is to rearrange (subsets of) inputs, where the inputs should generally be regarded as inequivalent.  Furthermore, switch networks are also classified on the basis of the photonic degree of freedom distinguishing their inputs. Schemes based on space and time are the most common, but the use of frequency \cite{GrimauPuigibert17, Joshi18, Hiemstra20}, orbital angular momentum \cite{Liu19} and combinations of multiple degrees of freedom have also been proposed~\cite{miller2013selfconfiguring}.

\subsection{Building blocks}
\begin{figure}
	\center
	\includegraphics[width=0.45\textwidth]{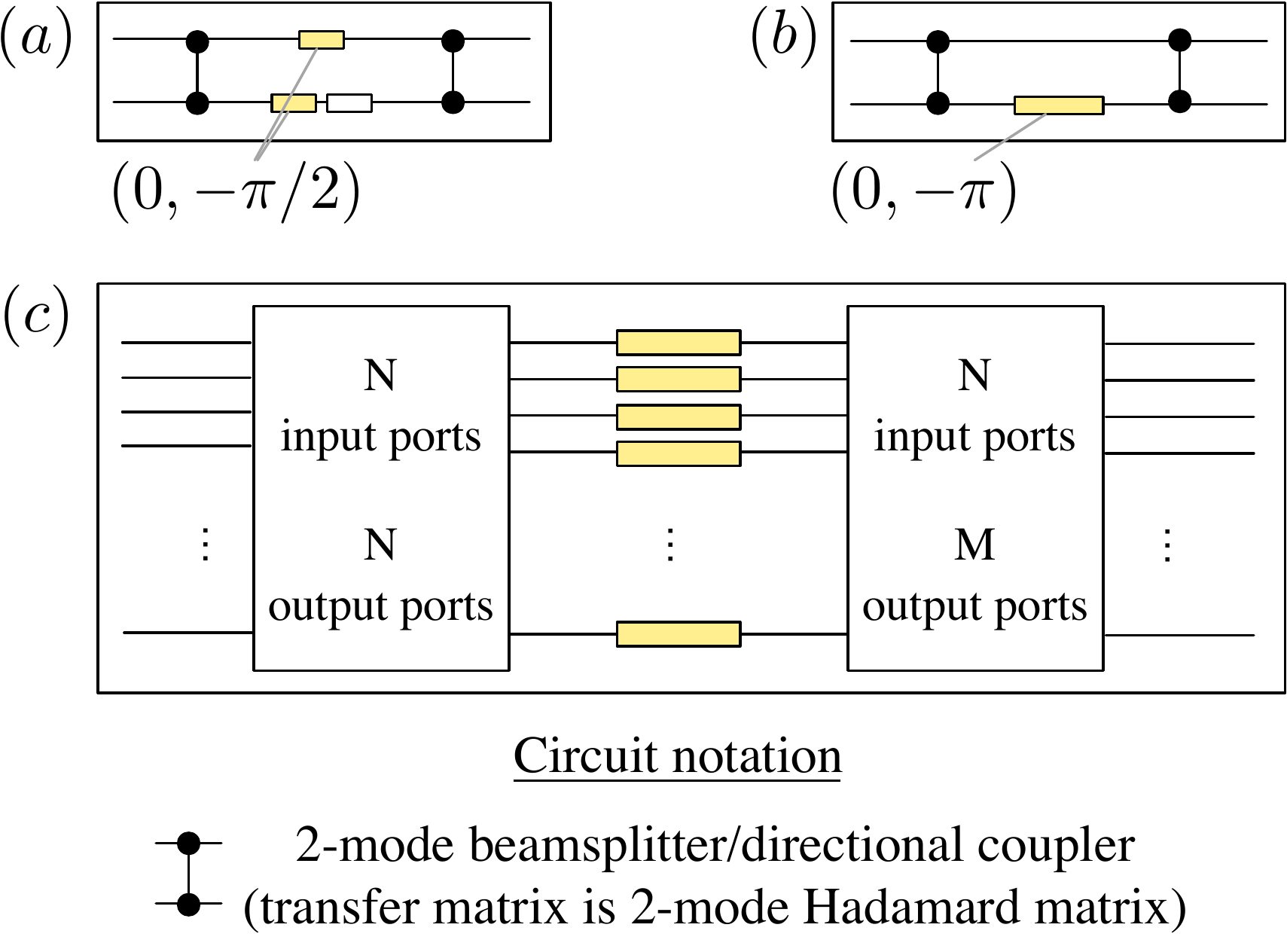}
	\caption{{\it Building blocks of composite switch networks.} (a, b) $2$-to-$2$ MZIs implement identity or swap operations on the inputs. The circuits consist of two directional couplers with an active phase shifter (yellow) on one or both arms between them. The push-pull configuration (a) also has a fixed passive $-\pi/2$ phase shift (white) on one arm and selects between the two operations by setting the top or bottom active phase to $-\pi/2$. Configuration (b) uses a 0 or $-\pi$ active phase to select the operation. (c) $N$-to-$M$ GMZI made of two passive balanced splitter networks (white) and a layer of $N$ active phase shifters (yellow)\cite{Lagali00}. Varying the settings of the active phases selects specific permutations of the $N$ inputs and routes them to $M>1$ output ports.\label{fig:buildingblocks}}       
\end{figure}

Mach-Zehnder Interferometers (MZIs) \cite{Lahiri16} are networks that implement identity or swap operations on two inputs.  For example, to switch between transfer matrices which are pairs of Pauli operations using active phase shifters, we can note that 
\begin{eqnarray}
\label{eq:MZIops}
I        \,{\rm or }\, X &=& 
h (I     \,{\rm or }\, Z ) h = 
S h_c (Z \,{\rm or }\, I) h_c S \nonumber \\ 
I        \,{\rm or }\, Y &=& 
S h (I   \,{\rm or }\, Z) h S^\dagger = 
Z h_c (Z \,{\rm or }\, I) h_c,
\end{eqnarray}
where 
$h = \left( \begin{smallmatrix}
1 & 1 \\ 1 & -1 
\end{smallmatrix} \right)/\sqrt{2}$, 
$h_c = \left( \begin{smallmatrix}
1 & -i \\ -i & 1 
\end{smallmatrix} \right)/\sqrt{2}$,
$S = \left( 
\begin{smallmatrix}
1 & 0 \\ 0 & i \\
\end{smallmatrix} 
\right)$, 
and 
$X = \left( 
\begin{smallmatrix}
0 & 1 \\ 1 & 0 \\
\end{smallmatrix} 
\right)$,
$Y = \left( 
\begin{smallmatrix}
0 & -i \\ i & 0 \\
\end{smallmatrix} 
\right)$, $Z = \left( 
\begin{smallmatrix}
1 & 0 \\ 0 & -1 \\
\end{smallmatrix} 
\right)$ are the Pauli matrices.  Two possible realisations of this type of circuit are shown in Fig.~\ref{fig:buildingblocks}(a) and (b), for which the active part generates -$iI/Z$ and $I/Z$ respectively. Many switch network architectures are built by connecting multiple MZIs to form various topologies.

The Generalised Mach-Zehnder Interferometer (GMZI) \cite{Ulrich78} is an extension of an MZI with $N>2$ inputs and $M\geq1$ outputs, shown in Fig.~\ref{fig:buildingblocks}(c). This configuration allows a set of permutations to be performed on the inputs, as discussed in detail in Sec. \ref{sec:SwitchingTheory}, making this device a powerful block for the construction of composite $N$-to-1 and $N$-to-$M$ switch networks.

\subsection{$N$-to-1 switch networks}
 
\begin{figure}
	\center
	\includegraphics[width=0.9\columnwidth]{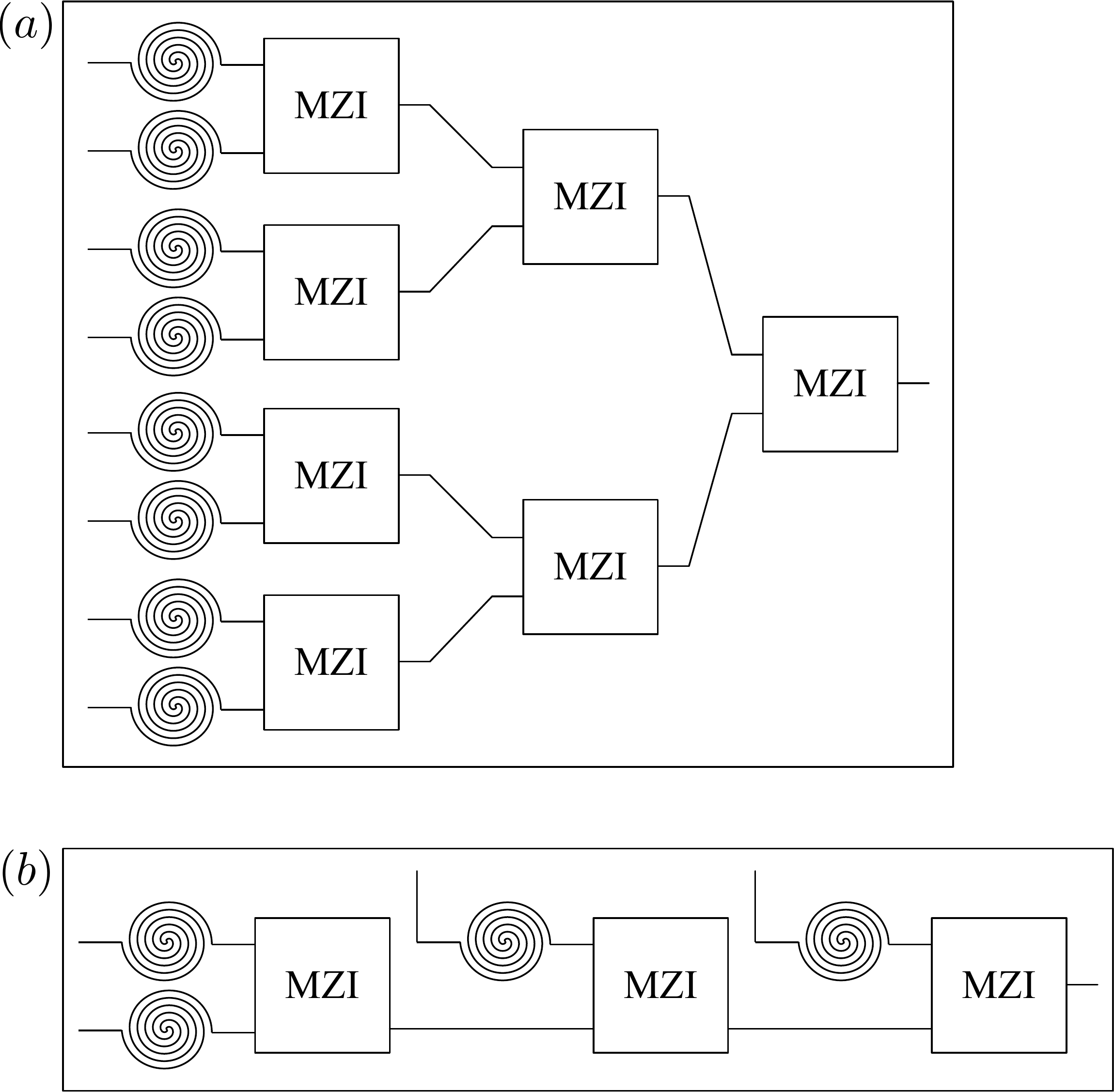}
	\caption{{\it Spatial $N$-to-1 muxes with inputs at $N$ spatially-distinct locations (ports).} (a) Log-tree mux ($N=8$ example). $2$-to-$1$ MZIs form a tree structure with $2\left(2^{\lceil\log_2(N)\rceil}-1\right)$ active phase shifters arranged in $\lceil\log_2(N)\rceil$ layers. (b) Chain mux ($N=4$ example). $(N-1)$ MZIs are connected through one output and input to form a line. The active phase shifter count is the same as for the log-tree, but the depth varies between 1 and $(N-1)$.\label{fig:nx1spatial}}       
\end{figure}

There are a number of spatial mux schemes that select one of multiple inputs from distinct locations in space.  A simple $N$-to-$1$ GMZI can be used as a mux, since it allows routing of any input to a single output port (see Sec.~\ref{sec:SwitchingTheory}). The main advantages of this scheme are its low constant active phase shifter depth (1) and count ($N$). However, the total propagation distance and the number of waveguide crossings increase rapidly with $N$. This downside of the monolithic GMZI structure is obviated by constructing composite switch networks of $2$-to-1 MZIs, at the cost of increasing the component depth and count. There are two common $N$-to-1 schemes of this kind, both of which can be built with no crossings. In a ``log-tree'' \cite{Shapiro07, Ma11, FrancisJones16}, the MZIs form a converging symmetric tree of degree two, where the chosen input is routed from one of the leaves to the root, as shown in Fig.~\ref{fig:nx1spatial}(a). An asymmetric variant of this scheme \cite{Mazzarella13}, known as a ``chain'', consists of MZIs cascaded to form a linear topology in which each block selects either the output of the previous block or the new input, as shown in Fig.~\ref{fig:nx1spatial}(b). For the chain scheme, the depth of the network traversed by the output depends on the chosen input, which can worsen the interference of resources from different chains, due to imbalanced losses and errors. The switching logic of this scheme presents an interesting advantage: while being very simple and entirely local to each individual MZI, it minimizes the amount of error by selecting the input available closest to the output. Analysis of these three schemes in the context of single-photon muxing \cite{Bonneau15} showed that all three architectures require components of similar high performance to achieve a muxing efficiency high enough for use in linear-optical quantum computing.

\begin{figure}
	\center
	\includegraphics[width=\columnwidth]{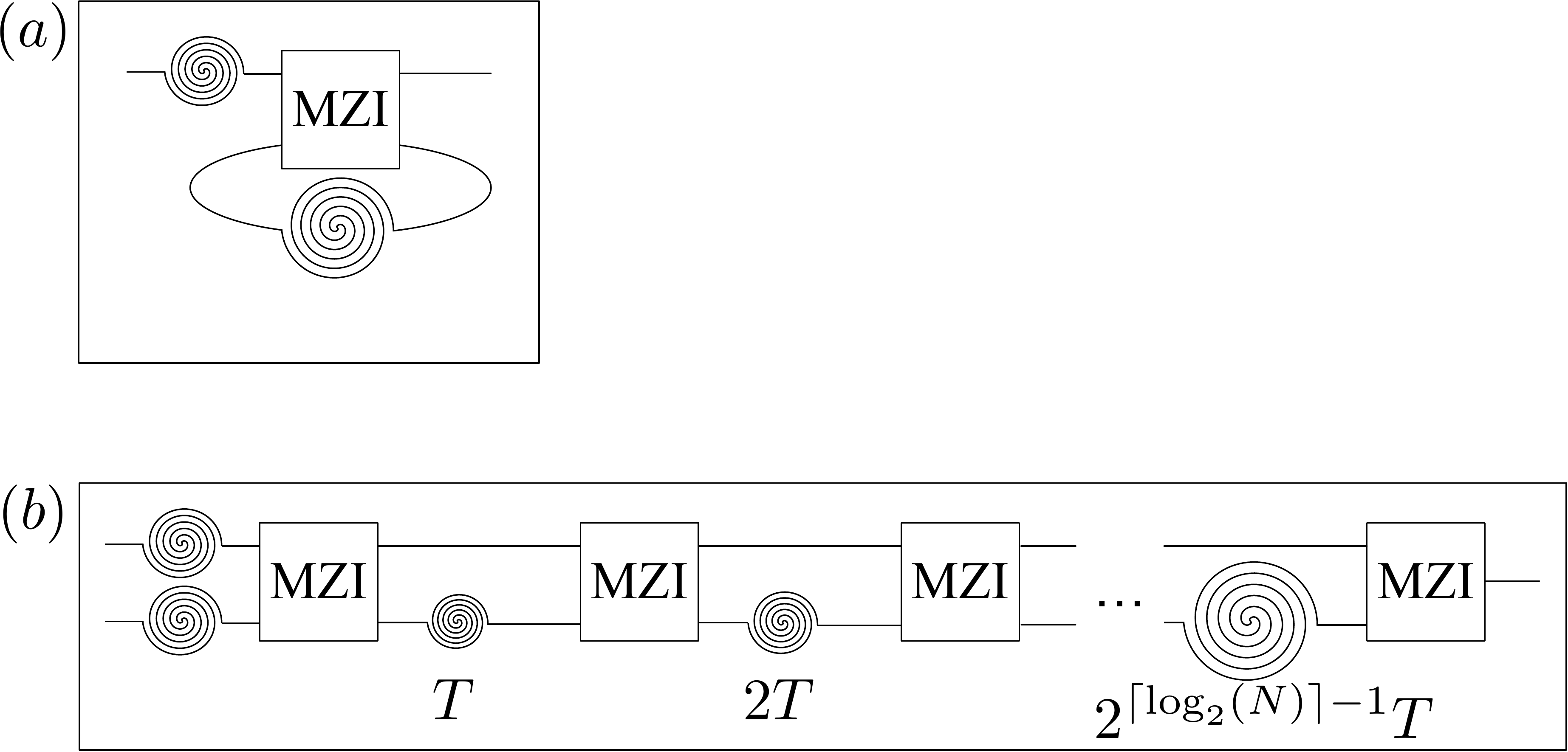}
	\caption{{\it $N$-to-$1$ temporal muxes with inputs in $N$ distinct time bins.} (a) Storage loop scheme (time chain). A MZI receives one resource per time bin $T$ and routes it to a storage device (a delay line here) or discards it. After $N$ time bins, the chosen input is output. The number of active phase shifters in the path of the chosen input varies between 1 and $N$. (b) Binary delay network (time log-tree). The scheme comprises a series of $\lceil\log_2(N)\rceil+1$ MZIs with delays of lengths $2^nT$ between them, where $T$ is the duration of a time bin at the input and $n=0,\ldots\lceil\log_2(N)\rceil-1$. The active phase shifter depth scales as with the number of input time bins as $\lceil\log_2(N)\rceil$.\label{fig:nx1temporal}}       
\end{figure}

In temporal muxing, resources are input at the same spatial location but at different times, and the aim is to produce an output in a specific time bin. This requires networks with fewer components, but the output time bins become longer. There are two main kinds of temporal schemes: designs with storage devices such as cavities or fiber loops \cite{Pittman02, Jeffrey04, Glebov13, Rohde15, Kaneda15, Hoggarth17, Kaneda19}, and designs based on networks of delays \cite{Mower11, Broome11, Schmiegelow14, Xiong16, Magnoni19}. The former simply consist of a storage device and a single MZI used to  choose whether to store or output each input, as shown in Fig.~\ref{fig:nx1temporal}(a). This can be thought of as the temporal version of a chain mux, and it presents the same advantage in terms of switching logic. The log-tree also has a temporal equivalent known as a ``binary-division delay network''. This scheme consists of a series of MZIs with delays of different lengths between them, as illustrated in Fig.~\ref{fig:nx1temporal}(b).

\begin{figure*}
	\includegraphics[width=\textwidth]{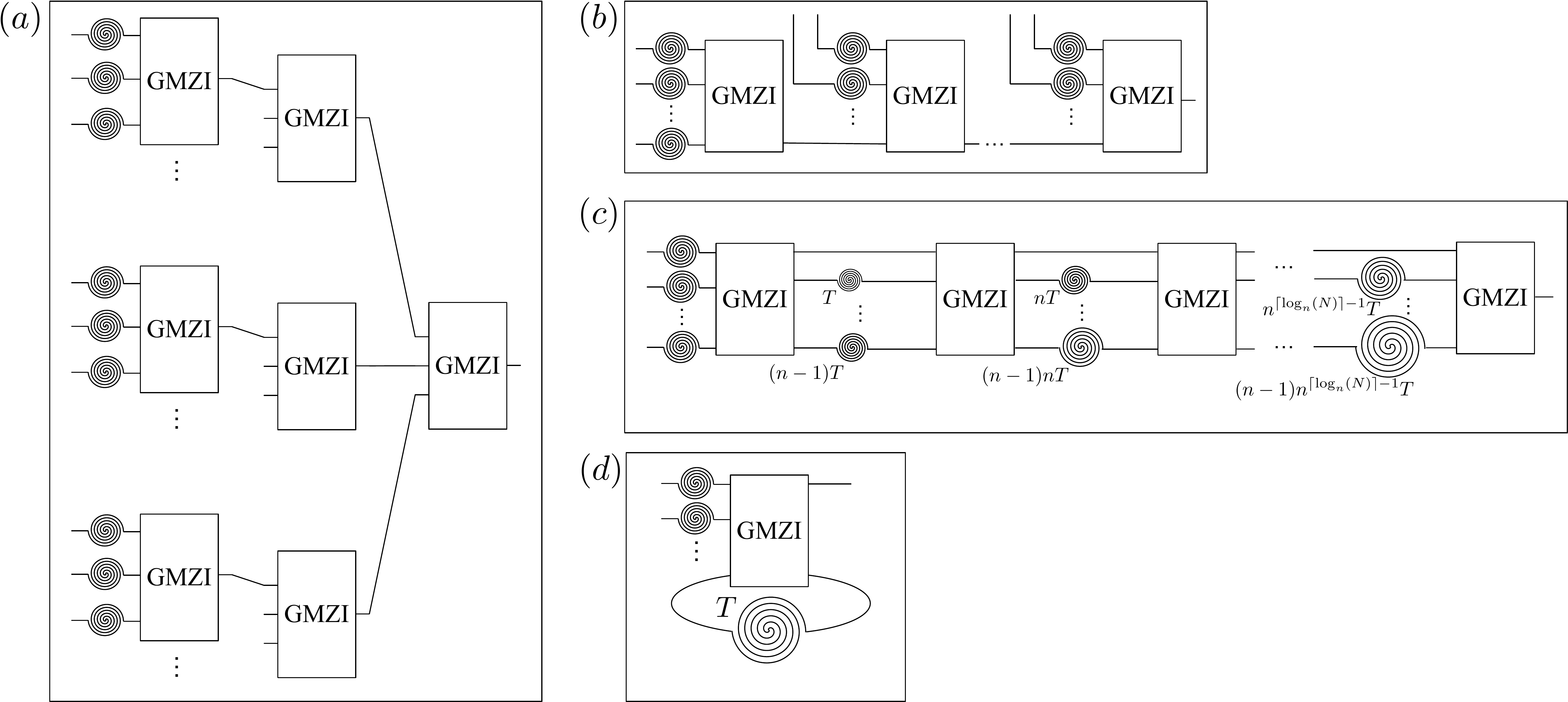}
    \caption{{\it $N$-to-1 composite muxes using $n$-to-$1$ GMZI sub-blocks.} (a) Generalised spatial log-tree. The degree of the tree is $n$ and the depth is $\lceil\log_n(N)\rceil$ ($n=3$ shown with some subblocks omitted). (b) Generalised spatial chain. Stages beyond the first take $n-1$ inputs, and the network depth varies between 1 and $\lceil (N-1)/(n-1)\rceil$. (c) Generalised delay network. The subblocks enclose $\lceil\log_n(N)\rceil$ layers of $n-1$ delays with durations $n^i T, \ldots, (n-1)n^i T$, where layer index $i=0, \ldots, \lceil\log_n(N)\rceil-1$. The number of active phase shifters traversed is $\lceil\log_n(N)\rceil+1$. (d) Generalised storage loop. $n-1$ inputs enter in every time bin and the selected output exits after $\lceil N/(n-1)\rceil$ time bins. \label{fig:nx1general}}   	
\end{figure*}

The topologies described above can be generalised by replacing each MZI with a GMZI with $n$ inputs, as shown in Fig.~\ref{fig:nx1general}. This introduces a trade-off between the active phase shifter depth and count, which decreases with $n$, and the number of waveguide crossings and propagation distance within each block, which increases with $n$. In addition, this modification turns temporal schemes into hybrid networks, where multiple spatially-distinct resources are input in each time bin. The trade-offs introduced by the parameter $n$ can be exploited to optimize the structure of these schemes for different regimes of physical error rates.

Hybrid schemes that combine muxing based on more than one degree of freedom are also found in the literature. Schemes consisting of time-multiplexed resources input to spatial networks have been proposed and analysed in Refs.~\cite{Latypov15, Bodog16}, and a specific implementation of the reverse has been realised experimentally in Ref.~\cite{Mendoza16}. It has also been suggested that the performance of temporal muxing networks can be improved by exploiting the frequency degree of freedom of single photons \cite{Heuck18}.

In applications such as FBQC, which rely on the interference of multiplexed resources, muxing is used to produce synchronised outputs. All the schemes described so far achieve this by having a single predetermined output spatio-temporal bin. However, when large output probabilities are needed this leads to a large waste of resources, which can be understood as follows. The number of available resources for a network of size $N$ follows a binomial distribution with average value $\bar{N}=Np$, where $p$ is the probability of an input being populated. The probability of a network successfully producing an output is then $p_{\rm mux}=1-(1-p)^N$. For the typical situation with large $N$ and small $p$ values, the binomial distribution is well approximated by a Poissonian distribution, and so $p_{\rm mux}\simeq 1- e^{-Np}$. It follows that the average number of inputs scales as $Np = -\ln(1-p_{\rm mux})$, and so the number of available resources that are not used grows rapidly as $p_{\rm mux}$ approaches 1. An alternative approach that leads to major efficiency improvements is ``relative muxing'', introduced in Ref.~\cite{Gimeno_Segovia17}. Rather than routing resources to single pre-allocated outputs, this technique uses spatial or temporal log-tree networks to synchronise selected inputs in variable space-time locations, chosen depending on the resources available at any particular instant.

\subsection{$N$-to-$M$ switch networks}

$N$-to-$M$ muxes address the inefficiency of $N$-to-1 schemes by routing multiple inputs ($M$) to the output ports simultaneously. Designs for optical switch networks of this kind have been studied and used for decades, mainly in the field of telecommunications \cite{Hunter06}. Their implementation as integrated photonic circuits is an active field of research, and component-level performance improvements are still necessary to make them a viable technology \cite{Stabile16, Lee19}. Although most of the work on these schemes is motivated by the requirements of classical communication technologies, there is overlap with some metrics of interest for quantum applications, such as the accessible permutations, and so classical designs provide a useful starting point.

$N$-to-$M$ schemes in the literature are generally based on the spatial degree of freedom. The simplest of these is a GMZI with more than one output, which has the appealing feature of a single layer of $N$ active phase shifters. However, it only gives access to $N$ permutations, and therefore to limited combinations of inputs. Consequently, the $N$ input $M$ output GMZI is more useful when used as a permutation network or as a building block for larger schemes. More flexible routing is achieved by using smaller networks to build composite topologies, known as ``switch fabrics''. However, the component depth and count and the size of the crossing networks of these schemes tend to be large, and these downsides trade against each other, making the networks impractical for use in the field of quantum applications.

\begin{figure}
	\center
	\includegraphics[width=0.9\columnwidth]{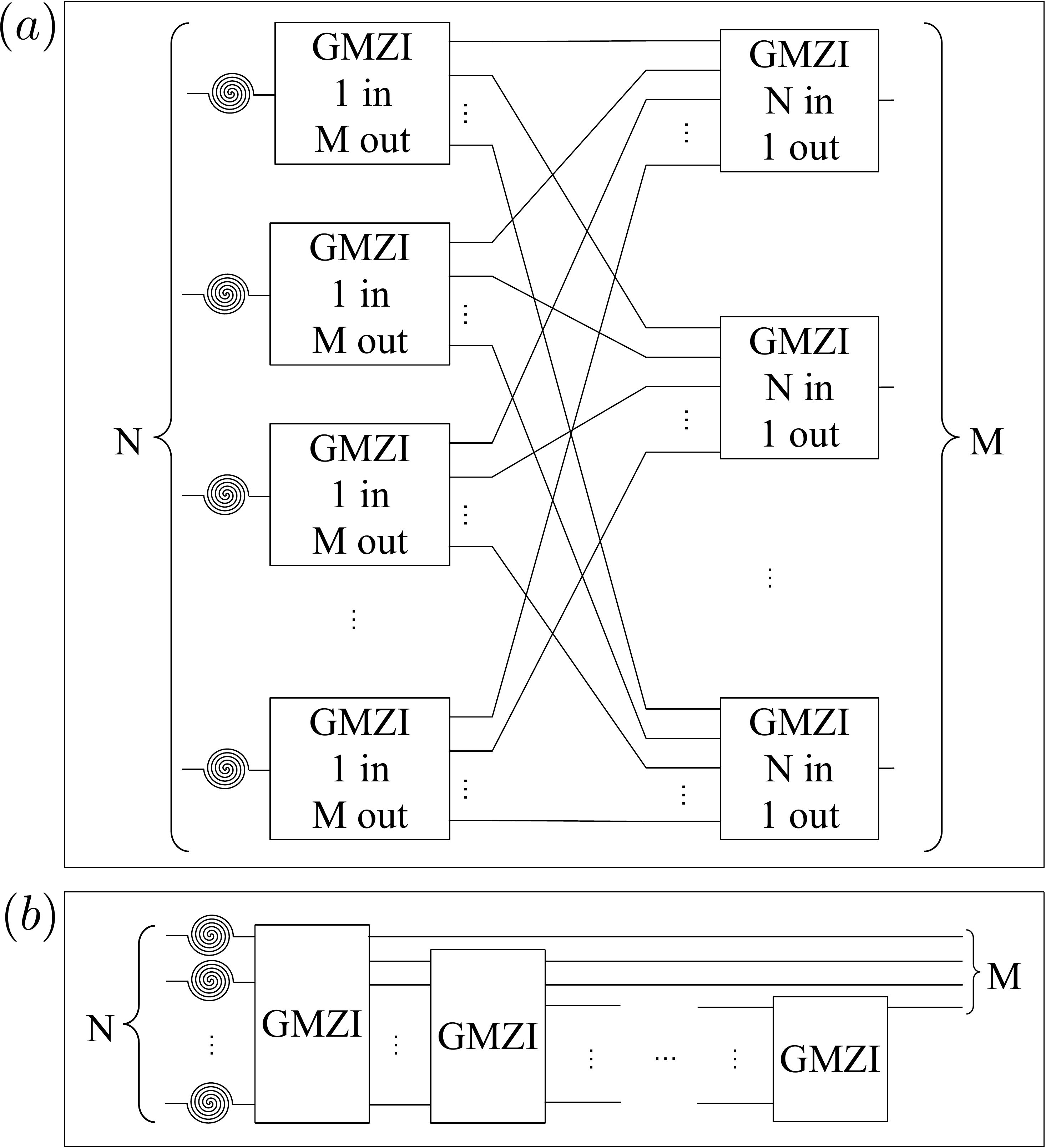}
	\caption{{\it Notable examples of $N$-to-$M$ switch networks enabling arbitrary routing.} (a) Spanke network using two layers of interconnected GMZIs.  The active phase-shifter depth is just two, but the numbers of active phase shifters and crossings scale as $\mathcal{O}(NM)$ posing challenges for large network sizes. (b) Concatenated GMZIs with progressively fewer outputs. No complex crossing networks are required between the GMZIs, but the $\mathcal{O}(NM)$ active phase shifter count and variable depth up to $M$ limit the feasible network size.\label{fig:nxm}}       
\end{figure}

As an example, Spanke's tree network \cite{Spanke86}, shown in Fig.~\ref{fig:nxm}(a), allows arbitrary rerouting of the inputs with a constant active switch depth of two, at the cost of a large number of active phase shifters and waveguide crossings. However, the number of active phase shifters and waveguide crossings scales as $O(NM)$. On the other hand, the scheme shown in Fig.~\ref{fig:nxm}(b) avoids large crossing networks, but has an active phase shifter count $O(NM)$ and depth that varies between 1 and $M$, resulting in variable error rates on the outputs \cite{Lagali00}.

For quantum applications, where low error rates are required, $N$-to-$M$ muxes need to be simplified to reduce the number of active phase shifters, both in total and along the path to the output, as well as the complexity of the crossing networks. The routing algorithms associated with these networks also need to be simplified, to avoid the need for unfeasibly long delays for the inputs. The complexity of the logic is largely determined by its generality, so restricting the operation of the networks to specific tasks is helpful to reduce processing times. These provide guiding principles for the design of the new schemes presented in the rest of this paper.

%% file: sections/gmzis.tex
\label{sec:GMZIs}

A general switch network implements a set of unitary transfer matrices $\{U_k\}$, where each unitary routes light between a subset of input and output ports.  If $U_k$ routes light from port $t$ to port $s$, then its $s^{\text th}$ row and $t^{\text th}$ column must be zero apart from $\vert U_{s,t} \vert = 1$, and similarly for other pairings of input and output ports.  
The aim of this section is to elucidate the sets of routing operations that are achievable using the simplest form of a many-mode switching network, which is to say one corresponding to transfer matrices $U_k = W D_k V^\dagger$, where the unitary matrices $W$, $V^\dagger$ describe passive interferometers, and the $D_k$ form a set of diagonal phase matrices.  The phase matrices are implemented physically using a single layer of fast phase shifters acting on every mode, and for simplicity, we will write $D$ in terms of a phase vector $\mathbf d$, $D_{s,t}=d_s\delta_{s,t}$.  The discussion below provides a comprehensive treatment of these switch networks and presents several new constructions.

\subsection{\label{sec:Commuting} Theory of Generalized Mach-Zehnder interferometers}

An important class of switch networks is obtained by considering sets of permutation matrices $\{U_k=W D_k V^\dagger\}$. By adding the \emph{fixed} passive network corresponding to e.g. $U_1^{-1}$ (so, the inverse of an arbitrary permutation from that set), we obtain a new set $\{U_k U_1^{-1}\}=\{W D'_k W^\dagger\}$ of pairwise commuting permutation matrices. So it makes sense to restrict the discussion to the case where the $\{U_k\}$ are commuting.
Switch networks of this type were introduced in Sec.~\ref{sec:SwNetReview} as ``generalized Mach-Zehnder interferometers'' (GMZIs), and were previously studied in works including \cite{Lagali00, Bonneau15}.  Here we need a more precise definition for GMZIs, and we will define them as switch networks having the following specific properties:

\begin{enumerate}[(i)]

\item  $\{U_k =W D_k W^\dagger\}$ is a set of transfer matrices corresponding to commuting permutations of $N$ modes.  The entries of $D_k$ are given by roots of unity (up to an overall global phase factor $e^{i\phi_k}$ which can be chosen at will). 
	
\item  The GMZI switch setting $D_k$ routes light from input port $1$ to output port $k$.
	
\end{enumerate}

From these properties it is possible to prove that \textit{the GMZI must have exactly $N$ settings, and that for any choice of input and output port, there is exactly one setting which routes light between the ports}.

From a mathematical standpoint, the set of operations implemented by a GMZI on $N$ modes forms an abelian group of order $N$.  This fact is very helpful here as it allows us to characterize the entire family of GMZIs defined by (i), (ii) using well-known results from group theory (namely the basis theorem for finite abelian groups \cite{GroupTheory}).  In particular, for any GMZI, $\{U_k\}$ must be isomorphic to a direct sum of cyclic groups, where the order of each of the cyclic groups is a power of a prime number.

To be more concrete, we define groups of commuting permutations  $\mathcal{G} ([n_1,n_2, \cdots, n_r])$ generated by matrices $C^{(n_1)}\otimes I^{(n_2)} \otimes I^{(n_3)}\cdots, I^{(n_1)}\otimes C^{(n_2)} \otimes I^{(n_3)} \cdots,  I^{(n_1)}\otimes I^{(n_2)} \otimes C^{(n_3)} \cdots$ ,  where $\bigl(C^{(n)}\bigr)_{i,j}=\delta_{i,(j + 1 \text{ mod } n)}$ is a cyclic permutation matrix of size $n$, $I^{(n_l)}$ is the $n_l\times n_l$ identity matrix, $\otimes$ is the Kronecker product on matrices \footnote{The Kronecker product here acts at the level of linear-optical transfer matrices and should not be confused with tensor product operations on quantum state spaces.}, and the group operation is matrix multiplication.  Then, any GMZI on $N$ modes, satisfying properties (i), (ii) above, must implement a set of permutation operations which corresponds to one of the possibilities for $\mathcal{G} ([n_1,n_2, \cdots, n_r])$ with $N=\Pi_{l=1}^{r} n_l$ (up to fixed mode permutations at the input and output).

The different types of GMZIs of fixed size can now be determined using the fact that $\mathcal{G} ([n_1,n_2])$ and $\mathcal{G} ([n_1n_2])$ are isomorphic if and only if $n_1$ and $n_2$ are coprime (see Ref.~\cite{GroupTheory} for details).  For example, for $N=8$, we can identify three fundamentally different types of GMZI:

\begin{itemize}
	
\item $\mathcal{G} ([2,2,2])$, permutations are generated by Pauli matrices $X\otimes I^{(2)}\otimes I^{(2)}$, $I^{(2)}\otimes X\otimes I^{(2)}$, $I^{(2)}\otimes I^{(2)}\otimes X$.
		
\item $\{\mathcal{G} ([4,2])\}$, permutations are generated by matrices

$C^{(4)}\otimes I^{(2)}$ where		
$C^{(4)}=\begin{pmatrix} & & & 1 \\ 1 & & & \\ & 1 & & \\ & & 1 & \end{pmatrix}$, and $I^{(4)}\otimes X$
		
\item ${\mathcal{G} ([8])}$, permutations are generated by matrix
		
$C^{(8)}=
\begin{pmatrix} & & & & & & & 1 \\ 1 & & & & & & & \\ & 1 & & & & & & \\ & & 1 & & & & & \\ & & & 1 & & & & \\ & & & & 1 & & & \\ & & & & & 1 & & \\ & & & & & & 1 & \\\end{pmatrix}$.
		
\end{itemize}

We refer to GMZIs implementing $\mathcal{G} ([2,2,\ldots,2])$, i.e. permutations of the form of swaps on subsets of modes, as ``Hadamard-type'' GMZIs due to the type of passive interferometer which is used (explained below).  Similarly, we refer to GMZIs implementing $\mathcal{G} ([N])$ as ``discrete-Fourier-transform (DFT)-type''.

The discussion above characterizes the routing power of linear-optical circuits using one-layer of fast phase shifters in the switch network.  In particular, a GMZI on $N$ modes is limited to $N$ routing operations, which is obviously small compared to the $N!$ possible mode rearrangement operations.  However, the possibility of implementing different sets of permutation operations is exploited by some of designs for spatial and temporal muxes which are discussed in Sec.~\ref{sec:NewSpatialMUXes} and Sec.~\ref{sec:NewTemporalMUXes}.   Strictly speaking the limitation to $N$ operations originates in property (ii) above -- i.e. the ability to route light from any input port to any output port.
More general constructions using a single stage of active phase shifts can be trivially obtained by acting with separate GMZIs on subsets of modes.
The resulting transfer matrices are given by the direct sum of the individual GMZIs' transfer matrices. For example, using three MZIs in parallel
results in a switch network on $6$ modes, allowing $8$ different settings. Such a construction can implement abelian groups of permutations of maximum order,
which are given in Ref.~\cite{Burns89} with the number of operations scaling to good approximation as $\sim3^{N/3}$.

\subsection{\label{sec:GMZIimplementation} Physical implementation}

We now turn to linear-optical circuits that can implement the GMZIs defined above.  In particular, a circuit that can implement the routing operations $\mathcal{G} ([n_1,n_2, \cdots, n_r])$ on $N=\Pi_{l=1}^{r} n_l$ modes must enact transfer matrices of the form,
\begin{equation}
	P_{\mathbf k} = \left(C^{(n_1)}\right)^{k_1} \otimes \left(C^{(n_2)}\right)^{k_2} \otimes \cdots \otimes \left(C^{(n_r)}\right)^{k_r},
	\label{eq:gmziperms}
\end{equation}
with settings vector $\mathbf k$ where $0\leq k_l < n_l$ with $l=1,\cdots,r$.  This can be achieved using a circuit with transfer matrices $W D_{\mathbf k} W^\dagger$ as follows:
\begin{eqnarray}
\label{eq:gmziV}
W&=&W^{(n_1)}\otimes W^{(n_2)} \otimes \cdots \otimes W^{(n_r)} 
 \\
&& \text{with }
\bigl(W^{(n_l)}\bigr)_{s,t} = \frac{\me^{\imath 2\pi st /n_l}}{\sqrt{n_l}},
\end{eqnarray}
where the $W^{(n_l)}$ are DFT matrices;
the ${\mathbf k}^{\text th}$ setting of the fast phase shifters is given by,
\begin{eqnarray}
\label{eq:gmziD}
D_{\mathbf k}&=&D_{k_1}^{(n_1)}\otimes D_{k_2}^{(n_2)} \otimes \cdots \otimes D_{k_r}^{(n_r)}, \\
&& \text{with }
\bigl({\mathbf d}_k^{(n)}\bigr)_{s}=\me^{-\imath 2\pi k s / n}
\text{ for }
D_{k}^{(n)}.
\end{eqnarray}

One route to constructing practical interferometers for $W$ and $W^\dag$ is to reduce them to networks of beam-splitter and phase-shifter components using generic unitary decompositions from Reck et al.~\cite{Reck94} or Clements et al.~\cite{Clements16}.  These decompositions have optical depth (number of optical elements encountered on the longest path through the interferometer) scaling as $2N-3$ and $N$ respectively. This means that the transmittance along the longest path will scale with an exponent which is proportional to the size parameter $N$ -- which presents a severe experimental limitation for scaling to large GMZI sizes.

\begin{figure}	
\includegraphics[width=\columnwidth]{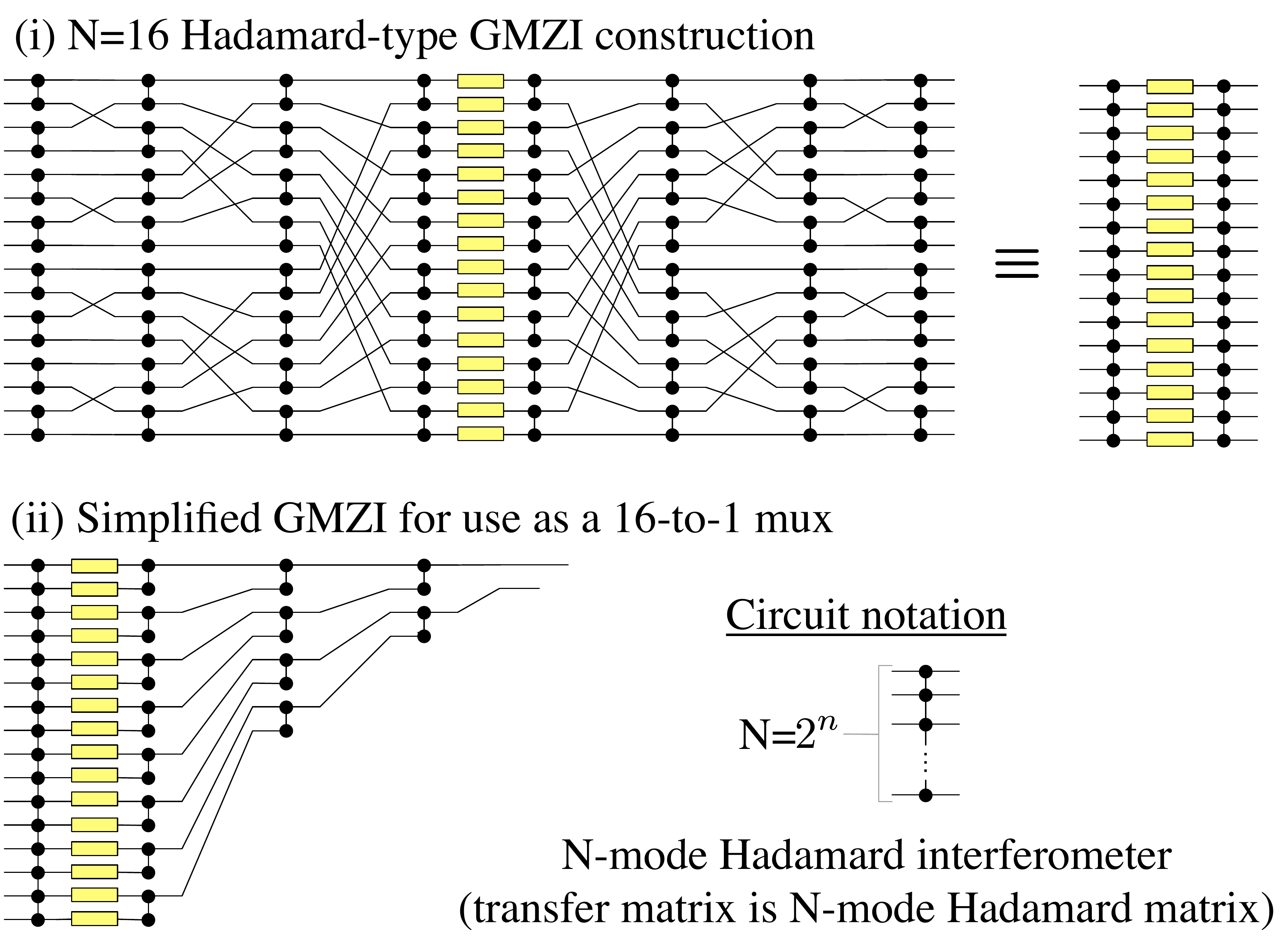}
\caption{{\it Hadamard-type GMZI constructions.} (i) illustration of a linear-optical circuit for a GMZI on $N=16$ modes, for which the fast phase shifters are set to configurations of $0$ and $\pi$ to select one of 16 operations from  $\mathcal{G} ([2,2,2,2])$; (ii) possible simplification of the circuit when only one output port is required --- as is the case when the GMZI is used as a $N$-to-$1$ mux.  The passive interferometers are constructed following the decomposition of Eq.~(\ref{eq:crossingexpansion}) with stages of interference using 50:50 beam-splitters or directional couplers on pairs of adjacent modes, separated by crossings networks.  Note that the phases in the physical interferometer generally differ from the constructions given in the main text, and this implies minor modifications for the transfer matrices and phase-shifter settings. 
\label{fig:gmziconstructions}}	
\end{figure}

GMZI networks however have a lot of special structure, and can exploit decompositions which scale to large $N$ with log-depth stages of interference (decompositions of this type were first explored theoretically in Ref.s~ \cite{Torma96,Barak07}).  The specific type of decomposition needed here is:
\begin{widetext}
\begin{eqnarray}
W
&=& \left ( W^{(n_1)}\otimes I^{(N/n_1)} \right) \left ( I^{(n_1)}\otimes W^{(n_2)} \otimes I^{(N/(n_1 n_2)} \right ) \cdots\left ( I^{(N/n_r)}\otimes W^{(n_r)}\right) \nonumber \\
&=& \left(S_{N/n_1,n_1} \, I^{(N/n_1)} \otimes W^{(n_1)} \, S_{N/n_1,n_1}^t \right)   \left(I^{(n_1)} \otimes S_{N/(n_1 n_2),n_2} \, I^{(N/n_2)} \otimes W^{(n_2)} \, I^{(n_1)} \otimes S_{N/(n_1 n_2),n_2}^t \right)\cdots \nonumber \\
&& \left(I^{(N/n_r)} \otimes W^{(n_r)}\right) \label{eq:crossingexpansion} 
\end{eqnarray}
\end{widetext}
where the matrices $S_{\cdot,\cdot}$ correspond to crossing networks which reorder modes within the interferometer.  Since  the subexpressions of the form $I^{(N/n_l)} \otimes W^{(n_l)}$  correspond to repeated blocks of modes interfering according to unitary $W^{(n_l)}$, Eq.~(\ref{eq:crossingexpansion}) can be seen to describe stages of local interference separated by crossing networks.  Note also that since the bracketed expressions in the decomposition commute  there is some freedom in the configuration of the crossing networks, and some of them can be treated as relabelings of modes rather than physical circuit elements.
Fig.~\ref{fig:gmziconstructions} illustrates the construction of a Hadamard-type GMZI using this decomposition, as well as a simplification which is possible when the GMZI is used as a $N$-to-$1$ mux.  For more general GMZI types, we note that the unitary matrices $W^{(n_l)}$ can be decomposed into elementary beam-splitter and phase-shifter operations using the generic decomposition methods mentioned above.  Alternatively, since the $W^{(n_l)}$ are assumed to be discrete Fourier transforms, they can be recursively decomposed into smaller discrete Fourier transforms acting on sets of local modes $I^{n_l/(n_l^\prime)} \otimes W^{(n_l^\prime)}$, $I^{n_l/(n_l^{\prime\prime})} \otimes W^{(n_l^{\prime\prime})}$ (for any sizes satisfying $n_l = n_l^\prime \times n_l^{\prime \prime})$ together with crossings networks and additional phase shifts~\cite{Cooley65}.  It should be noted that the optical depth of networks constructed using the recursive method is highly dependent on actual hardware implementation, as the depth of the crossing networks must be accounted for in addition to the stages of local interference (e.g. the depth in crossings of the largest crossing network scales with $N/2-1$).  

\subsection{\label{sec:Multimultiplexing}Combining switching stages using enlarged GMZIs}

One more subtle feature of the GMZI constructions that was remarked on in Sec.~\ref{sec:Commuting} is that the matrices $D_k$ for the GMZIs are determined up to a setting-dependent global phase factor $e^{i \phi_k}$.  In principle these global phases can be freely set over a range of $2\pi$ (provided the active phase shifters themselves are configured with sufficient phase range). For an application such as single-photon muxing, the global phase factors have no role in the operation of the switch network.  However, they can be useful if the switch network is applied to only some part of the input states (e.g. single rails from dual-rail qubits) or if it is incorporated in larger interferometers. In these cases, additional functionality can be absorbed into the operation of the switch network without adding extra layers of switching.

\begin{figure}	
\includegraphics[width=0.8\columnwidth]{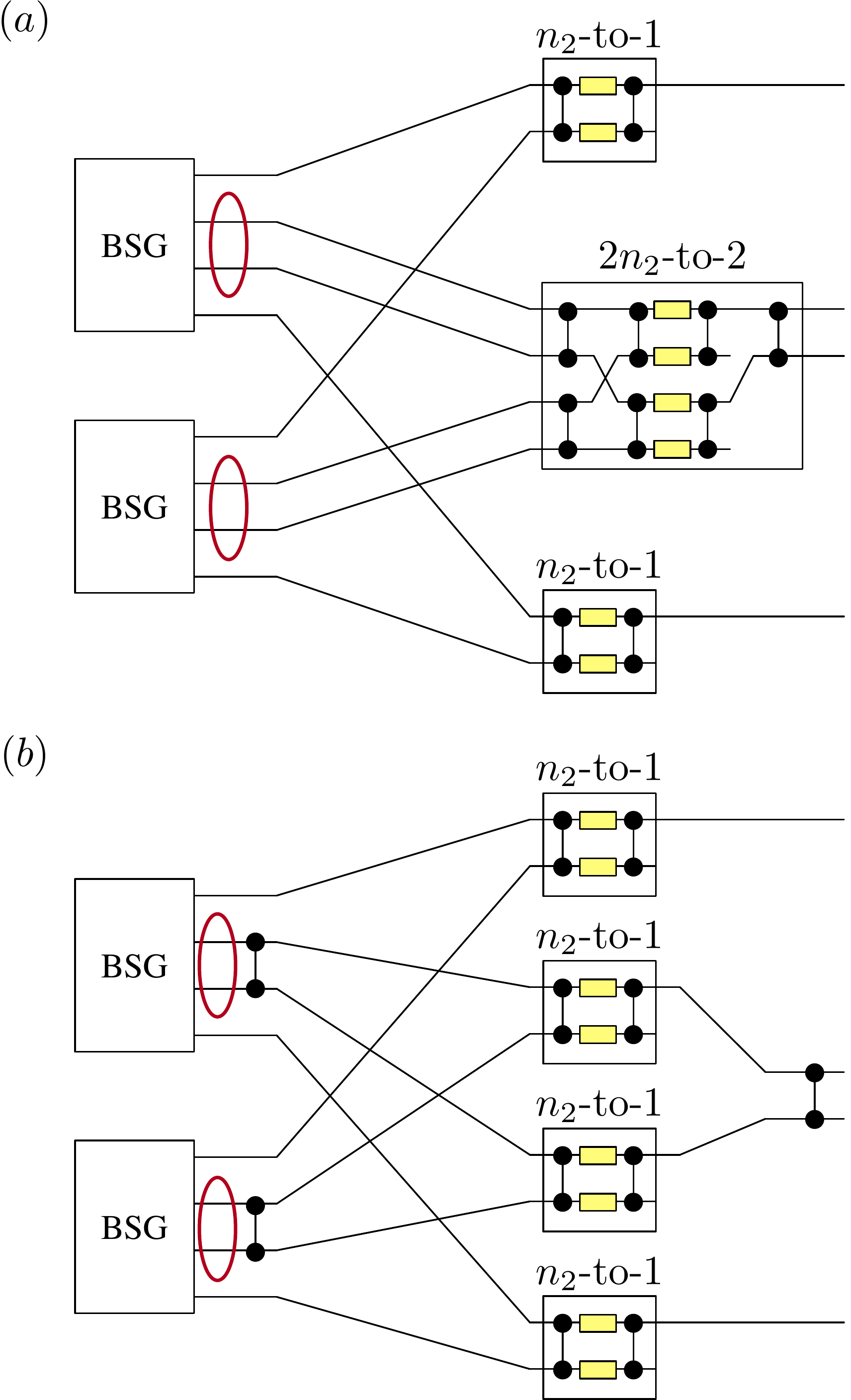}
\caption{{\it Use of a larger GMZI to implement adaptive swaps of rails while muxing Bell states generated with $n_2$ standard BSGs.} (a) Sending the two rails that might need to be swapped (circled in red) through a single GMZI of size $N=n_1n_2$ ($n_1=n_2=2$ in this diagram) allows muxing and permutation operations to be combined while avoiding the need for an additional switching stage. (b) The modular structure of the GMZI can be exploited to apply portions of the circuit at different locations and to optimize the physical implementation. In this example, the network which incorporates the swap operation can be decomposed into two $2$-to-$1$ GMZIs with extra directional couplers applied at the output of the BSGs and between the two output rails.\label{fig:bsg_multimulti}}
\end{figure}

This idea is very useful for photonic FBQC, where it is often necessary to mux some circuit which generates entangled states, whilst also applying internal adaptive corrections to its output. An example of this occurs when muxing Bell states from a standard Bell-state generator (BSG) circuit. This circuit produces a Bell state across four modes with probability $3/16$ \cite{Browne05, Zhang08}, but the Bell states do not conform to dual-rail qubit encoding (i.e. with qubits allocated to fixed pairs of modes) in a third of cases. Although this problem can be addressed using an additional MZI at the mux output to perform an optional mode-swap operation, a more elegant solution is presented in Fig.~\ref{fig:bsg_multimulti}(a,b). In this approach, a mux on $n_2$ copies of the BSG implements muxing and swap operations, using a size $N=n_1 n_2$ GMZI on $n_1=2$ inner rails from each BSG, and regular $n_2$-to-$1$ muxing for the outer rails. The ability to permute the rails increases the success probability for generating a dual-rail encoded Bell state from $1/8$ to $3/16$, and thereby decreases the amount of muxing needed to reach any particular target output probability by a factor of $\sim1.55$.

More generally from Eq.~(\ref{eq:gmziperms}), the transfer matrices associated with a GMZI that implements the routing operations $\mathcal{G} ([n_1,n_2])$ are
\begin{align}
	P_{(k_1,k_2)} &= \left(C^{(n_1)}\right)^{k_1} \otimes \left(C^{(n_2)}\right)^{k_2}\nonumber \\
	&= \left(C^{(n_1)} \otimes I^{(n_2)}\right)^{k_1} \left(I^{(n_1)} \otimes C^{(n_2)}\right)^{k_2}.
\end{align}
This can be interpreted as $n_1$ separate copies of $n_2$-to-$1$ GMZIs (second term) with an additional set of permutations of the $n_1$ outputs also available (first term). So, permutations of $n_1$ rails can be implemented while muxing each one $n_2$ times by sending all $N=n_1n_2$ inputs through a single larger GMZI rather than smaller separate ones. The key advantage of this method is that the depth and total number of active phase shifters do not change (1 and $N$ respectively).

Using a larger GMZI comes at the cost of increasing the optical depth of the circuit, particularly in terms of waveguide crossings. As seen from Eq.~(\ref{eq:crossingexpansion}), the passive interferometers in a GMZI can be decomposed into smaller networks connected by layers of crossings. This modular structure can be exploited to distribute parts of the circuit across different locations and avoid large on-chip crossing networks. In the BSG example, the implementation shown in Fig.~\ref{fig:bsg_multimulti}(b) highlights how the first layer of crossings can be realised in a different way, e.g. using long distance phase-stable optical routing, to mitigate the impact of the largest crossing network in the interferometer.

\subsection{\label{sec:AltGMZI} Alternative GMZI constructions}

The discussion so far presented a large family of GMZIs and explained their key properties, taking an approach focused on achievable sets of permutations which is different to earlier works.  As well as $N$-to-$1$ muxing (potentially with extra functionality as explained in Sec.~\ref{sec:Multimultiplexing}), these GMZIs have assorted applications as building blocks for spatial and temporal muxes, a variety of which are discussed in Sec.~\ref{sec:NewSpatialMUXes} and Sec.~\ref{sec:NewTemporalMUXes}.  Alternative constructions of GMZIs are also possible, and it is valuable to explore them with a view to minimizing practical requirements on fast phase shifters. However, it is not feasible to exhaust all possible GMZI designs, as even the problem of finding all complex Hadamard matrices of a given order is still open \cite{Tadej06}.  Instead we will highlight some specific new constructions with useful properties.

One simple observation is that phase swing requirements (where the swing is defined per phase shifter as the difference between the maximum and minimum phase shifts across all GMZI settings) can sometimes be reduced by introducing fixed phase-shift offsets.  For the constructions in Sec.~\ref{sec:Commuting}, the phase shifter settings correspond to complete sets of roots of unity, and the phase swing is $\pi$ for Hadamard interferometers and  $>\pi$ for the other GMZI types.  Table~\ref{tab:gmzi_phase_swing} shows examples of reduced swing for GMZI sizes $N=2,3,4$.

\begin{table}[b]
\center
\begin{tabular}{| p{1.8cm}| p{1.8cm} | p{3.2cm} |}
\hline
GMZI type & Fixed phase offsets & Comment \\
\hline
Hadamard $N=2$ & $(-3\pi/2,0)$ &  Swing reduced from $\pi$ to $\pi/2$, coinciding with MZI variant in Fig.~\ref{fig:buildingblocks}(a). \\ 
DFT $N=3$ &  $(-4\pi/3, 0, 0)$ & Swing reduced from $4\pi/3$ to $2\pi/3$. \\  
Hadamard $N=4$ & $(-\pi,0,0,0)$ & Swing unchanged at $\pi$, but for each setting only one phase shifter is set to $\pi$ and the others to 0. \\ 
\hline
\end{tabular}
\caption{\label{tab:gmzi_phase_swing} {\it Examples of GMZIs with reduced phase swing using fixed phase-shift offsets.}  It is assumed that all the fast phase shifter components are identical and access the same range of phase shifts (which is minimized).  Note that the use of offsets necessitates modification of the GMZI transfer matrices by additional phase factors --- corresponding to setting-dependent ``global'' phases at the output.}
\end{table}

To find some more subtle constructions, we can consider general constraints on GMZIs implementing transfer matrices $U_k=W D_k V^\dag$ on $N$ modes, which are required to act minimally as $N$-to-$1$ muxes.  It is possible to prove a lemma stating that (a), $V$ in this case must be proportional to a complex Hadamard matrix (i.e. $V$ must satisfy $\vert V_{s,t}\vert=1/\sqrt{N}$ as well as being unitary), and (b) the phase vectors ${\mathbf d}_k$ must be orthogonal (refer to Appendix~\ref{sec:GMZIAppendix} for a proof of the lemma) .  A simple consequence of this result is that \textit{it is never possible to construct any GMZI for which the phase-shifter swing is less than $\pi/2$} (since it is never possible to achieve 0 for the real part of $\langle {\mathbf d}_k, {\mathbf d}_{k^\prime}\rangle$).  Similarly, when the phase-shifter values are restricted to $\{0,\pi/2\}$ it is not possible to find more than 2 orthogonal vectors ${\mathbf d}_k$ for any even value of $N$ (and never more than 1 for odd values of $N$), which is to say that it is not possible to do better than a 2-to-1 mux.

As another application of this lemma, one can look for sets of orthonormal phase vectors $\{{\mathbf d}_k\}$ and construct a GMZI which uses these as phase settings for a $N$-to-$1$ mux, by choosing $V$ to have row vectors ${\mathbf v}_k={\mathbf d}_k$, and any unitary $W$ with first row vector ${\mathbf w}_1 = (1,1,\cdots,1)/\sqrt{N}$.  An interesting and non-trivial example of such a set of phase vectors is given in Table~\ref{tab:exotic_gmzi}.  A $N=6$ GMZI constructed using these settings can implement a 4-to-1 mux which has phase swing of only $2\pi/3$ (by restricting to the first four phase-shifter settings).  Furthermore, it is easily seen that this example is not related to the constructions from Sec.~IIIB since the only possibility would be the GMZI implementing  $\mathcal{G} ([6]) \cong \mathcal{G} ([3,2])$, for which individual phase settings range on six values (compared to three in Table~\ref{tab:exotic_gmzi}).

\begin{table}[t]
\center
\begin{tabular}{|p{7cm}|}
\hline
Settings for a $N=6$ GMZI acting as a $6$-to-$1$ mux \\
\hline
${\mathbf d}_1 = \left(1,1,1,e^{-2 \imath \pi/3},e^{-2 \imath \pi/3},e^{-2 \imath \pi/3}\right)/\sqrt{6} $ \\
${\mathbf d}_2 = \left(1,e^{-2 \imath \pi/3},e^{-2 \imath \pi/3},1,e^{-2 \imath \pi/3},1\right)/\sqrt{6} $ \\
${\mathbf d}_3 = \left(e^{-2 \imath \pi/3},1,e^{-2 \imath \pi/3},e^{-2 \imath \pi/3},1,1\right)/\sqrt{6} $ \\
${\mathbf d}_4 = \left(e^{-2 \imath \pi/3},e^{-2 \imath \pi/3},1,1,1,e^{-2 \imath \pi/3}\right)/\sqrt{6} $ \\ 
${\mathbf d}_5 = \left(1, e^{-2 \imath \pi/3}, e^{-4 \imath \pi/3}, e^{-2 \imath \pi/3}, 1, e^{-4 \imath \pi/3}\right)/\sqrt{6}$ \\
${\mathbf d}_6 = \left(e^{-2 \imath \pi/3}, 1, e^{-4 \imath \pi/3}, 1, e^{-2 \imath \pi/3}, e^{-4 \imath \pi/3}\right)/\sqrt{6}$ \\
\hline
\end{tabular}
\caption{\label{tab:exotic_gmzi} {\it Example of six orthogonal phase vectors with a subset
${\mathbf d}_1,\cdots,{\mathbf d}_4$ having a reduced phase swing of $2\pi/3$
(compared to $4\pi/3$ for the entire set).} A GMZI using these phase settings can be used to enable a 4-to-1 MUX where only the first four modes are populated by inputs. }
\end{table}

Finally, we turn to a new way of using GMZIs when phase settings are modified from those
connecting single input and output ports.  Taking Hadamard-type GMZIs with transfer matrices $U_k = W D_k W^\dag$ on $N$ modes, consider first when the phase vector ${\mathbf d}_{k}$ for $D_{k}$ is modified so that $-\pi$ phases are set to a (common) value $-\phi$, while the $0$ phases are unchanged.  In this case $U_{k}$ is modified to,
\begin{equation} 
\tilde{U}_{k}(\phi) = e^{-i \phi/2} \left[ \cos\left(\frac{\phi}{2}\right) I^{(N)} + i \sin\left(\frac{\phi}{2}\right) U_{k} \right].
\end{equation}
This unitary maps a single photon incident at one input port to a superposition across the mode at the input and the output under the permutation $U_k$, with weighting controlled by the value of $\phi$.  Further modification of the phase settings can achieve mappings from one input to arbitrary pairs of output ports --- suppose it is desired to map from input port $p_1$ to output ports $q_1$ and $q_2$, then this can be implemented by finding the (unique) settings $k_1, k_2$ with $U_{k_{1(2)}}=W D_{k_{1(2)}} W^\dag: p \mapsto q_{1(2)}$, and choosing phase vector
\begin{equation} 
\tilde{\mathbf d}=e^{-i \phi/2} \left[\cos \left( \frac{\phi}{2} \right) {\mathbf d}_{k_1} + i \sin \left( \frac{\phi}{2} \right) {\mathbf d}_{k_2}\right].
\end{equation}
The transfer matrix for the GMZI is then
\begin{equation}
\label{eq:switchable_bs}
\tilde{U}(\phi) = e^{-i \phi/2} \left[ \cos\left(\frac{\phi}{2}\right) U_{k_1} + i\sin\left(\frac{\phi}{2}\right) U_{k_2} \right].  
\end{equation}
where the individual phase settings are taken from the set $\{0,-\phi,-\pi,-\pi-\phi\}$.  Note that a second input port $p_2$ is also mapped to the pair $q_1$ and $q_2$, where 
$U_{k_1} U_{k_2}:p_1 \mapsto p_2$.  We call a GMZI used according to Eq.~(\ref{eq:switchable_bs}) a {\it switchable pairwise coupler} and it can be useful in  spatial and temporal muxes (with the proviso that one of the paired ports receive the vacuum state to avoid contamination of the intended input). As an example, the technique can be used to integrate space-to-time qubit encoding into a mux without adding an extra stage of switching, as is described in Sec.~\ref{sec:rastermux}.

Another example is provided by direct modification of the MZI operations in Eq.~(\ref{eq:MZIops}) with half the range for the active phase shifters:
\begin{eqnarray}
\label{eq:halfMZI}
I \,{\rm or}\, e^{-i \pi/4} h_c^\dagger 
&=& h \left(I \,{\rm or}\, D[{\mathbf d}=(1,-i)] \right) h \nonumber \\
&=&
S h_c Z \left( I \,{\rm or}\, D[{\mathbf d}=(1,-i)]\right) h_c S.
\end{eqnarray} 
A push-pull implementation of this would modify the active part to phase vectors ${\mathbf d}=(\exp(-i \pi/4),1)$ and  ${\mathbf d}=(1,\exp(-i \pi/4))$, together with a fixed phase offset corresponding to $(-7\pi/4,0)$.  This gives operations $I/h_c^\dagger$ (without a setting-dependent global phase factor), and since $h = S^\dag h_c\dag S^\dag$, $Z/h$ can be obtained instead using additional fixed phase shifts at the input and output.  Remarkably then, two-mode Hadamard operations can be made switchable using active phase depth of only $\pi/4$.  One application for this is implementing controllable coupling of photons into a subcircuit, which is relevant for example to some of the schemes discussed in Sec.~\ref{sec:RandomInput}.

%% file: sections/spatial_muxes.tex
\label{sec:spatial_muxes}

In this section we consider the problem of designing practical and efficient mux schemes which route groups of photons from a bank of heralded single-photon sources (HSPSs) into one or multiple entanglement generation circuits (where we denote the number of circuits by $g$).  We focus on muxing groups of four photons for Bell-state generators (BSGs), and extensions to six photons for GHZ generation circuits \cite{Zhang08,Varnava08,PsiEntGen}, which is of critical importance for photonic FBQC \cite{Bartolucci21}.  Typical experimental constraints, such as the need to limit multi-photon contamination, mean that values for heralding probability $p$ for the sources are typically much smaller than the theoretical maximum (which is $p=0.25$ for photon-pair generation using spontaneous parametric downconversion or spontaneous four-wave mixing) and so a large number of sources $N$ must be muxed to achieve useful rates of photon generation.  

The simplest type of muxing for entanglement-generation circuits uses $m\times$ $N/m$-to-$1$ single-photon muxes to supply the circuit inputs separately, but this strategy wastes photons.  In principle, it is possible to use switch networks which implement all possible permutations of their inputs to supply one or multiple entanglement-generation circuits with mux efficiency which is limited only by the statistical properties of the HSPSs.  In this section we call muxes {\it optimal} if they attain theoretical bounds on mux efficiency dictated by the binomial distribution for the total number of heralded photons.  These bounds are discussed in Sec.~\ref{sec:Metrics} for different scenarios.  However, the optimal muxes described in Sec.~\ref{sec:SwNetReview}, i.e. Spanke networks and concatenated GMZIs, typically have unacceptable hardware costs, e.g. loss due to device depth, due to the requirement for large $N$ \footnote{A $N$-to-$m$ Spanke network requires $N\times$ $1$-to-$m$ muxes followed by $m\times$ $N$-to-$1$ muxes. From exchange symmetry, the muxes sizes in the first layer can be reduced to $1$-to-$1$, $1$-to-$2$, ..., $1$-to-$m$.  A concatenated GMZI scheme needs GMZI sizes between $N$ and $N\!-\!m\!+\!1$, and the output is subject to variable switch depth between $1$ and $m$.}.  Below we present a number of new schemes which are more efficient than the $m\times$ $N/m$-to-$1$ mux strategy, with much reduced hardware costs compared to na{\"i}ve switching networks.  These gains can be especially significant when there are experimental constraints, such as restrictions on the maximum sizes of GMZIs or on $N$, or when $p$ varies across a large range of values.
  
In Sec.~\ref{sec:hugmux}, we argue how we can exploit additional inputs to a standard BSG to improve mux efficiency with mimimal additional switching compared to the $4\times$ $N/4$-to-$1$ mux strategy, and we summarize the results of applying the same ideas for GHZ-state generation.  We also describe mux schemes which achieve optimality by incorporating simple switch networks before standard $m\times$ $N/m$-to-$1$ muxes.   Next, in Sec.~\ref{sec:RandomInput}, we explore mux strategies for a family of BSG circuits introduced in \cite{PsiEntGen}, which extend the standard BSG on eight modes to $N=2^n$-modes.  These new circuits offer many new possibilites since for $N>8$ there is a big increase in the number of four-photon input patterns that can be used, but with the disadvantage that the success probability for generating a Bell state is reduced from 3/16 to 3/32 for the majority of the cases (the  success probability is 3/16 for all patterns for $N=8$).  In Sec.~\ref{sec:bnmux}, we consider a family of two-layer switch networks for supplying photons to multiple entanglement-generation circuits, since the most robust improvements in mux efficiency are possible when input photons can be assigned to several groups of outputs (referred to below as {\it sharing} strategies).  A final consideration for all the schemes discussed is the classical control which is required to enable routing in real time.  The complexity of this control impacts delay (and therefore optical loss) and varies considerably between mux designs, and so a review of the requirements is given in Appendices~\ref{sec:logic_bsg8_muxes}, \ref{sec:logic_rnd_inp_muxes} and \ref{sec:bn_mux_alg} for the various schemes, together with a general method for simplifying routing logic in Appendix~\ref{sec:logic_reduction}.

\subsection{\label{sec:Metrics} Metrics and bounds on mux efficiency}

\begin{figure*}
	\centering
	\includegraphics[width=0.75\textwidth]{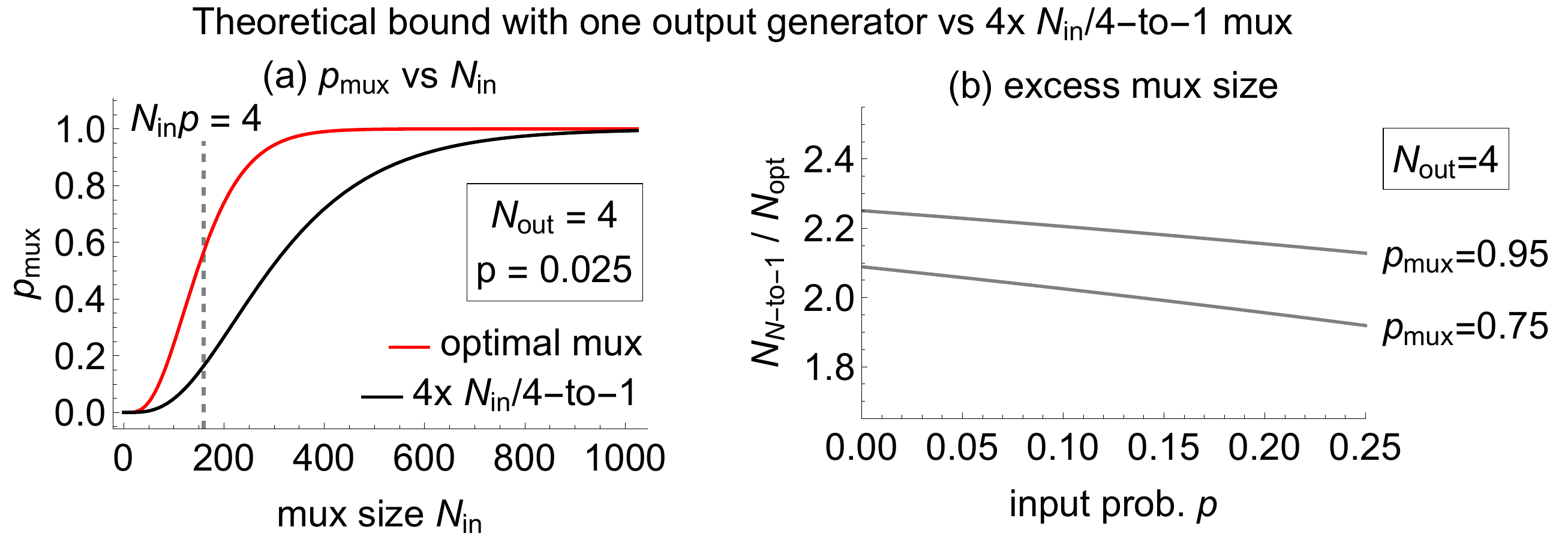}
	\caption{{\it Muxing four photons for one BSG.} (a) Comparison of the performance of $4\times$ $N/4$-to-$1$ muxes and an optical mux with perfect $N$-to-$4$ routing capability; (b)  ratio of the number of required inputs for $4\times$ $N/4$-to-$1$ versus an optimal mux, as a function of input probability $p$ for two target $p_{\rm mux}$ values.
	\label{fig:muxadv}}       
\end{figure*}

\begin{figure*}
	\centering
	\includegraphics[width=0.8\textwidth]{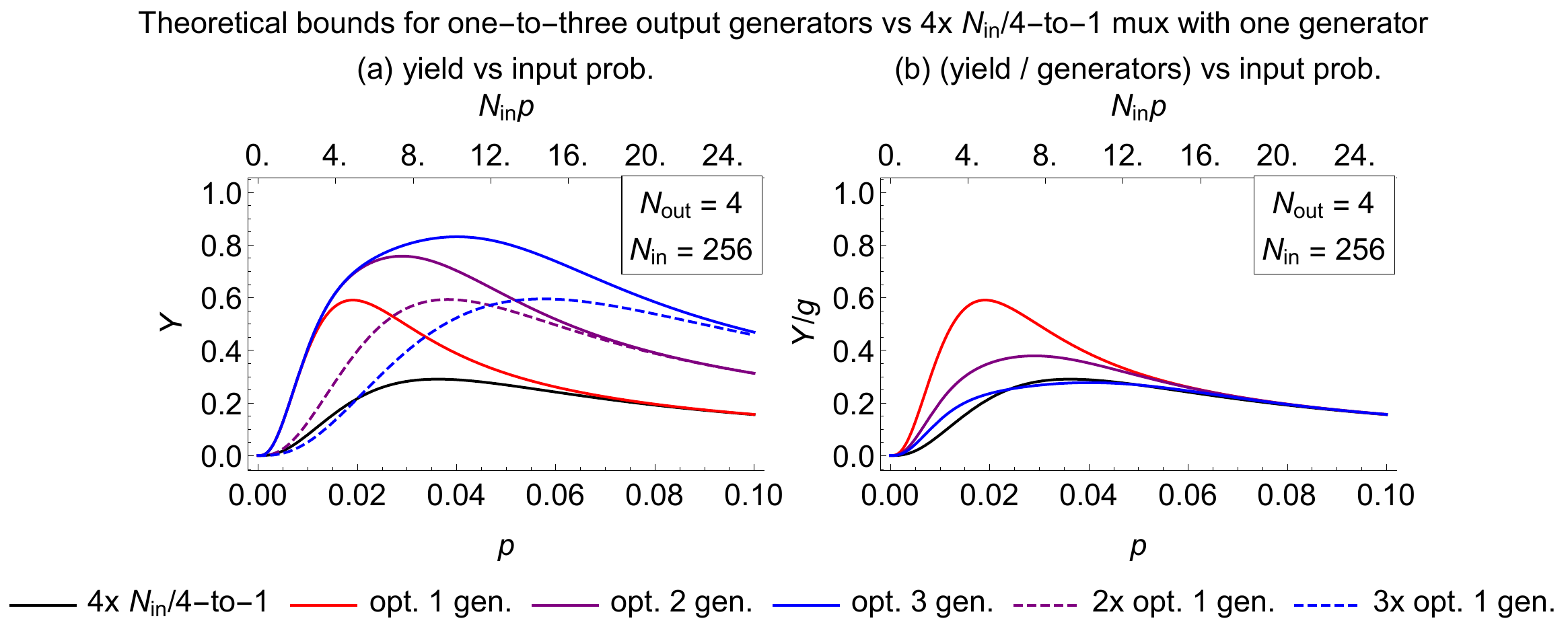}
	\caption{{\it Comparison of best-achievable yield for muxing groups of four photons for different strategies with the same total number of HSPSs.}  The strategies are: $4\times$ $N/4$-to-$1$ with one generator (black), optimal muxes feeding one-three generators jointly (solid, colored), and one-three independent muxes supplying generators individually at the optimal level (dashed, colored).  (a) shows overall yield of groups of four photons, and (b) shows the overall yield divided by the number of output circuits. 
	\label{fig:yieldadv}}
\end{figure*}

A simple metric to evaluate spatial muxes on identical HSPS inputs is the probability $p_{\rm mux}$ of obtaining the target number of photons (e.g. four photons for a BSG) as a function of the number of inputs $N$ and the probability $p$ of each input being occupied.  The $4\times$ $N/4$-to-$1$ strategy is compared with an optimal $N$-to-$4$ mux in Fig.~\ref{fig:muxadv}(a).  It can be seen that $p_{\rm mux}$ tends to 1 a lot quicker using the optimal strategy. Furthermore large relative improvements in $p_{\rm mux}$ are possible when the mux strategies are compared for fixed $N$, when $N$ is small.  To give a different perspective, the two strategies are compared in Fig.~\ref{fig:muxadv}(b) on the basis of the relative sizes required to attain a target $p_{\rm mux}$ for a given value of input probability $p$.  As a general rule, slightly-more than doubling $N$ allows the $4\times$ $N/4$-to-$1$ strategy to achieve the same output probability as an optimal mux with the original number of HSPSs.  (The behaviour for $N_{\rm out}=6$ is similar, see Appendix~\ref{sec:MetricsExtra} for details.) Overall, Fig.~\ref{fig:muxadv} reveals that it is important to improve mux efficiency when size or loss considerations make it impractical to scale up $N$-to-$1$ muxes --- which for example is a realistic scenario for hardware built using integrated quantum photonics due to fabrication limits on photonic and electical chip sizes (namely the reticle size limit).  

Another important metric for comparing mux schemes is yield.  For the case of muxes outputting groups of $m$ photons, we define the mux yield as:
\begin{equation}
\label{eq:yield_def}	
Y = \frac{\langle \text{\rm number of photons output in groups of } m \rangle}{\langle \text{\rm photons generated at input} \rangle},
\end{equation} 
where $\langle \cdot \rangle$ denotes the average value.  This definition extends to schemes where a single mux feeds multiple entanglement-generation circuits, which can be more efficient due to the possibility of exchanging extra photons between generators.

The achievable mux yield using different strategies is shown in Fig.~\ref{fig:yieldadv}.  Note that in the regime of interest, i.e. large $N$ and small $p$, the underlying probability distribution for the total photon number generated in a single run is very well approximated by a Poissonian distribution with mean $Np$, and consequently the yield in all cases is also effectively parametrized by $Np$, rather than $N$ and $p$ separately (and hence the behaviour with varying $N$ can be inferred directly from Fig.~\ref{fig:yieldadv}).  Again we can observe in Fig.~\ref{fig:yieldadv}(a) an improvement in yield for an optimal mux strategy feeding one generator versus $4\times$ $N/4$-to-$1$ muxing in the ``source-poor'' regime, which might be roughly characterized as average input number of photons $N p < 8$, and especially $N p < 4$.  It is also clear from Fig.~\ref{fig:yieldadv}(a) that, when using optimal muxing, it makes sense to output to one, two, or three generators when $N p \simeq 4$, $8$, or $12$ respectively, since this maximizes the yield (and the maximum yield increases with the number of generators).  Note however that the high maximum yield is accompanied by a reduction in yield per generator compared to outputting to a single generator, as shown in Fig.~\ref{fig:yieldadv}(b).  In particular, if the outputs of the entanglement generating circuits are also to be muxed, then those circuit muxes must be enlarged to accommodate higher numbers of circuits to achieve high yield first at the single-photon mux stage.  
 
Finally, it is interesting to compare the relative optimal performance of strategies which can share photons between entanglement generators versus those which feed generators independently, which is shown in Fig.~\ref{fig:yieldadv}(a).  The best achievable overall yield without sharing is 0.59, while the optimal sharing strategy can attain 0.76 (with two generators) and 0.83 (with three generators).  Perhaps more important than the gains in peak yield (which are fairly modest) is the fact that a \textit{multi-generator strategy incorporating sharing can potentially deliver improved yield over a broad range of input probability $p$}, which is important experimentally to mitigate against unexpected low $p$.

\subsection{\label{sec:hugmux} Multiplexing groups of single photons for a standard Bell or GHZ-state generating circuit}   

\begin{figure*}
	\centering
	\includegraphics[width=0.8\textwidth]{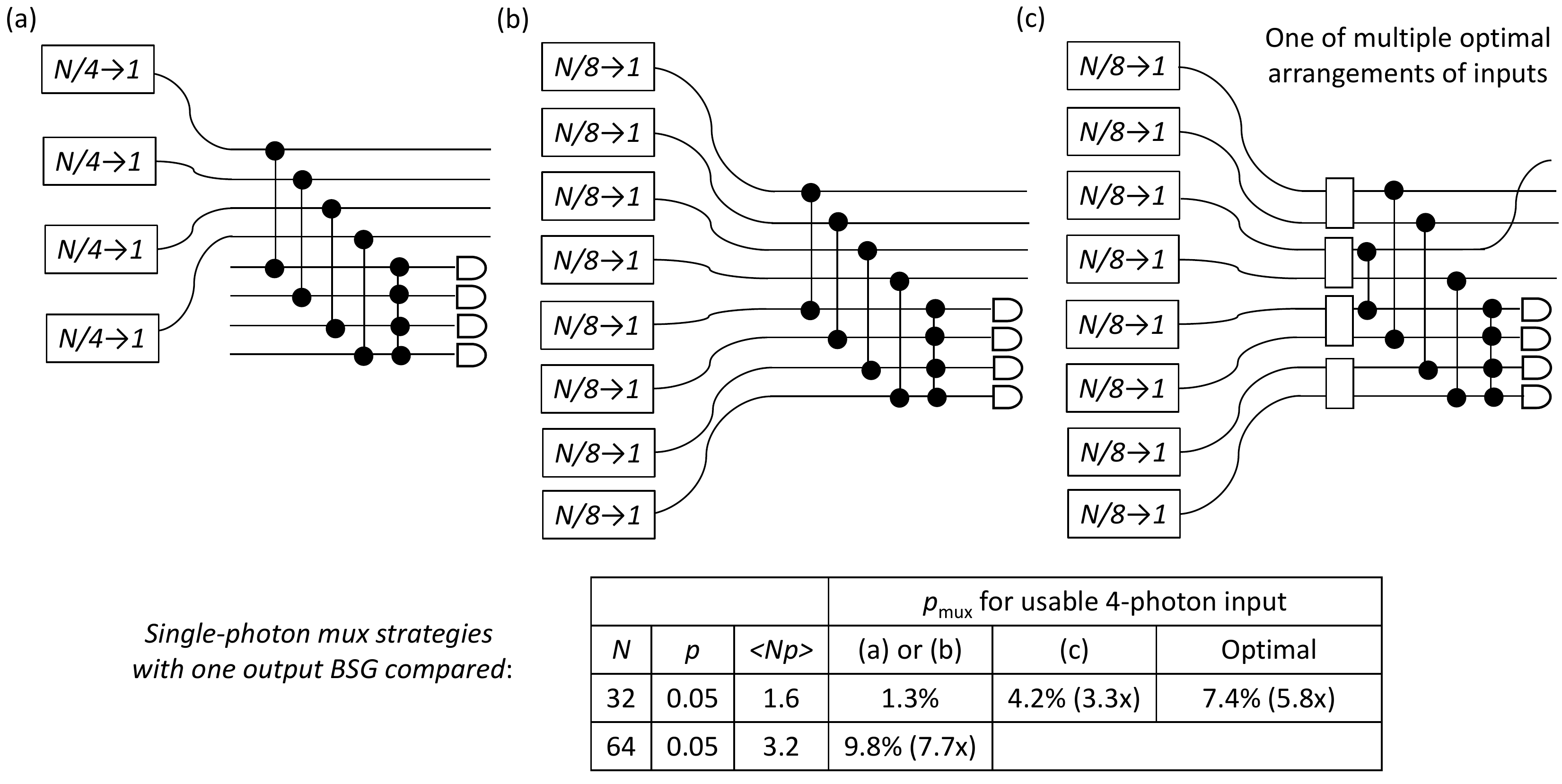}
	\caption{\label{fig:regbsgmux} {\it Input muxing configurations for a regular BSG circuit.} (See Ref.~\cite{PsiEntGen} for further explanation of the circuit.) (a) $4\times$ $N/4$-to-$1$ muxes; (b) $8\times$ $N/8$-to-$1$ muxes for which 16 out of 70 patterns of four input photons are usable; (c) $8\times$ $N/8$-to-$1$ muxes together with a single layer of MZIs (white boxes) for which 66 out of 70 patterns of four input photons are usable (multiple MZI configurations achieve 66/70).  (b) and (c) assume 50:50 power splitting for initial beam-splitter/directional coupler operations, but this is not required for (i).  Digital logic requirements for each of the schemes is reviewed in Appendix~\ref{sec:logic_bsg8_muxes}. The notation used here is as described in Fig.~\ref{fig:buildingblocks} and Fig.~\ref{fig:gmziconstructions}.}       
\end{figure*}

One approach to achieving more efficient muxing of groups of photons for BSG circuits is to exploit the precise structure of the circuit itself.  The standard BSG circuit generates Bell states with qubits in dual-rail encoding, and is an eight-mode device with four modes typically used for the input and output, as shown in Fig.~\ref{fig:regbsgmux}(a).  (See Ref.~\cite{PsiEntGen} for a detailed explanation of how the BSG works.)  To achieve more efficient muxing of single photons at the input,  additional modes can be used to inject the photons.  As shown in Fig.~\ref{fig:regbsgmux}(b), each photon can enter a pair of inputs: first or fifth, second or sixth, third or seventh, fourth or eighth  inputs from the top.  Note that the beam-splitter/directional coupler operations at the input should generally be restricted to 50:50 power splitting ratios in this case \footnote{For the configuration in Fig.~\ref{fig:regbsgmux}(a) it can be useful to consider unbalanced splitting ratios.}.  The scheme in Fig.~\ref{fig:regbsgmux}(b) can utilize $2^4=16$ patterns of input photon --- each time the circuit operates, four out of the eight muxes need to supply a photon in one of 16 patterns, while the remaining muxes must supply a vacuum state.  The vacuum inputs can originate either from sources that don't herald successful photon generation, which have a small probability of containing photons, or from dedicated ``clean'' vacuum inputs.  It should be noted that there can be additional phase changes for the output states depending on the pattern of input photons: the setup in Fig.~\ref{fig:regbsgmux}(a) outputs three types of dual-rail encoded Bell states (including one with qubit rails exchanged) whereas the setup in Fig.~\ref{fig:regbsgmux}(b) outputs six Bell-state types with additional phase flips. This is a situation where we can take advantage of the additional operations that a mux can enable, as described in Sec.~\ref{sec:Multimultiplexing}, to correct those phase flips without the need for additional switching devices. The new scheme has some practical advantages as it has improved robustness to hardware device failure (since the BSG can still work as long as a subset of the eight input muxes can supply usable patterns of input photons between them) \footnote{In the absence of hardware failures, $p_{\rm mux}$ is the same for Fig.~\ref{fig:regbsgmux}(a) and Fig.~\ref{fig:regbsgmux}(b).}.  
 
To achieve increased muxing efficiency, without the high cost of using a Spanke network, or similar, we can consider what improvements are available using only a few MZIs in addition to the $N/8$-to-$1$ muxes in Fig.~\ref{fig:regbsgmux}(b).  MZIs are in general expected to cause much less optical loss than GMZIs, partly because they do not use large interferometers, and partly because the required range of the fast phase shifters is only $\pi/2$ and so the phase-shifter devices themselves can be shorter than those in Hadamard-type GMZIs (which need a phase swing of $\pi$).  Furthermore, the extra MZIs can be added to the existing $N/8$-to-$1$ muxes without impacting the electronics delay significantly.  Using exhaustive search, it can be shown that a single layer of MZIs in an optimal configuration can be used to rearrange 66 out of 70 possible patterns of four photons across eight modes to one of the 16 usable by a regular BSG, and one of the configurations that achieves this is shown in Fig.~\ref{fig:regbsgmux}(c).  (The four input patterns that cannot be used occupy two MZIs and cannot be rearranged.)  For every case in which the $8\times$ $N/8$-to-$1$ muxes can supply $>4$ photons, a subpattern of four photons can always be found which is usable by the BSG, assuming that the muxes can freely route vacuum for the remaining four inputs.  The scheme in Fig.~\ref{fig:regbsgmux}(c) therefore allows an improvement in yield which approaches $66/16 \approx 4\times$ compared to Fig.~\ref{fig:regbsgmux}(a,b)  when $Np\ll4$, assuming the schemes are compared with the same $N$, $p$.  

These arguments can be extended to muxing six photons for a three-qubit GHZ generator.  Instead of the simplest mux configuration using $6\times$ $N/6$-to-$1$ single-photon muxes, $12\times$ $N/12$-to-$1$ muxes can be used to supply six photons in any one of 64 patterns which are usable (in analogy to the setup in Fig.~\ref{fig:regbsgmux}(b)).  In fact, there are a total of 924 possible arrangements of six photons over 12 modes and using a single layer of MZIs in one of several optimal configurations enables 666 patterns of input photons to be used.  The use of an additional layer of MZIs therefore allows a $666/64 \approx 10\times$ improvement in yield when $Np\ll6$.  Taking an operating point with $N=48,p=0.05$, the achievable improvement using the scheme with the added MZIs compared to using only $6\times$ $N/6$-to-$1$ muxes is $7.0\times$, compared to $22\times$ using an optimal mux or $21\times$ by doubling the number of sources to $N=96$.    

\begin{figure*}
\begin{center}
\includegraphics[width=0.8\textwidth]{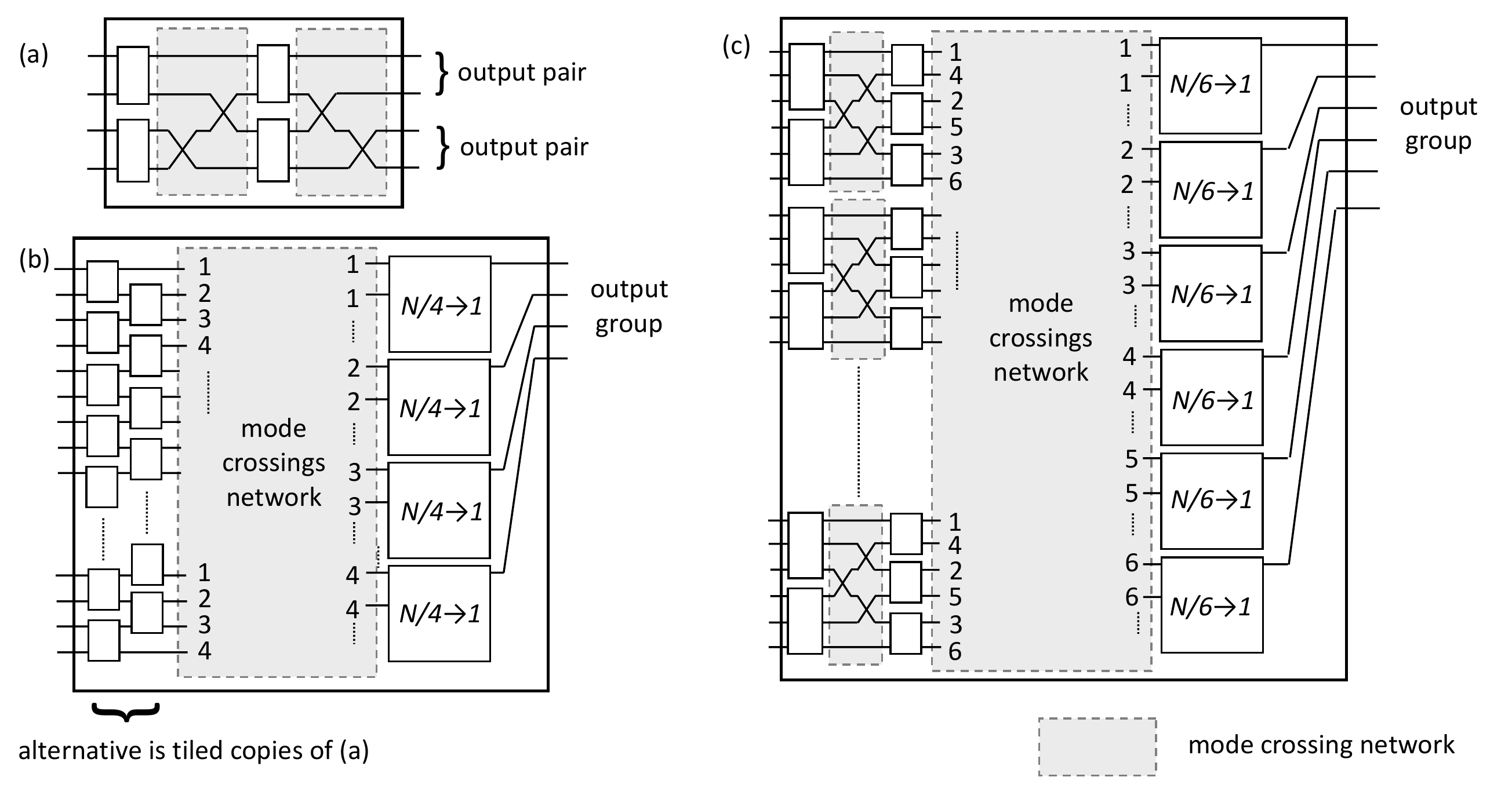}
\caption{\label{fig:tutuMZI} {\it Switch networks with optimal mux efficiency.}  The schemes are: (a) network routing arbitrary patterns of two inputs to a pair of output ports (either the top pair or the bottom pair), with the option to swap the inputs; (b) an optimal mux for outputting groups of four photons --- two layers of MZIs assign photons to each of four sets of modes (as numbered) and GMZIs route one output for each set;  (c) an optimal mux for outputting groups of six photons --- one layer of three-mode GMZIs and one layer of MZIs assign photons to each of six sets of modes (as numbered) and GMZIs route one output for each set.}
\end{center}
\end{figure*}

The question arises as to what are the simplest switch networks that achieve optimal mux efficiency.  To answer this, it is useful to start by considering the arrangement of MZIs shown in Fig.~\ref{fig:tutuMZI}(a).  This network provides 16 permutations using two layers of switching and total active switch depth $2\times \pi/2$ (assuming push-pull type MZIs), compared to only four permutations using a Hadamard-type GMZI on four modes with the same switch depth.  As well as being able to route from any one input port to any output port, this network can map arbitrary pairs of inputs to the top (or bottom) pair of outputs.  Furthermore, when the two inputs are considered distinguishable, the network can controllably swap their ordering at the output. 

Fig.~\ref{fig:tutuMZI}(b) shows a related scheme for the task of muxing groups of four photons.  The network is designed to act on any random distribution of four input photons, so that one photon is routed to each of four sets of output ports (labelled $1-4$) using two layers of MZIs. (If there are more than four photons then the extras are randomly assigned.)  A $4\times$  $N/4$-to-$1$ mux can route the rearranged photons to four specific output ports, and dump any excess input photons.  A routing algorithm which works for every possible distribution of (four) input photons is as follows: MZIs in the first layer implement swaps to ensure that pairs of photons are routable to outputs $\{1,4\}$ and $\{2,3\}$; this is always possible, as a MZI with one input can assign its photon freely, while a MZI with two inputs assigns one photon to $\{1,4\}$ and one to $\{2,3\}$ by default.  Next, MZIs in the second layer associated with the output pairs $\{1,4\}$ assign one photon to the $1$'s and one to the $4$'s, and similarly for the MZIs associated with $\{2,3\}$.  It might be observed that the network of MZIs works similarly to multiple copies of Fig.~\ref{fig:tutuMZI}(a) but with the mode crossings eliminated, and with each copy outputting between zero and four photons.  Remarkably the scheme overall achieves optimal mux efficiency using a switch network which is four times smaller than the equivalent Spanke network (measured in terms of the total number number of optical modes, input ports, or output ports) and equivalent total active switch depth (i.e. $2\pi$ assuming Hadamard-type GMZIs).  The scheme also represents only a minor increase in hardware complexity compared to Fig.~\ref{fig:regbsgmux}(c), which has suboptimal mux efficiency.  

The ideas above can also be directly extended to the task of muxing groups of six photons using the switch network illustrated in Fig.~\ref{fig:tutuMZI}(c).  This network uses one layer of three-mode GMZIs and one of MZIs to assign input photons to every one of six sets of outputs, labelled $1-6$, and a $6\times$ $N/6$-to-$1$ mux can be used at the end.  For the routing algorithm, the three-mode GMZIs at the start implement permutations so that two photons are routable to each of $\{1,4\}$, $\{2,5\}$, and $\{3,6\}$, and the MZIs allocate photons to each member of each of these pairs.  The network successfully routes any distribution of six photons at the input, including cases where two or three photons are incident at a single GMZI.  This can be verified by noting that a GMZI with a single input photon can freely assign to any of $\{1,4\}$, $\{2,5\}$, and $\{3,6\}$, while a GMZI with two inputs can freely assign to any pair of these, and a GMZI with three input photons assigns one photon to each by default, so that it always possible to ensure that each of $\{1,4\}$, $\{2,5\}$, and $\{3,6\}$ is allocated two photons.  Overall the switch network has total active switch depth which is comparable to the equivalent Spanke network (nominally the switch depth is $2\pi/3+\pi/2+\pi=13\pi/6$ including the final muxes versus $2\pi$ for the Spanke network), but is six times smaller.

\subsection{\label{sec:RandomInput} Switch networks for entanglement generation using random-input strategies}

\begin{figure*}
\begin{center}
\includegraphics[width=0.8\textwidth]{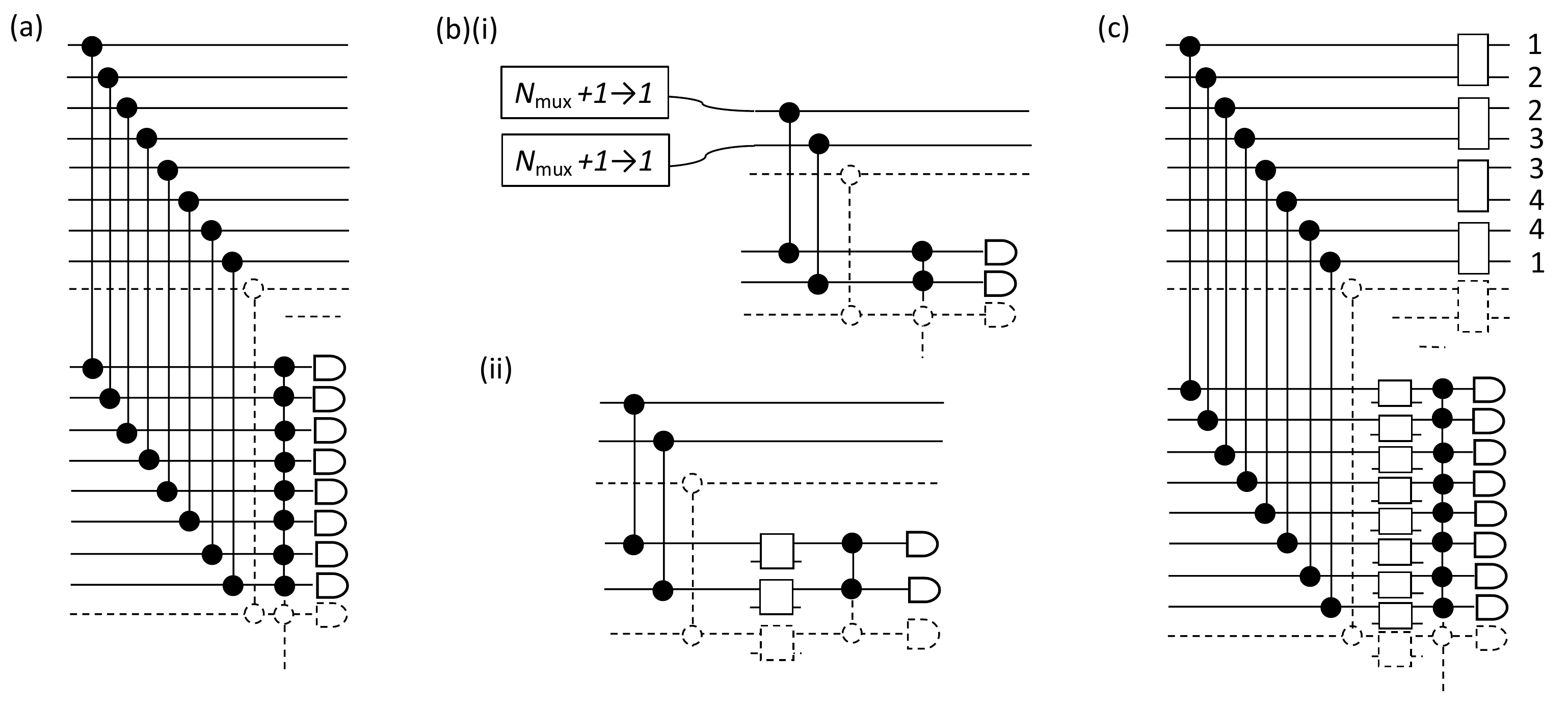}
\caption{\label{fig:rndinpschemes} {\it Bell-state generation using enlarged generator circuits.} (See Ref.~\cite{PsiEntGen} for detailed explanation of the BSG circuits.) (a) {\it Random-input} approach with ballistic generation of single photons. (b) Schemes discarding excess photons either: at the input, as in (i) combined with  $N_{\rm mux}$-to-$1$ muxing of the sources; or as in (ii) before the Hadamard interferometer on the detector modes using MZIs for blocking. (c) Scheme using a single-layer of MZIs at the output of the BSG to rearrange the rails of the Bell-states into four discrete sets of modes (following the numbering shown repeated across all outputs).  Digital logic requirements for each of the schemes is reviewed in Appendix~\ref{sec:logic_rnd_inp_muxes}.}      
\end{center}
\end{figure*}

We now turn to schemes based around a family of enlarged BSGs which have been introduced in Ref.~\cite{PsiEntGen} and which are illustrated in Fig.~\ref{fig:rndinpschemes}(a).  These BSGs generalise the standard eight-mode circuit in Fig.~\ref{fig:regbsgmux}(a), and are defined on $N_{\rm BSG}=2^n$ modes.  Similarly to the standard  BSG, these generalized circuits have two-mode ``down-coupling'' beam-splitter/directional coupler operations on pairs of input modes, and a measurement circuit based upon a Hadamard interferometer and photon detectors, which is extended to $N_{\rm BSG}/2$ modes.  Four photons are required at the input, and they must enter different down-coupling mode pairs (i.e. input photons cannot be in modes $N_{\rm BSG}/2$ apart in Fig.~\ref{fig:rndinpschemes}(a)).  The Bell states are generated on subsets of four modes from the $N_{\rm BSG}/2$ modes at the output, as determined by the pattern of input photons.  The success probability of the BSG itself is $3/16$ for some input patterns and $3/32$ for others --- given any three down-coupling mode pairs, there is a unique fourth pair for which the higher probability is obtained.

The generalized BSGs are of special interest, as the number of usable patterns of input photons increases rapidly with $N$. In particular, this reduces the requirement for muxing at the level of single-photon generation compared to the standard BSG, which enables the use of smaller switch networks (with less optical loss) or even a completely ballistic approach with no switching at the single-photon level.  This is particularly advantageous for applications where there is no need to restrict the output Bell states to specific spatial modes, or where the task of routing the Bell states into four specific modes can be combined with muxing the outputs of multiple BSGs.  
 
To enable comparison of schemes using BSG circuits of different sizes, it is necessary to account for both the probability of generating four-photon inputs from a total of $N$ sources and the probability for generating a Bell state.  In a ballistic approach, using the circuit from Fig.~\ref{fig:rndinpschemes}(a) without any switching, the useful input states have four photons distributed across the $N/2$ pairs of inputs, and the vacuum state at the $(N-4)$ modes which do not provide single photons.  For simplicity, we assume here that the inputs can be modeled as perfect (lossless) sources of two-mode squeezed state, $\vert \psi_{\rm src}\rangle = 1/\cosh(r)\sum^\infty_{k=0}\tanh
^k(r)\vert kk\rangle$ parametrized by squeezing parameter $r>0$, with ideal heralding detectors which are fully number-resolving.  The probability for heralding a single photon is then $p=\tanh^2(r)/\cosh^2(r)$, while the probability for heralding the vacuum state is $p_{\rm vac}=1/\cosh^2(r)$, so that
$p_{\rm vac}=\left(1\!+\!\sqrt{1\!-\!4p}\right)/2$.  The probability for a four-photon input state is then given by,
\begin{equation}
\label{eq:prob4ball}
p_4^{\rm ball}(N,p) = 2^4 {N/2 \choose 4} p^4 \left(\frac{1 + \sqrt{1\!-\!4p}}{2}\right)^{N\!-\!4}.
\end{equation}
The incorporation of switches which transmit or discard input photons (which we refer to as {\it blocking} switches) allows for an improvement in the  probability for a four-photon input state, as it is possible to use cases where $\ge4$ sources herald photons.  The blocking switches themselves can, for example, be implemented using MZIs with one vacuum input and a discard port at the output.  Rather than placing the blocking switches at the input, they can also be be placed in front of the Hadamard interometer, as shown in Fig.~\ref{fig:rndinpschemes}(b)(ii). Alternatively, the down-coupling operations themselves can be made controllable, with 100:0 versus 50:50 splitting being selectable using a switch with active phase depth of only $\pi/4$, as explained in Sec.~\ref{sec:AltGMZI}. 
These last two approaches rely on switching after the BSG to discard excess photons, which typically comes for free with muxing of multiple BSGs. The probability for a four-photon input state when using blocking switches is given by,
\begin{equation}
\label{eq:prob4block}
p_4^{\rm blck} (N,p) = 
1 - \sum_{k=0}^{3} {N/2 \choose k}
\left(1 - (1\!-\!p)^2\right)^k
(1\!-\!p)^{N\!-\!2k}.
\end{equation}      
If small $N_{\rm mux}$-to-$1$ muxes are used at the input, so that the total number of single-photon sources is $N$ and the BSG circuit is defined on $N_{\rm BSG}=N/N_{\rm mux}$ modes, as shown in Fig.~\ref{fig:rndinpschemes}(b)(i), then the probability for a four-photon input state is modified to,
\begin{equation}
\label{eq:prob4simplemux}
p_4^{\rm mux} (N,p) = p_4^{\rm blck} \left(\frac{N}{N_{\rm mux}}, 1 - (1-p)^{N_{\rm mux}}\right). 
\end{equation} 
Each mux can also implement blocking functionality for excess photons by incorporating an additional vacuum input.  Assuming that the patterns of input photons are equally likely, the average success probability for the $N$-mode BSG is given by,
\begin{equation}
\label{eq:BSGprob}	
P_{\rm BSG} = 
\frac{1}{4} \frac{\binom{N/2}{3}}{\binom{N/2}{4}} \times \frac{3}{16} + 	
\left(1 - \frac{1}{4} \frac{\binom{N/2}{3}}{\binom{N/2}{4}} \right) \times \frac{3}{32}.
\end{equation}
This expression for $P_{\rm BSG}$ declines quickly from $3/16$ at $N=8$ to $3/32$ for large values of $N$, so practically it suffices to approximate $P_{\rm BSG}=3/32$ for $N>8$. 

Schemes can now be compared on the basis of the quantity $Y \times P_{\rm BSG}$ (where $Y$ is the yield for the four-photon input states) which gives the yield of Bell states, normalized by the average rate of photon generation at the sources.  Note that this quantity can be directly related to the total number $S$ of HSPSs used to achieve Bell state generation with a target probability $p_{\rm out}$, and so $S$ can be used as a measure of resources.  If it is required to generate at least one Bell state using $K$ generators, then $p_{\rm out} = 1 - (1 \! - \! Np Y P_{\rm BSG}/4)^K $ and,
\begin{eqnarray}
\label{eq:footprint}
S &=& NK \nonumber \\
&=& N\frac{\ln{\left(1 - p_{\rm out}\right)}}{\ln{\left(1-Np Y P_{\rm BSG}/4\right)}} \nonumber \\
&\simeq& -4\frac{\ln{\left(1 - p_{\rm out}\right)}}{p}\frac{1}{P_{\rm BSG} Y}.
\end{eqnarray}
It can be seen that minimizing $S$ is equivalent to maximizing the production of Bell-states as given by $P_{\rm BSG} Y$.
 
\begin{figure}
\begin{center}
\includegraphics[width=0.8\columnwidth]{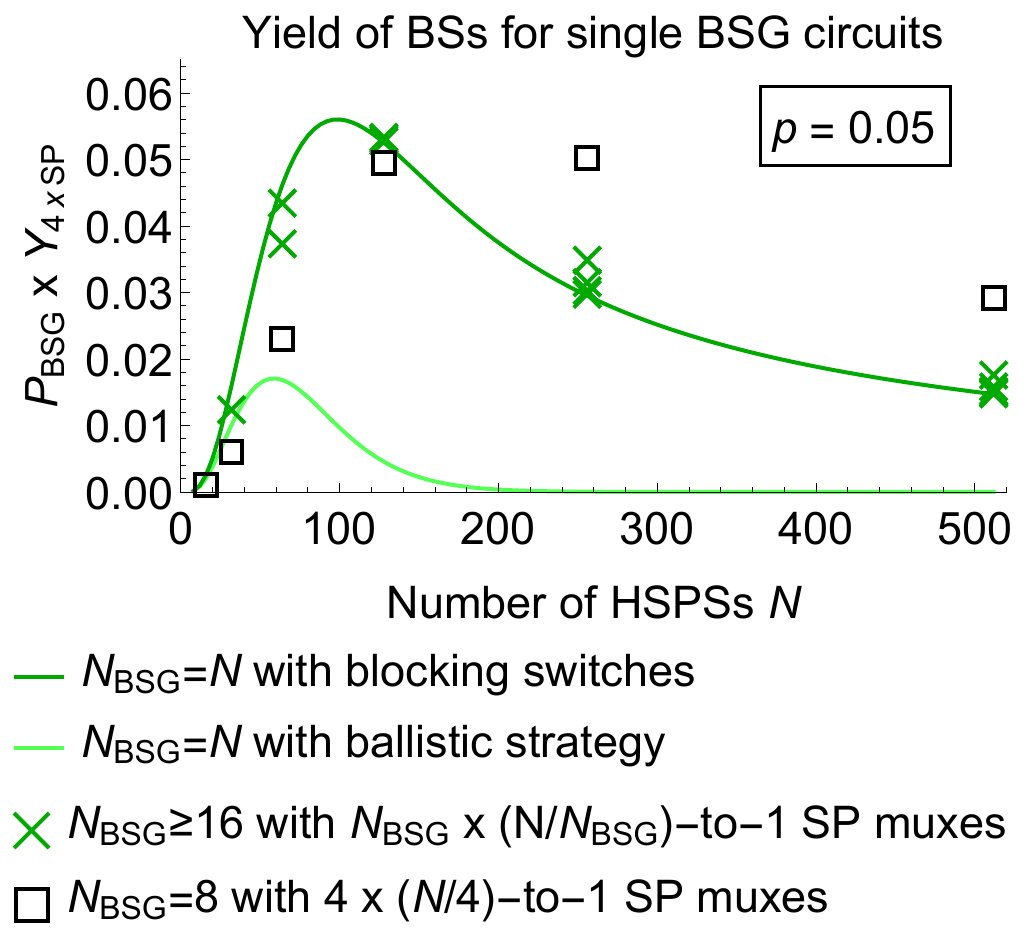}
\caption{{\it Yield of output Bell states $P_{\rm BSG} Y$ for different strategies using $N$ HSPSs.} (1) [light green line] Random-input BSG using the ballistic approach (refer Fig.~\ref{fig:rndinpschemes}(a)) for which usable input states comprise four photons incident at different down-coupling mode pairs. (2) [dark green, line] Random-input BSG using blocking switches (refer Fig.~\ref{fig:rndinpschemes}(b)(ii)) which can discard excess photons.  (3) [dark green, crosses] Random-input BSG using small single-photon muxes which can also be used to discard photons (refer Fig.~\ref{fig:rndinpschemes}(b)(i)).  (4) [black boxes] $4\times(N/4)$-to-$1$ single-photon muxing and an eight-mode BSG circuit.
\label{fig:mod_yield_plot}}
\end{center}
\end{figure}

\begin{figure*}
\begin{center}
\includegraphics[width=0.8\textwidth]{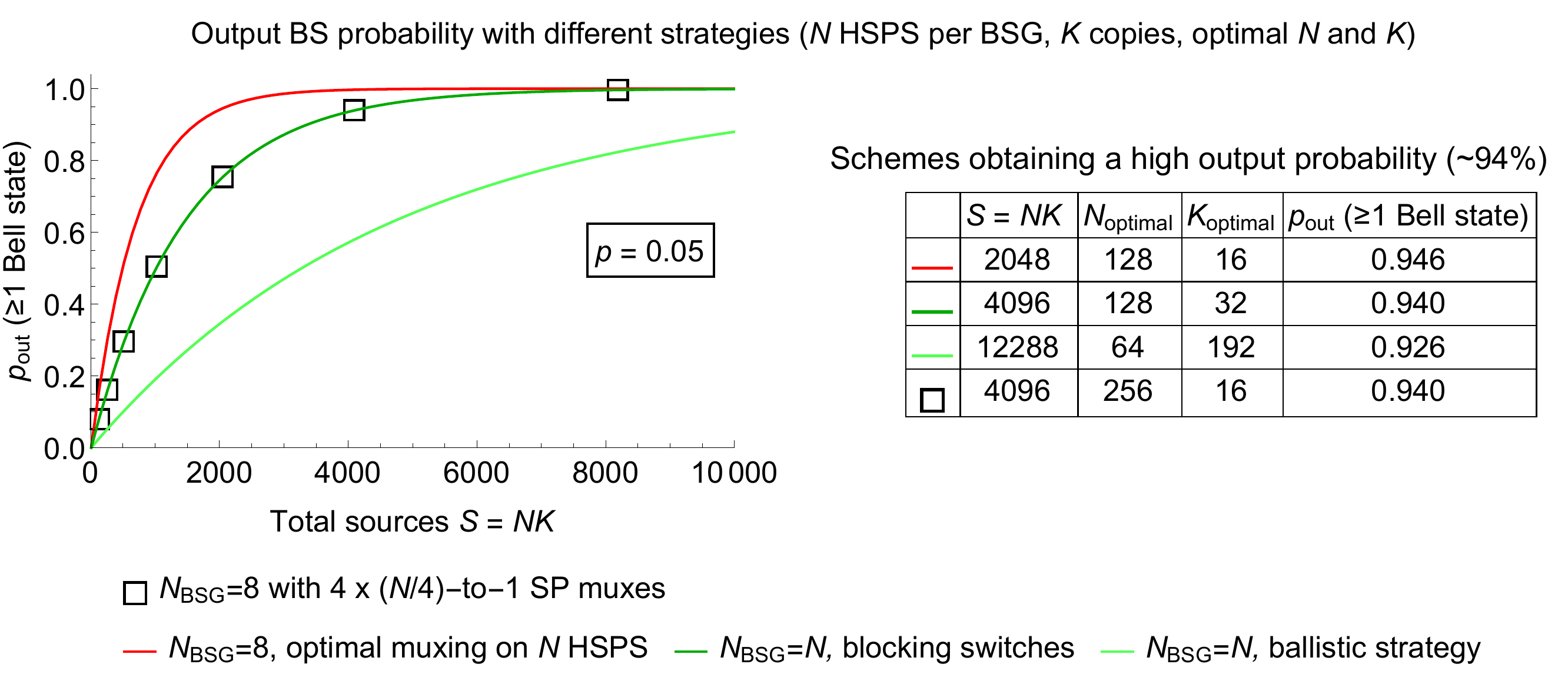}
\caption{\label{fig:rndinpcmp} {\it Comparison of schemes on the basis of the probability of generating at least one Bell state.}  It is assumed that multiple (independent) copies of each BSG circuit are used, and that any size parameters are optimised for the total number of HSPSs. The single-photon input probability is fixed at $p=0.05$.  Light green corresponds to a random-input ballistic strategy with an enlarged BSG (as shown in per Fig.~\ref{fig:rndinpschemes}(a)).  Dark green is for the case that the random-input strategy is modified using blocking switches (as in Fig.~\ref{fig:rndinpschemes}(b)(i) with $N_{\rm mux}\!=\!1$) so that cases with $>4$ input photons can be used as well as those with exactly four photons. Red shows the output probability assuming a standard eight-mode BSG (with success probability 3/16) and optimal muxing of single photon inputs, while the black squares assume $4\times N/4$-to-$1$ muxing instead (as in Fig.~\ref{fig:regbsgmux}(a)).}  
\end{center}
\end{figure*}

Comparison of the best-achievable values for the quantity $P_{\rm BSG} Y$ for different strategies in Fig.~\ref{fig:mod_yield_plot} reveals the following.  First, the best achievable values for $P_{\rm BSG} Y$ are almost identical for a random-input strategy as in Fig.~\ref{fig:rndinpschemes}(b)(i) with $N_{\rm mux}\!=\!1$ or Fig.~\ref{fig:rndinpschemes}(b)(ii) which discards excess photons beyond four, compared to a standard eight-mode BSG with $4\times$ $N/4$-to-$1$ single-photon muxes.  This is interesting as it shows that although the random-input strategy can exploit more patterns of input photons, this advantage is negated by lower success probability for the BSG circuit itself.  Furthermore, the random-input strategy might be considered worse in the sense that the output Bell states are randomly distributed over $N/2$ modes.  It also turns out that intermediate cases with small single-photon muxes and enlarged BSG circuits as in Fig.~\ref{fig:rndinpschemes}(b)(i) do not provide improvements in $P_{\rm BSG} Y$.

Turning to the performance of the ballistic strategy without any switches as in Fig.~\ref{fig:rndinpschemes}(a), the best values for $P_{\rm BSG} Y$ are approximately $3\times$ worse than when using blocking switches to discard excess photons.
It may be surmised that the ballistic strategy performs surprising well, and that the penalty it incurs in terms of requiring more HSPSs is an acceptable trade-off for avoiding the use of single-photon muxes.  However, it should be noted that the Bell states that are produced using the ballistic strategy are degraded by contamination from imperfect vacuum inputs (e.g. arising due to lossy source detectors), and this may be unacceptable for some applications.  A partial solution is to add blocking switches at the sources which use simple feedforward to  transmit light only when their respective HSPS heralds a photon.  This approach avoids the requirement for a complex mux controller which operates the blocking switches jointly, but the potential yield/footprint gain is lost as a consequence of not being able to use cases with $>4$ input photons --- as is also true for the ballistic approach.  The optical loss resulting from the addition of the blocking switching also represents a trade off for eliminating imperfect vacuum inputs.

A different perspective on all these results is given in Fig.~{\ref{fig:rndinpcmp}} which compares strategies on the basis of the probability of producing at least one Bell state using multiple (independent) copies of the BSG circuit.  In this comparison the number of copies of the circuits and their sizes is optimised.  However, no assumption is made about any muxing at the level of the Bell states which are generated, and the number of modes in which the output Bell states are located is not fixed.
	
The discussion so far has established that large $N$-mode BSGs can achieve efficient extraction of patterns of four input photons without necessarily resorting to single-photon muxing, although the Bell states at the output are randomly distributed across subsets of four modes.  Although a Spanke network or similar can be used to route these Bell states to four specific modes, a simpler switch network is highly desirable to minimize hardware costs.  The main task for this switch network is to route the rails of the Bell states into disjoint sets of modes, as four $N/8$-to-$1$ muxes can then finish the job.  It is not generally necessary to treat Bell states which are generated with internal rail swaps differently, as corrective mode swap operations can typically be implemented using Bell-state muxes, as explained in Sec.~\ref{sec:Multimultiplexing}.

One solution is simply to bin the modes, so that output states are only accepted if the rails lie in four disjoint sets of modes.  However, the efficiency of binning the output modes itself is rather low, since the probability of a Bell-state having rails in allocated bins is only $3/4 \times 2/4 \times 1/4 = 9.4\%$.  A more effective solution is shown in Fig.~\ref{fig:rndinpschemes}(c) which uses a single layer of MZIs to rearrange the outputs.  The idea of this scheme is to use the MZIs to place one rail from the output Bell state in each of the mode sets labelled 1, 2, 3 and 4, and it uses $N/16$ repeated blocks of MZIs with the outputs labelled as in the figure.  For the smallest case with $N=16$, the MZIs act in essentially the same way as for the scheme for single-photon muxing from eight modes to four, discussed in Sec.~\ref{sec:hugmux} above and shown in Fig.~\ref{fig:regbsgmux}(c), which works $66/70=94\%$.  The mux efficiency falls to $45/64=70.3\%$ cases for larger values of $N$, as there are more configurations of the output rails which cannot be rearranged by the MZIs (see Appendix~\ref{sec:reverse_hugmux_efficiency} for details).  However, it is interesting to note that the two-layer MZI network from Fig.~\ref{fig:tutuMZI}(b) can also be applied post Bell-state generation to assign Bell-state rails to distinct subsets of modes.  This network is optimal, and dual-rail qubit encoding can be achieved at the output for every input configuration using swaps in the second MZI layer.  Overall it may be concluded that a light-weight switch network is sufficient to route Bell states to specific modes when using random-input strategies.

\begin{figure*}
\begin{center}
\includegraphics[width=0.85\textwidth]{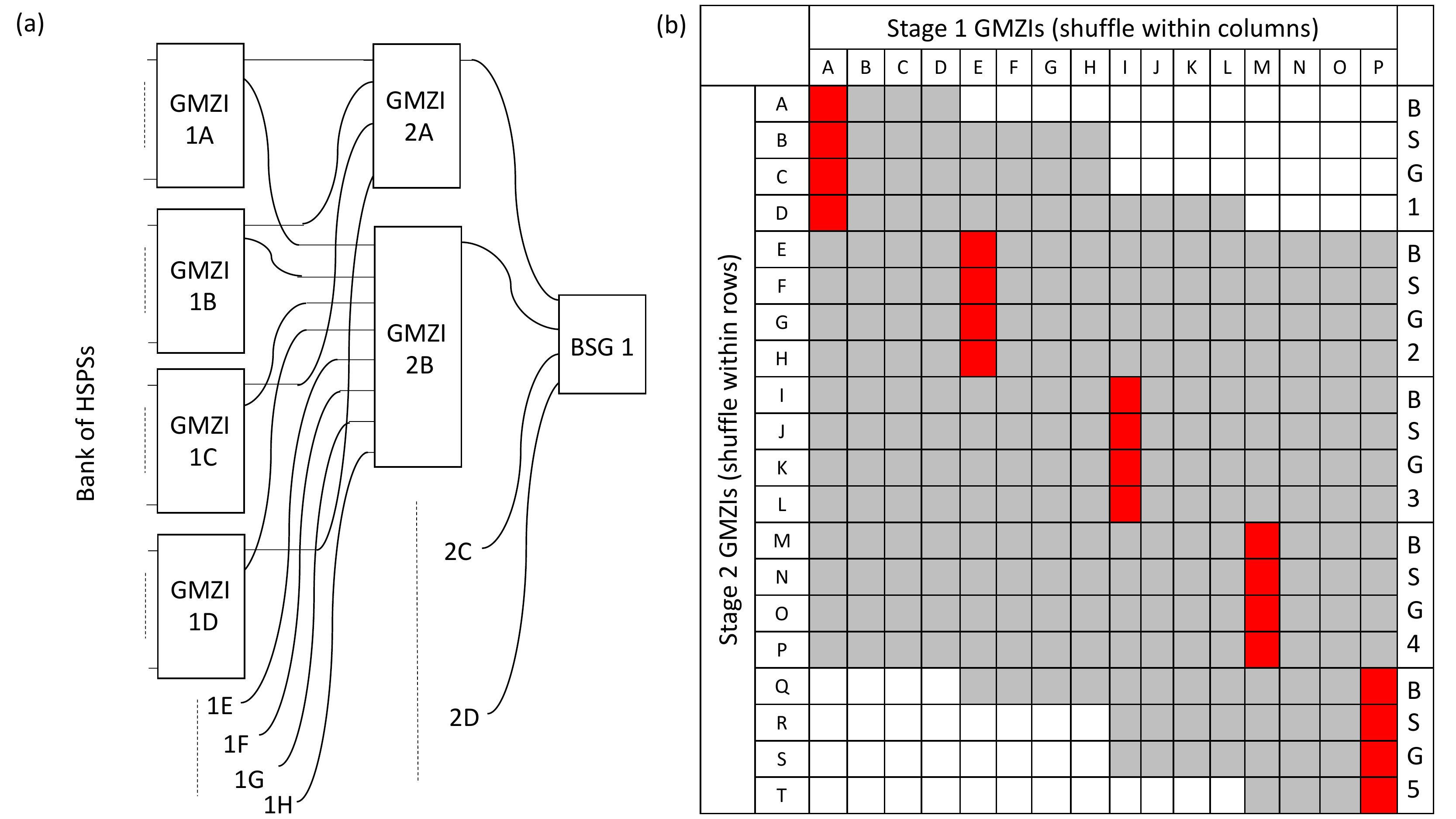}
\caption{\label{fig:bn_mux_eg} {\it Example of a mux configuration using two layers of switching to supply groups of photons to multiple circuits simultaneously.} (See description in Sec.~\ref{sec:bnmux} of the family of muxes which includes this one.)  There are $N=256$ HSPS at the input, $16\times$ $16$-to-$16$ GMZIs in the first switching layer, 20 $n$-to-$1$ GMZIs in the second switch layer of size $n=4$, $8$, $12$ and $16$, and $g=5$ BSG circuits at the output.  (a) illustrates connectivity for the first few GMZIs and the first BSG circuit, and (b) uses a grid notation to show the connectivity for the entire mux.  In this notation every gray square represents a mode which can receive an input photon; each layer 1 GMZI can rearrange the gray squares in a single column; each layer 2 GMZI can rearrange the gray squares in a single row.  The switch network attempts to fill the designated outputs in complete blocks (marked red).}  
\end{center}
\end{figure*}

\subsection{\label{sec:bnmux} Switch networks for multiple entanglement generators with shared heralded single-photon sources}

Muxes with a large number $N$ of HSPSs, which are designed to supply groups of photons to multiple circuits simultaneously, can in principle achieve high values for mux yield across a wide range of input probability $p$, as shown by the theoretical bounds in Sec.~\ref{sec:Metrics}.  Here we describe a family of muxes with two layers of switching, which are much more practical than Spanke networks for large values of $N$ and which can achieve high yield, albeit below optimal values.  The new family of two-layer muxes is defined as follows: 
\begin{itemize}
    \item[(i)] GMZIs in the first layer are connected to subsets of GMZIs in the second layer; 
    \item[(ii)] each GMZI in the second layer hosts a single mux output; 
    \item[(iii)] consecutive outputs are grouped and assigned to individual circuits.
\end{itemize}  
A huge number of mux configurations are possible with different connectivity between the switching layers and GMZIs of different sizes and types (see Sec.~\ref{sec:Commuting} for an explanation of different types of GMZI).  These options can be captured using a grid notation, which is illustrated for one example in Fig.~\ref{fig:bn_mux_eg}(a) and (b).  In this notation, a grayed square denotes a mode which can receive an input photon, and which can be rearranged by the stage 1(2) GMZIs that correspond to the label of the column(row) of the square.  GMZIs in the first switching layer rearrange photons within vertical blocks, and GMZIs in the second layer rearrange photons within horizontal blocks.  The GMZIs in the first layer implement permutations of the photons. The type of these GMZIs determines the possible permutations and therefore affects the operation of the switch network, whereas the type of the GMZIs in the second layer does not, since these only act as $n$-to-1 muxes. However, numerical studies reveal that the performance of the mux is mostly determined by the sizes and connectivity of the GMZIs and not by their type.  



\begin{figure}
\begin{center}
\includegraphics[width=0.8\columnwidth]{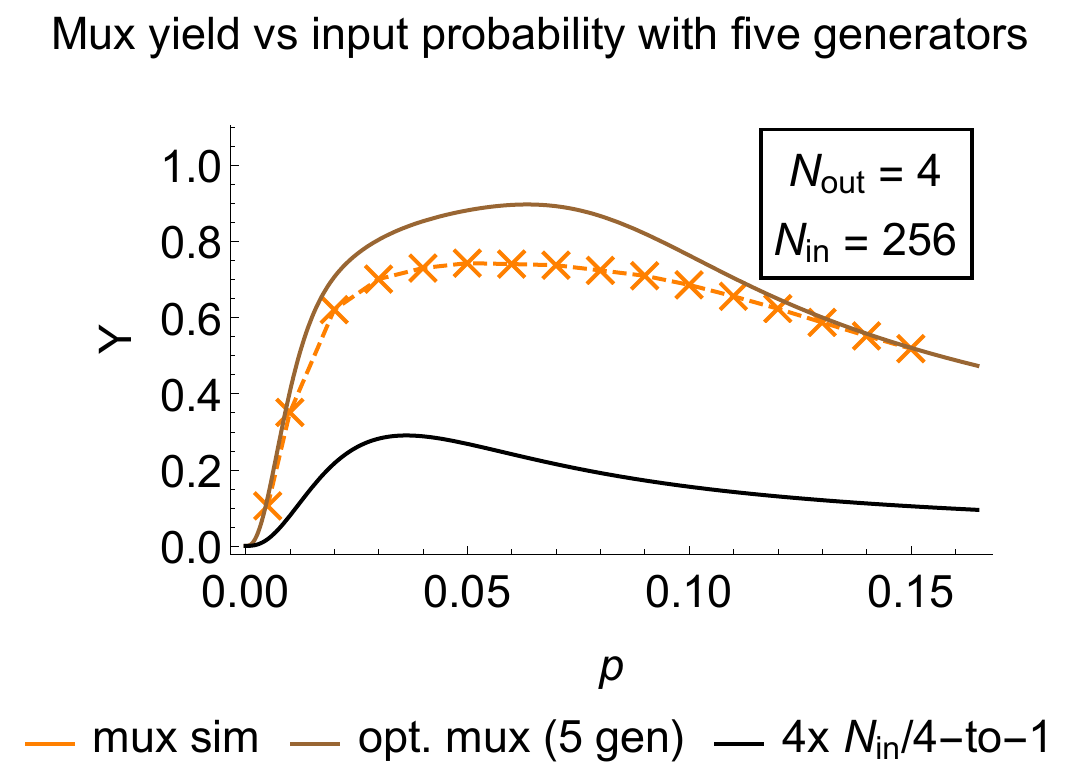}
\caption{\label{fig:bn_mux_sim} {\it Results of Monte-Carlo simulation (orange) of yield obtained using a simple routing algorithm for the mux configuration given in Fig.~\ref{fig:bn_mux_eg}.} (Assumes Hadamard-type GMZIs in the first switching layer.)  The brown line shows the theoretical bound assuming the same number of sources ($N_{\rm in}=256$), and the same number ($g=5$) and type ($N_{\rm out}=4$) of circuits at the output. The black lines shows the yield achievable using the same number of sources to feed just one circuit using a $4\times$ $N_{\rm in}$-to-1 mux strategy.  The algorithm used in the simulation attempts to fill groups of outputs consecutively using a nested loop, and can be used for the entire family of muxes described in Sec.~\ref{sec:bnmux} (a complete description is provided in Appendix~\ref{sec:bn_mux_alg}).}  
\end{center}
\end{figure}

An important consideration is the algorithm used for routing, which must select permutations in the first layer of switching to fill as many complete groups of outputs as possible.  The design of this algorithm is non-trivial as it must be simple enough to use for real-time operation while the number of combinations of possible routing operations is extremely large, meaning that it is not possible to access every permutation of the input.  The results of simulation of one simple algorithm is shown in Fig.~\ref{fig:bn_mux_sim} for the mux configuration in Fig.~\ref{fig:bn_mux_eg} (a description of the algorithm using pseudocode is provided in Appendix~\ref{sec:bn_mux_alg}). It can be seen that the yield using this algorithm is in general very much better than using a simple $4\times$ $N/4$-to-$1$ mux scheme, although it is significantly sub-optimal for intermediate values of $p$ compared to the theoretical bound with five output generators.  There are a large number of alternative mux configurations and algorithms that can be explored to achieve performance approaching the theoretical bounds for different numbers of output generators.

Overall, the schemes presented in Sec.~\ref{sec:spatial_muxes} demonstrate that the essential task of muxing groups of photons for entanglement generation can be implemented effectively in many different ways, all with low and constant switch depth, as the number of HSPS inputs increases --- which can be compared with log depth scaling that is typical for most previous proposals (see for example Ref.~\cite{Gimeno_Segovia17}).  On one hand, the addition in the schemes from Sec.~\ref{sec:hugmux} of one or two layers of MZIs to a standard setup, with $m\times$ $N/m$-to $1$ single-photon muxes and one standard BSG or GHZ generation circuit, achieves mux efficiency which is nearly or fully-optimal (as per Fig.~{\ref{fig:yieldadv}(a) and Fig.~{\ref{fig:yieldadv6}(a)).  The optimal mux efficiency enables an approximate halving of the number of required inputs (see Fig.~\ref{fig:muxadv}(b) and Fig.~\ref{fig:muxadv6}(b)), as well as an approximate halving of active switching power due to the use of smaller GMZIs.  On the other hand, the use of two layers of GMZI switching for the sharing schemes in Sec.~\ref{sec:bnmux}, where photons are allocated from a large bank of HSPS to multiple entanglement-generation circuits simultaneously, allows for even higher overall mux yield (approaching optimal values as per Fig.~{\ref{fig:yieldadv}(a) and Fig.~{\ref{fig:yieldadv6}(a) for multiple generators), but at a cost of significantly-increased optical connectivity and complexity for the mux routing logic.  In general, the sharing approach is most robust to varying input probability, although it improves yield overall at the cost of lowering the rate at which groups of photons are delivered output circuits individually (in line with Fig.~{\ref{fig:yieldadv}(b) and Fig.~{\ref{fig:yieldadv6}(b))).  Finally, fully ballistic schemes using random-input circuits are also shown to provide acceptable yield in Sec.~{\ref{sec:RandomInput}, but the elimination of switching comes at cost of increasing the number of sources by $\approx 3\times$, and is only useful when the rate of contamination from imperfect sources can be tolerated.  There are also many configurations combining random-input circuits with various amounts of muxing both before and after entanglement generation that yield approximately the same final probability for the output quantum states (Fig.~\ref{fig:mod_yield_plot} and Fig.~\ref{fig:rndinpcmp}), but with different implications for hardware implementation (such as due to optical losses occurring at different stages).

%% file: sections/temporal_muxes.tex
\label{sec:temporal_muxes}

Time mux schemes offer alternatives to spatial mux schemes which trade demands on spatial resources, such as number of single-photon sources, with longer time delays and requirements on hardware components for increased repetition rates.  In this section we present several new techniques and schemes for time muxing.  The underpinning motivation of the new schemes is to enable designs which have reduced hardware footprint and minimal layers of active switching, and which capitalize on differences in natural operating and reset speeds for different component types.

\subsection{Multiplexing using Rastering}
\label{sec:rastermux}

\begin{figure*}
\centering
\includegraphics[width=0.8\textwidth]{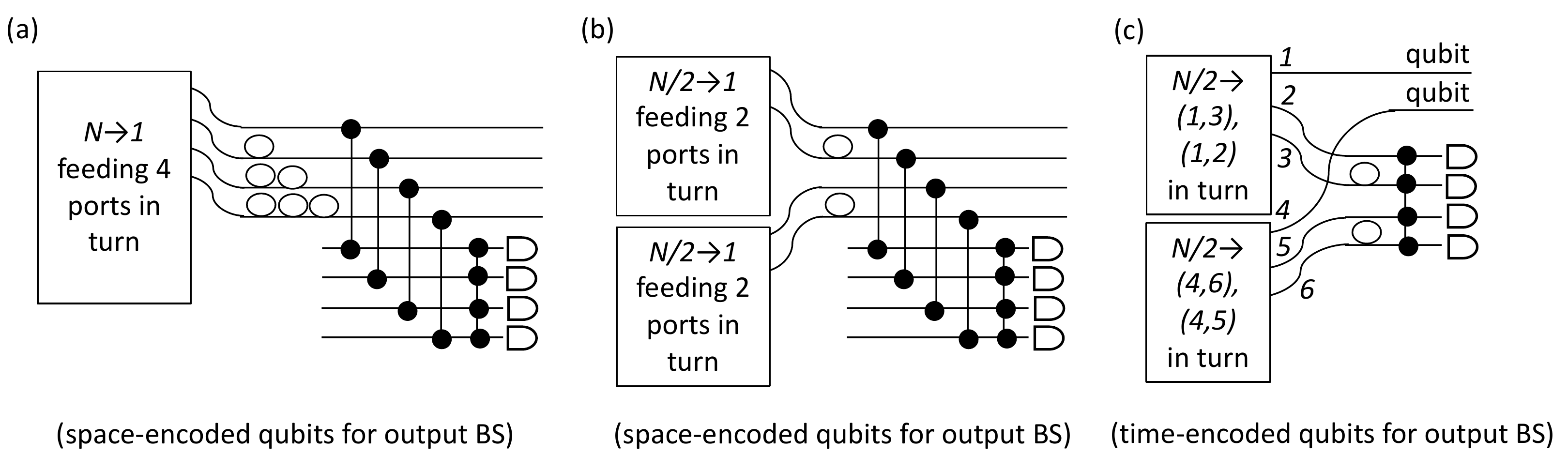}
\caption{\label{fig:raster_muxes} {\it Examples of muxes using rastering, which exploits correlations between time bins and spatial modes.} (a) A single $N$-to-$1$ mux supplies photons to each input of a BSG circuit over four steps of operation, using ``catchup delays'' to ensure that the muxed photons arrive at the BSG circuit at the same time; (b) a variation of (a) using two $N/2$-to-$1$ muxes and two steps of operation; (c) a variation of (b) and the BSG circuit which integrates temporal encoding of output qubits into  single-photon muxing, by using GMZI settings which output photons in superposition states across pairs of modes.}   
\end{figure*}

A simple but powerful muxing concept that can be applied in diverse scenarios is repurposing $N$-to-$1$ muxes using GMZIs as $N$-to-$R$ muxes by careful sequencing of the inputs, output ports and switching --- a technique that we call ``rastering''.  Two examples of rastering are shown in Fig.~\ref{fig:raster_muxes}(a,b) for muxing groups of four photons for a BSG. In Fig.~\ref{fig:raster_muxes}(a), a single $N$-to-$1$ mux supplies four photons to a BSG one-at-a-time, using delays to ensure that the photons arrive to the circuit simultaneously.  This rastering scheme uses only a quarter of the number of single-photon sources compared to using four copies of the mux in parallel, but incurs a cost in time as it takes four times as long to generate the input photons sequentially.  A variation of the idea is shown in Fig.~\ref{fig:raster_muxes}(b), using two $N/2$-to-$1$ muxes and two stages of rastering.  

\begin{figure}
\centering
\includegraphics[width=0.9\columnwidth]{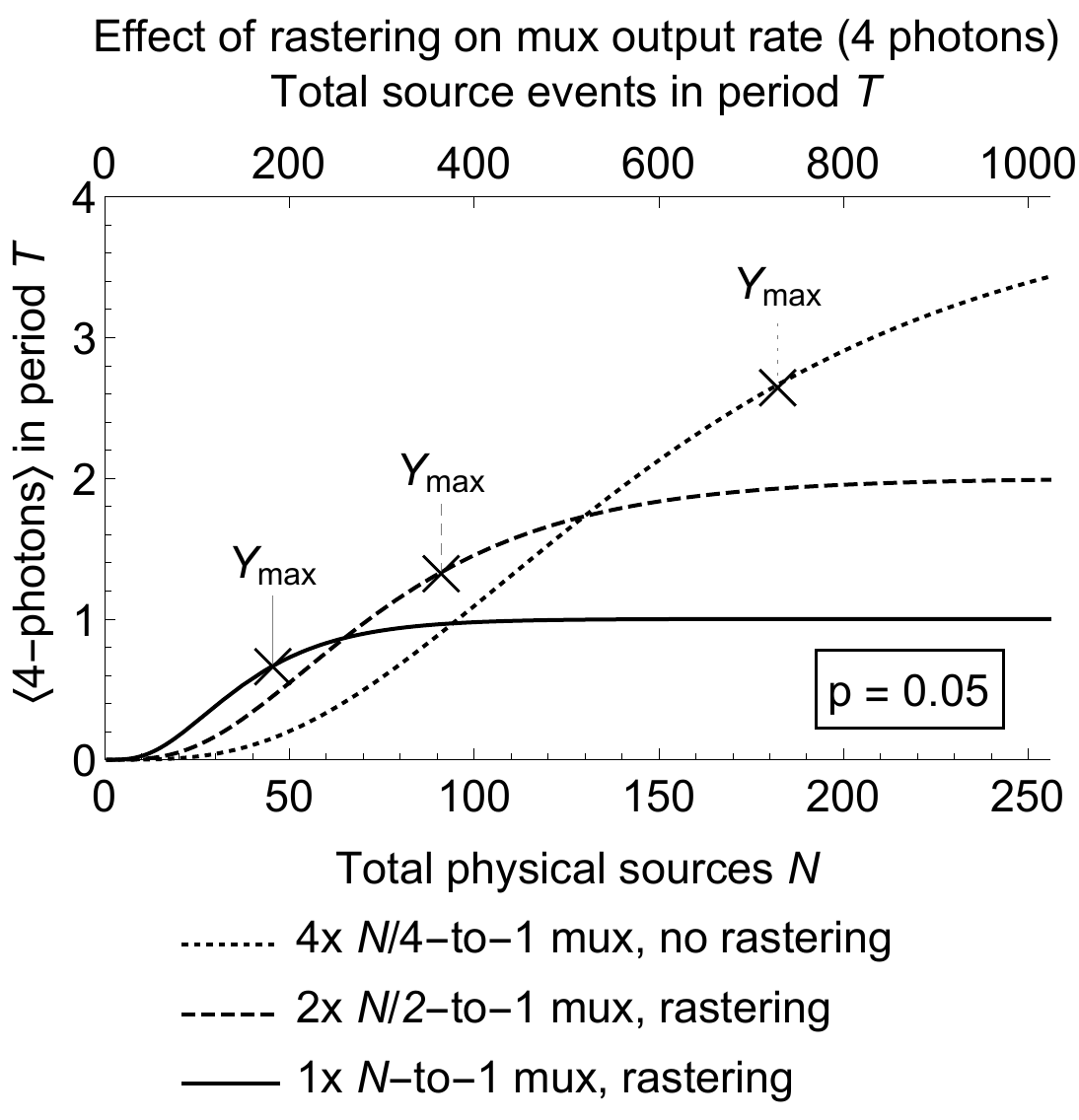}
\caption{\label{fig:raster_comparison} {\it Comparison of average rates for generating groups of four photons using muxes with either two or four rastering steps, or none at all.}  The output of the muxes is considered over time intervals of duration $T$, while the sources are assumed to operate over short time bins of duration $T/4$ with probability $p$ for heralding a single photon in each one.  The maximum achievable mux yield is the same for all mux types ($Y_{\rm max}=0.29)$.}       
\end{figure}

The way in which rastering exchanges space and time resources is slightly subtle.  The resources used for Fig.~\ref{fig:raster_muxes}(a,b) and spatial muxes are compared in Table~\ref{tab:rastering_resources}, where the number of sources and the frequency at which they are operated are the same for all schemes.  The average rates at which the strategies generate groups of photons are compared in Fig.~\ref{fig:raster_comparison}.  By taking the rates to be defined over a time interval $T$, the notion of yield in Eq.~(\ref{eq:yield_def}) also extends to cases with rastering. The maximum yield must be the same for all the strategies in Table.~\ref{tab:rastering_resources}, as they are all based around repeated ``$N$-to-$1$ type muxing'', but $Y_{\rm max}$ occurs for different numbers of physical sources and output rates, as shown in Fig.~\ref{fig:raster_comparison}.  Although the rastering and nonrastering schemes can be observed to offer the same overall efficiency when space and time resources are freely exchangeable, \textit{the rastering strategies are dramatically more efficient than spatial muxing when the number of sources $N$ is limited to small values}, which can be an important consideration experimentally.  Futhermore, even at the point where the rastering strategy (Fig.~\ref{fig:raster_muxes}(a) say) is overtaken by standard spatial muxing  (around $N = 96$), the rastering strategy maintains an advantage as it achieves the same yield --- i.e. efficiency of extraction of groups of generated photons --- but in one long time bin rather than randomly across four shorter bins so that $p_{\rm mux}$ is $\approx 4\times$ greater.  From a different perspective, the scheme in Fig.~\ref{fig:raster_muxes}(a) is more efficient than a spatial mux operated with low output probability $p_{\rm mux} < 0.25$,  and the scheme in Fig.~\ref{fig:raster_muxes}(b) is more efficient when $0.25< p_{\rm mux}<0.5$.

\begin{table*}
\centering
\begin{tabular}{|p{3.6cm}|p{2cm}|p{0.7cm}|p{1.55cm}|p{1.55cm}|p{0.8cm}|p{2.6cm}|}
\hline
Muxing strategy & GMZIs & HSPS & Input time bin duration (HSPS) & Output time bin duration (BSG) & BSGs & Average number of \hspace{0.5cm} four-photon events in period $T$ \\
\hline
(i)  $1\times$ $N$-to-$1$, rastering  &  $1\times$ size $N$  & $N$ & $T/4$ & $T$ & $1$ & $[1-(1-p)^N]^4$ \\
(ii)  $2\times$ $N/2$-to-$1$, rastering & $2\times$ size $N/2$ & $N$  & $T/4$ & $T/2$ & $1$ & $2[1 -(1-p)^{N/2}]^4$ \\
(iii)  $4\times$ $N/4$-to-$1$ & $4\times$ size $N/4$ & $N$ & $T/4$ & $T/4$ & $1$ & $4[1-(1-p)^{N/4}]^4$ \\
(iv)   Four small copies of (i) & $4\times$ size $N/4$ & $N$ & $T/4$ & $T$  & $4$ & $4[1-(1-p)^{N/4}]^4$ \\
\hline
\end{tabular}
\caption{\label{tab:rastering_resources} {\it Resources for comparable mux schemes for generating groups of four photons.}  These schemes correspond to: (i) Fig.~\ref{fig:raster_muxes}(a); (ii) Fig.~\ref{fig:raster_muxes}(b); (iii) standard spatial muxing i.e. Fig.~\ref{fig:regbsgmux}(a); (iv) multiple copies of Fig.~\ref{fig:raster_muxes}(a).  The heralded photon probability at the inputs is $p$ for a $T/4$ time bin, and output rate corresponds to the average number of groups of four photons generated over a time interval of duration $T$.}
\end{table*}

\begin{figure}
\centering
\includegraphics[width=0.8\columnwidth]{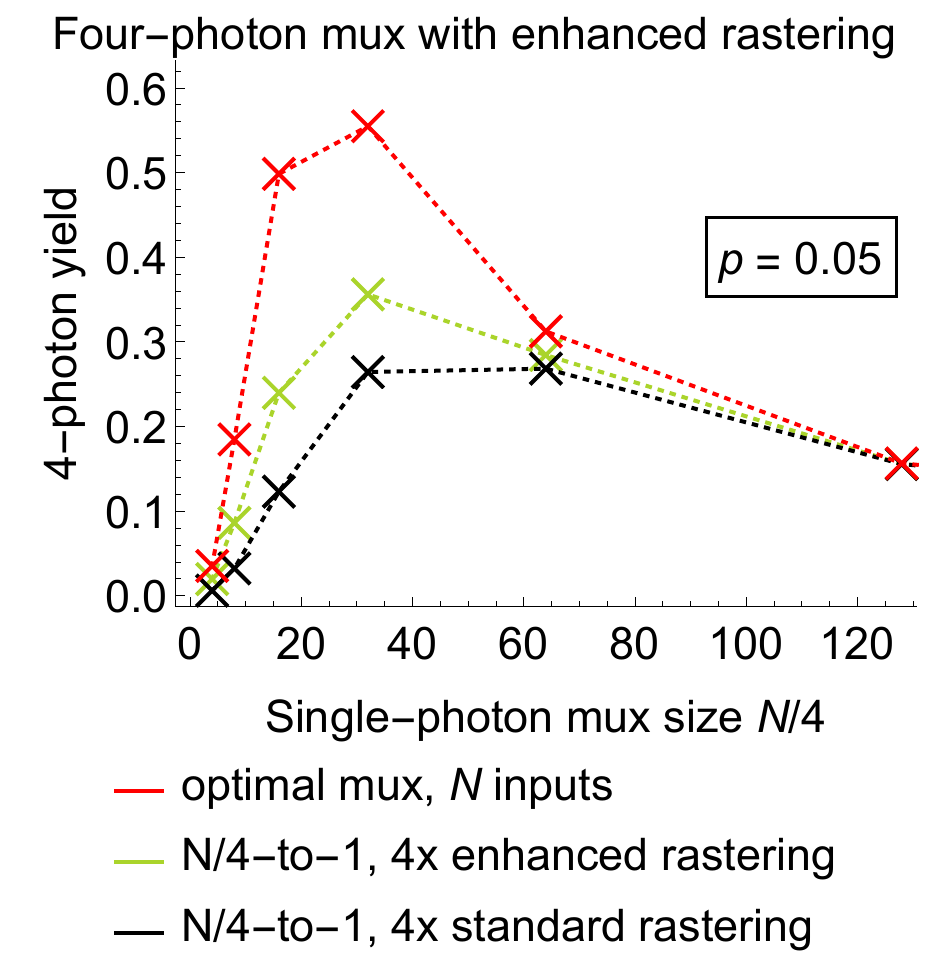}
\caption{\label{fig:enhancedrastering}
{\it Comparison of the performance of the enhanced-rastering strategy.}  The schemes compared are: (a) an optimal mux scheme using $N$ source events over a period $T$ (red); (b)
the four-step scheme in Fig.~\ref{fig:raster_muxes}(a) using regular rastering with $N/4$ sources firing four times during $T$ (black); (c) the same as (b) but using enhanced rastering (green).}
\end{figure}

Interestingly, the sequential nature of rastering can enable an improvement in mux efficiency using a strategy which we call ``enhanced-rastering''.  Differently from above, where it is assumed that the rastering schemes always cycle through all their outputs, \textit{the idea of the enhanced strategy is simply to restart the rastering process after any step where the single-photon mux fails to provide a photon}. Mux efficiency is improved in this approach, since attempts to generate four-photon groups are terminated as soon as it is known that an insufficient number of photons can be generated.  The improvement is shown in Fig.~\ref{fig:enhancedrastering} for the scheme in Fig.~\ref{fig:raster_muxes}(a) with input probability $p=0.05$.  Although the improvement in maximum yield using enhanced rastering can be seen to be fairly modest relative to an optimal mux strategy, \textit{compared to the regular rastering scheme significant gains are available at smaller mux sizes} --- which could for example enable the use of a smaller single-photon mux.  Assuming (as above) that source time bins are of duration $T/4$ and that a complete raster cycle takes time $T$, the main cost of enhanced rastering is that the output photons are located randomly in any of four output time bins of duration $T/4$, versus one of duration $T$ for regular rastering (although the output photons are always temporally aligned).  

Rastering presents key advantages compared to standard temporal mux schemes using variable delay networks (as described in Sec.~\ref{sec:SwNetReview}) as it does not need extra layers of switching. In addition, the amounts of optical loss and dispersion due to delays of different length are constant at fixed mux outputs (unlike variable delay networks), which makes it easier to compensate for the differences.  A potential cause for concern is that rastering may lead to contamination of non-target output ports from inputs with either heralded or unheralded photons.  Taking Fig.~\ref{fig:raster_muxes}(a) as an example, it can be noted that excess photons can enter the delays, but they enter at different times from the intended output.  Hence, these excess photons do not necessarily cause a problem, providing the timing resolution of the detectors in the BSG circuit is sufficient. For example, the timing resolution should accord with time bins of duration of $T/4$ for all the schemes in Table~\ref{tab:rastering_resources}, and detections in $3/4$ of these time bins should be ignored.  Alternatively, problems caused by excess photons can be minimised by operating the GMZI using a ``sliding'' window approach, so that heralded photons do not enter delays early.  This can be implemented most-easily using a DFT-type GMZI which implements cyclic permutations, and taking care to route  a heralded input to the target output only when adjacent inputs to other rastering delays do not have heralded photons (which occurs generally with high probability). This approach might be helpful when the detectors have long reset delays for instance, although it cannot prevent some unheralded photons from triggering detectors outside of the target time bin.

A final notable aspect of rastering schemes is that they provide opportunities for incorporating interconversion of spatial and time encoding of qubits for logical states, which traditionally requires added stages of switching.  An example of this is illustrated in Fig.~\ref{fig:raster_muxes}(c), which removes the usual downcoupling operations from the BSG circuit, and uses Hadamard-type GMZIs as  switchable pairwise couplers instead (as explained in Sec.~\ref{sec:AltGMZI}).  The GMZIs are configured so as to populate single output modes for time-encoded qubits, and alternating inputs of the BSG measurement circuit, using catchup delays to ensure that inputs for the measurement circuit arrive simultaneously.  As noted previously, heralded photons from the sources must be paired with heralded vacuum states to prevent unwanted (heralded) photons from affecting the measurement.
 
\subsection{Permutation networks using rastered switching}
\label{sec:rastering_permutation_networks}

\begin{figure*}
\centering
\includegraphics[width=0.8\textwidth]{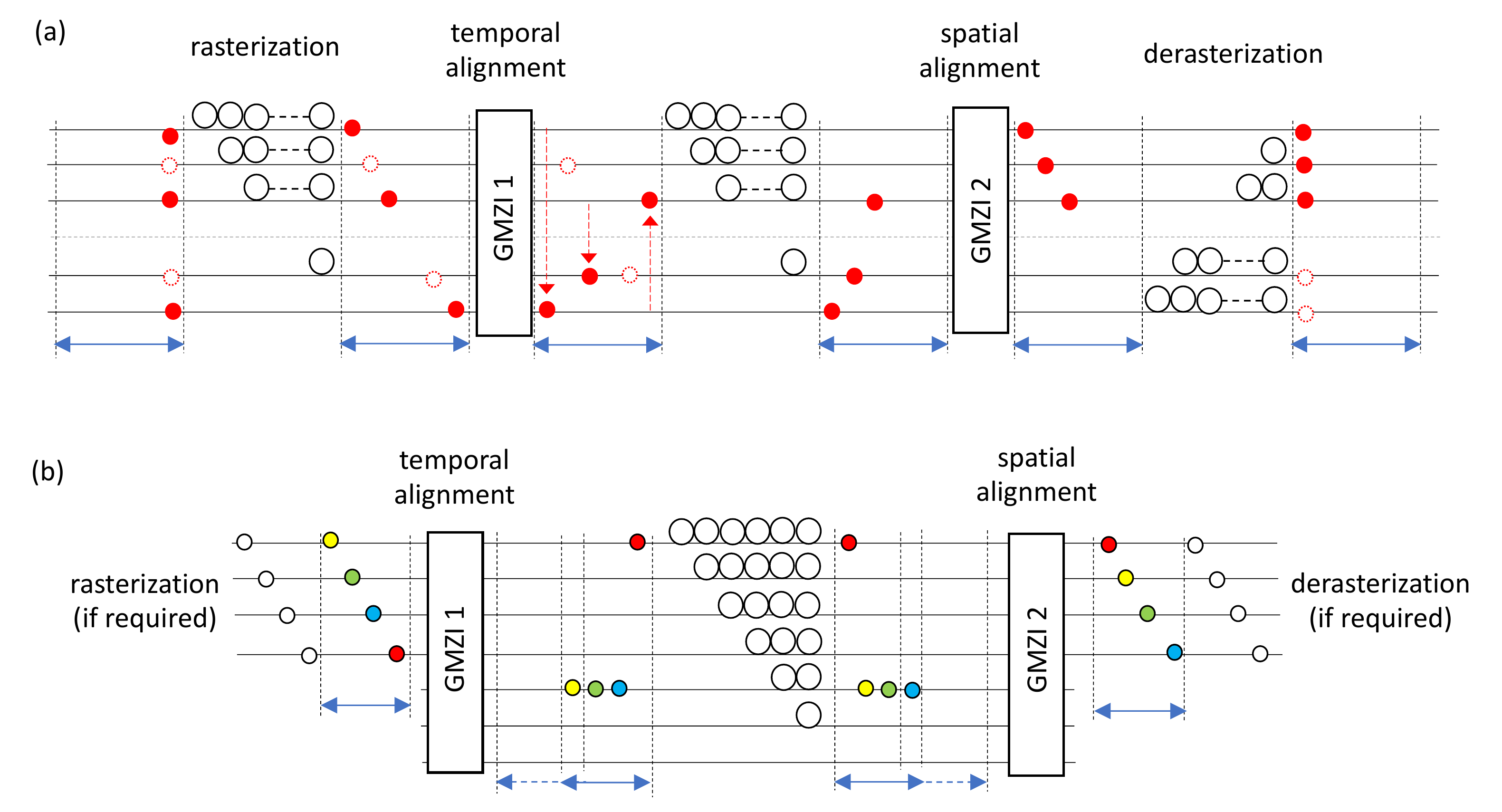}
\caption{{\it $R$-to-$R$ permutation networks.} The tasks performed are: (a) rearrangement of $R$ inputs to fill outputs consecutively from the top; (b) arbitrary rearrangement of the  inputs.  (a) requires GMZIs of size $R$ and delays between $0$ and $R-1$.  (b) requires GMZIs of size $2R-1$ and extra modes to accommodate delays between 0 and $2(R-1)$.  In addition,  to ensure the timing of the output is independent of the permutation operation, delays must align to the back of a timing window defined over $2R-1$ bins (see dashed lines) although the input state itself is spread over $R$ bins (the same as for (a)).
\label{fig:temporal_spanke}}       
\end{figure*}

Another form of rastering is useful for enabling functionality equivalent to Spanke networks but in compact switch networks.  The essential ideas are shown in Fig.~\ref{fig:temporal_spanke}(a) and (b): (a) is designed to gather all occupied inputs (e.g. single photons) into consecutive outputs, and (b) implements arbitrary permutations, typically for distinguishable inputs.  In both schemes the input state, which is assumed to arrive synchronously in $R$ spatial modes, is spread over $R$ time bins to enable the GMZIs to route the inputs independently, and \textit{arbitrary permutations are possible using two stages of temporal and spatial rearrangement}.  From a resource-counting perspective, the cost of these schemes amounts to the need for expanded time bins that can accommodate $R$ switching operations at the GMZIs.  Note however that the size of these expanded time bins (and also the sizes of the delays that are used) are dictated only by the time it takes to reconfigure the GMZIs with different switch settings and by the photon transit time.  The schemes in Fig.~\ref{fig:temporal_spanke} are likely to be especially useful when switching is fast compared to the speed of other hardware components (such as the detector deadtime and the repetition rate of single-photon sources).



\subsection{de Bruijn switch networks}
\label{sec:de_Bruijn}

The switch network schemes discussed above exploit time muxing either to fill a sequence of individual outputs using rastering (Sec.~\ref{sec:rastermux}), or to access an increased number of permutation operations by rearranging mux inputs individually rather than all-at-once (Sec.~\ref{sec:rastering_permutation_networks}).  Another diverse category of time mux schemes rearrange subsets of inputs in each time step. One straightforward implementation of such schemes can be the use of two GMZI per spatial mode, with a set of $(0,1,..,L)$ delays sandwiched in between, as described in figure \ref{fig:intro_temporal_muxes}. This allows the synchronization of events in groups by appropriately choosing the delay for each mode. A different approach is to use a single switching device to switch all spatial modes together, here we will give one example of such a scheme which is designed to exploit the mathematical properties of de Bruijn sequences \cite{deBruijn46}. 

\begin{figure*}
\centering
\includegraphics[width=0.8\textwidth]{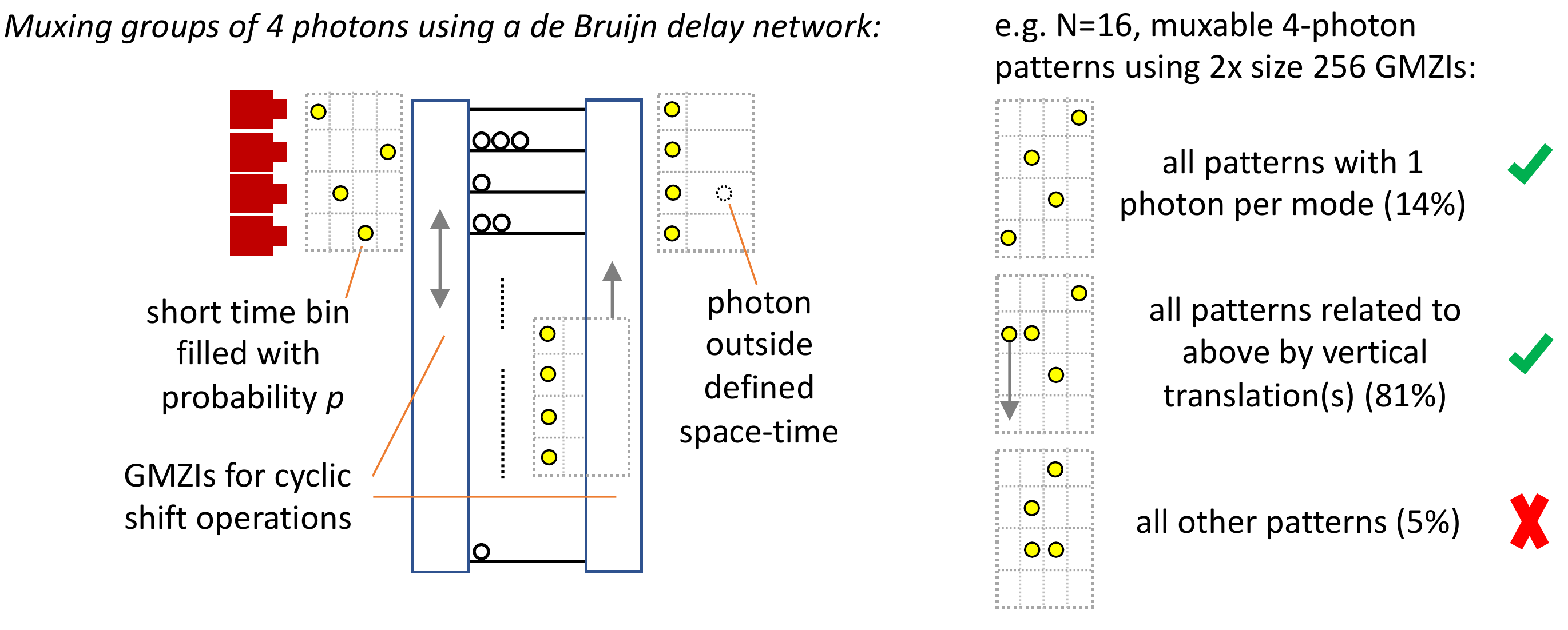}
\caption{\label{fig:deBruijn}
{\it Example of a time mux using a de Bruijn-sequence delay network.}  Every pattern of input space-time bins with one photon per mode has a corresponding subsequence of delays which can be used to achieve temporal realignment of the input photons.  The subsequence is selected by the first GMZI using a (cyclic) translation operation, and the second GMZI is used to shift the photons to the mux output ports.  Improved mux efficiency can be achieved by changing the setting of the first GMZI for different input time bins. For the specific mux configuration illustrated, almost all patterns of four input photons can be muxed by (see right).  Similar mux networks can be defined for other values of the number of sources and time bins, and the output group size.}
\end{figure*}

The general scheme consists of two DFT-type GMZIs connected by a set of delays; $m$ sources are connected to the first GMZI and fire once per time bin, producing a photon with probability $p$. The switching network acts on groups of events in the muxing time window, which consists of $N/m$ time bins and $m$ spatial modes, and contains $N$ source events in total. The specific example of $N/m = m = 4$ is illustrated in Fig. 23. Once the muxing time window is complete, there are $\left(\frac{N}{m}\right)^m$ possible time bin configurations with one photon per spatial mode. The first GMZI selects one of the available configurations and performs a (cyclic) permutation of the inputs such that the chosen photons are sent through a sequence of delays that aligns them temporally. So, the network needs to contain all the sequences of $m$ adjacent delays needed to synchronize all the possible time bin configurations. It is desirable for the total number of delays to be as small as possible, since this determines the size of the GMZIs in the scheme. An optimally small set of delays with all the required sequences can be found by setting the delay lengths according to a de Bruijn sequence with alphabeth $A=\{0, \ldots,m-1\}$ and length $L=m$. The number of delays required is $\left(\frac{N}{m}\right)^m$ and can be reduced to $\left(\frac{N}{m}\right)^m-\left(\frac{N}{m}-1\right)^m$ if the photons do not need to be output in the last time bin of the muxing time window. In cases when it's unimportant which output spatial modes contain the groups of photons, the second GMZI can be removed.

An important observation is that greater mux efficiency can be attained with the de Bruijn network using a ``Tetris-like'' strategy, where the first GMZI can be reconfigured during the muxing time window to cyclically shift  photons between modes in specific time bins.  For the example in  Fig.~\ref{fig:deBruijn}  with four sources and four time bins, almost all input configurations with four photons can be successfully muxed this way.  As an example, with input probability $p=0.25 $, the output probability is only $p_{\rm mux}=0.22$ using a single configuration of the network (equivalent to using four standard depth-two temporal delay networks in parallel), but as high as $p_{\rm mux}=0.56$ using the extra input patterns --- and the relative  improvement in $p_{\rm mux}$ is even greater for lower values of $p$, see Fig.~\ref{fig:yieldadv}(a).

\begin{figure*}
\centering
\includegraphics[width=0.8\textwidth]{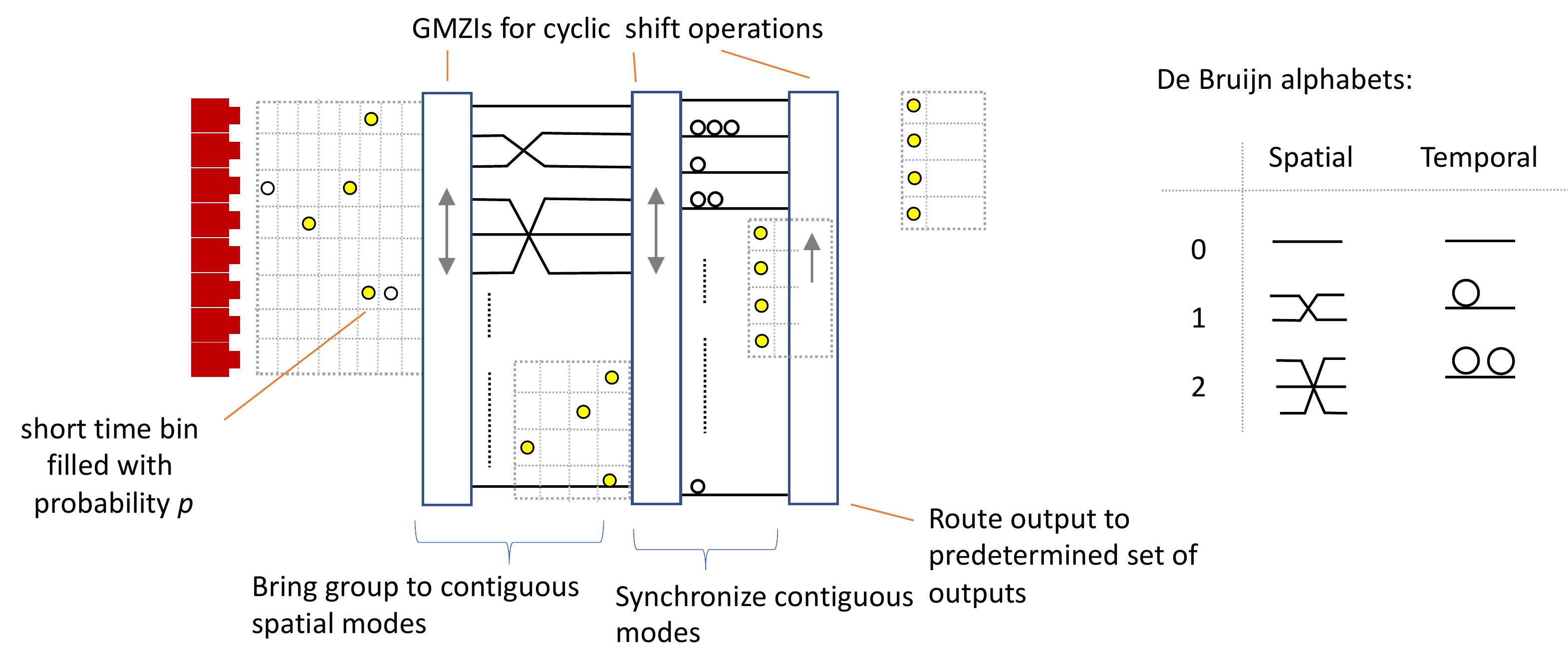}
\caption{\label{fig:full_deBruijn}
{\it Design of a de Bruijn-based spatio-temporal switch network.} Groups of $n$ photons, four in the case shown, are found in a larger set of modes and routed to a predetermined output. this switch network uses one cyclic GMZI and a set of crossings to first bring a group of photons (filled circles) to contiguous modes, a second cycle GMZI followed by a set of delays synchronizes the photons, and a final cyclic GMZI brings the selected spatial modes to a predetermined output. If there is multiple outputs which could serve (such as in some cases mentioned in Ref.~\cite{PsiEntGen}), the last GMZI can be omitted. On the right of the figure, we show the translation of the de Bruijn alphabet to mode swaps and delays.}
\end{figure*}

However, to enable reasonable values for $p_{\rm mux}$ for small values of $p$, it is necessary to increase the number of source events $N$. In order to avoid increasing the size of the GMZIs beyond what is practical, this can be achieved, for example, by doing some spatial pre-muxing of the sources (using any of the muxes presented in previous sections), or by using more than $m$ sources and extracting photons from variable groups of spatial modes. The latter strategy can be implemented using a spatial version of the de Bruijn network, where the set of delays is replaced by a network of crossings where the de Bruijn sequence determines the distances between pairs of modes to be swapped \footnote{In this case, the length of the de Bruijn sequence in no longer the same as the number of modes in the GMZI since larger mode swaps require more modes, leading to worse scaling of the network size.}. This results in a spatio-temporal de Bruijn network, like the one shown in Fig. 24. This consists of three cyclic GMZIs, a set of crossings which allow to place $n$ photons from $m>n$ sources into adjacent spatial modes, and a set of delays which allow to synchronize the photons. The last GMZI can be removed if the photons do not need to be output in specific spatial modes. The de Bruijn sequences for the delays and the crossings need not be the same (see Appendix~\ref{sec:dB_sequences}).

The yield of the network can be optimized while keeping a fixed optical depth by applying the enhanced rastering technique in Sec.~\ref{sec:rastermux} to the de Bruijn network. In this case the muxing time window is divided into smaller batches, i.e. groups of time bins and spatial modes,  which are multiplexed simultaneously and routed to different groups of output ports by the final GMZI, e.g. to feed different generators.


\begin{figure*}
    \centering
    \includegraphics[width=0.9\textwidth]{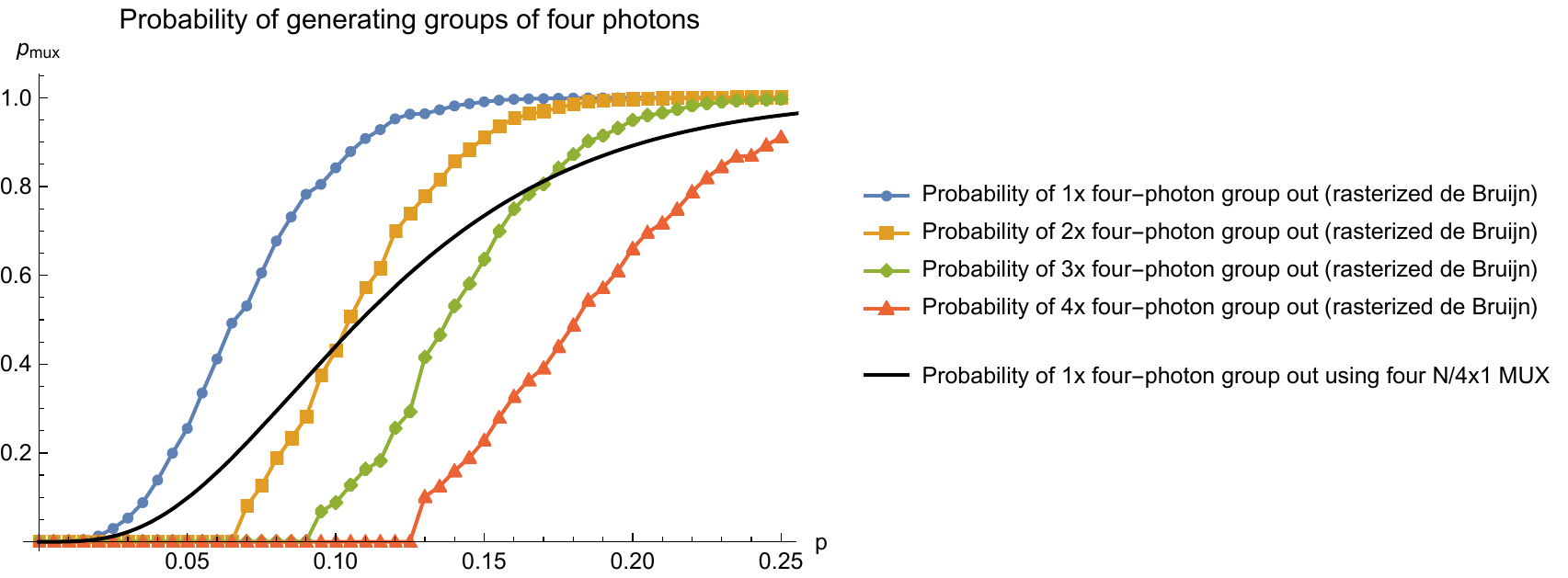}
    \caption{\label{fig:rasterized_deBruijn} {\it Success probabilities for extracting multiple groups of photons from a batch of $N_{\rm total}=256$ spatio-temporal bins.} The scheme uses a rasterized version of the de Bruijn spatio-temporal switch network. In this case the maximum delay and maximum crossing is two. }
\end{figure*}

Fig.~\ref{fig:rasterized_deBruijn} shows the probability of obtaining one, two, three and four groups of four photons from a total of $N_{\text{total}}=256$ spatio-temporal bins ($p_\text{mux}$) against the probability of each input being populated ($p$). The scheme used in the simulation has 16 spatial modes and optimizes the yield to determine whether a single group of four should be extracted from all the modes or whether it is better to divide them into smaller batches to output a higher number of photons. Plotted for comparison is the probability of generating a single group of four photons from four independent $N$-to-$1$ muxes with $N=N_{\text{total}}/4=64$.

The schemes presented in this section demonstrate that using time as a degree of freedom when muxing photons is highly advantageous, as they can achieve high yield while reducing the number of physical resources needed dramatically. Furthermore, time muxing can be achieved with a low constant active phase-shifter depth regardless of the number of modes, a significant improvement over previous proposals and demonstrations~\cite{Kaneda15,Gimeno_Segovia17} which relied on schemes with logarithmic depth of phase shifters in the number of modes.


%% file: sections/conclusions.tex
Switch networks built from connected GMZIs can play the same role for optical routing in a photonic quantum computer as switch fabrics based on cross-bar switches do for many classical communications tasks.  Therefore Sec.~{\ref{sec:GMZIs}} explored in depth the properties of GMZIs which are pertinent to strategies for generating resource states.  We have:
\begin{itemize}
	\item fully characterized all permutation operations enabled by one layer of GMZIs (Sec.~\ref{sec:Commuting}), and proved that non-permutation operations can be performed on single-photon states to enact operations with fewer steps than using classical permutation switching (Sec.~\ref{sec:AltGMZI}). 
	
	\item shown how native GMZI operations can be used to integrate additional functionality into muxes (Sec.~\ref{sec:Multimultiplexing}) e.g. to recover entangled states generated in a non-standard form using controllable swaps, or to incorporate time encoding of qubits.  

	\item provided natural physical circuit constructions for the passive networks in GMZIs in terms of beam-splitters/directional-couplers and crossing networks, with log-depth stages for Hadamard-type GMZIs (Sec.~\ref{sec:GMZIimplementation}).  In addition, we have proven a minimum bound on phase-swing for active phase shifters ($>\pi/2$ for $>2$ permutation operations) and given examples of GMZIs with reduced phase-swing requirements (Sec.~\ref{sec:AltGMZI}).
\end{itemize}

In Sec.~\ref{sec:spatial_muxes} we presented several new mux strategies for preparing groups of photons with intrinsic benefits compared to na{\"i}ve muxes with the same number $N$ of HSPSs and single-photon herald probability $p$.  Entanglement-generation circuits can typically receive input photons in a larger number of patterns than are typically used, and we have shown that:
\begin{itemize}
	\item	simple switch networks (e.g. one layer of MZIs) can vastly increase the probability of usable patterns of input photons e.g. leading to approximately $10\times$ improvement in yield for a standard GHZ-state generator when $Np \ll 6$ (Sec.~\ref{sec:hugmux}). 

	\item	Bell-state entanglement can be generated at the same rate as a $4\times N/4$-to-$1$ strategy but without any single-photon muxing, just with blocking switches to dump excess input photons. A fully ballistic strategy without any switching imposes a cost of approximately $3\times$ in the number of sources, but with the considerable advantage of reduced feedforward operation (Sec.~\ref{sec:RandomInput}).

	\item Bell states generated in random modes can be switched to pre-assigned output mode bundles with low cost e.g. $>70\%$ of attempts using only one layer of MZIs, and $100\%$ of attempts using two layers of MZIs (Sec.~\ref{sec:RandomInput}). 
\end{itemize}

On the other hand, we have described simple switch networks which can be added to standard $4\times N/4$-to-$1$ ($6\times N/6$-to-$1$) mux strategies to make them optimal, with the entire muxes being being four (six) times smaller than Spanke networks while having comparable active switch depth (Sec.~\ref{sec:hugmux}).  In addition, we have exhibited the advantages of ``sharing'' mux strategies, taking a specific example of a mux which outputs groups of four photons to five entanglement-generation circuits simultaneously.  Compared to a $4\times N/4$-to-$1$ strategy, this new type of mux achieves a minimum relative improvement in yield of $2.5\times$ at $p= 0.03$, with much greater gains for smaller and larger values of $p$, up to $5\times$ for large values (Sec.~\ref{sec:bnmux}). 

In Sec.~\ref{sec:temporal_muxes}, we introduced powerful new time-mux strategies showing specifically that:
\begin{itemize}
	\item rastering strategies can be used to maintain any given target mux success probability with a reduction in the number of sources by a factor equal to the number of output photons (Sec.~\ref{sec:rastermux}).
	\item switch networks can be made perfectly efficient by incorporating rastering (Sec.~\ref{sec:rastering_permutation_networks}).
	\item for an example using a de-Bruijn-type delay network, sequencing of switch operations enables a $\simeq6\times$ increase in the number of usable patterns of input photons compared to a na{\"i}ve mux strategy (Sec.~\ref{sec:de_Bruijn}).
\end{itemize}

%% file: sections/appendices.tex
\newpage
\onecolumngrid

\section{\label{sec:GMZIAppendix} Generalized GMZI constructions} 

{\it Lemma:} For a GMZI with transfer matrices of the form $U_k = WD_kV^\dagger$ acting as a $N$-to-$1$ mux, $V$ must be a complex Hadamard, and the phase vectors ${\mathbf d}_k$ corresponding to $D_k$ for different settings must form an orthonormal set.

{\it Proof:} For a GMZI with $U_k=W D_k V^\dagger$ acting as a $N$-to-$1$ mux, without loss of generality assuming routing into the first output,
we have
\[
  \bigl(U_k\bigr)_{1,t} = \langle {\mathbf w}_1, D_k^\ast {\mathbf v}_t \rangle = \delta_{k,t} \label{eqn:gmzi:permutation}
\]
with row vectors of $W$ and $V$, ${\mathbf w}_i$ and ${\mathbf v}_i$, respectively.

Due to $D_k$ being a diagonal phase matrix, we get $\vert {\mathbf v}_t \vert = \vert {\mathbf w}_1 \vert$
(with $\vert\cdot\vert$ being the element-wise absolute value).
By the Hadamard inequality for determinants,
\[
  \det(V) \leq \prod_{k=1}^N \sqrt{N\vert W_{1,k}\vert^2}=N^{N/2} \sqrt{\prod_{k=1}^N \vert W_{1,k}\vert^2} \,.
\]
Equality holds due to orthogonality of $V$'s column vectors.
The inequality of arithmetic and geometric means states
\[
  \sqrt{\prod_{k=1}^N \vert W_{1,k}\vert^2} \leq = \bigl(\|w_1\|^2/N\bigr)^{N/2} = N^{-N/2} \,,
\]
so $\det(V) \leq 1$. Due to unitarity we know equality must hold.
According to the inequality above, equality holds if and only if $\vert W_{1,1}\vert^2 = \ldots = \vert W_{1,N}\vert^2$
i.e. $V$ is a complex Hadamard matrix ($\vert V_{s,t}\vert=N^{-1/2}$).

Given that $\langle {\mathbf w}_1, D_k^\ast {\mathbf v}_t \rangle = \delta_{k,t}$ by assumption, it is also true that
$\langle {\mathbf d}_k, \diag\{{\mathbf w}_1^\ast\}{\mathbf v}_t \rangle = \delta_{k,t}$.
Now, orthonormality of the ${\mathbf v}_t$ implies orthonormality of the $\diag\{{\mathbf w}_1^\ast\}{\mathbf v}_t$
and thus orthonormality of the ${\mathbf d_k}$.

{\it Comment:} One way to search for sets of orthogonal phase  $\{ {\mathbf d}_{k^\prime} \}$ is to search for largest cliques on graphs with vertices corresponding to all possible phase-shifter vectors defined from a set of values of interest, and edges defined between vertices corresponding to orthogonal vectors.  This method was used to find the example in Table~{\ref{tab:exotic_gmzi}}.

\onecolumngrid

\section{\label{sec:MetricsExtra} Theoretical bounds for single-photon muxes outputting groups of six photons}

\begin{figure*}
\begin{center}
\includegraphics[width=0.8\textwidth]{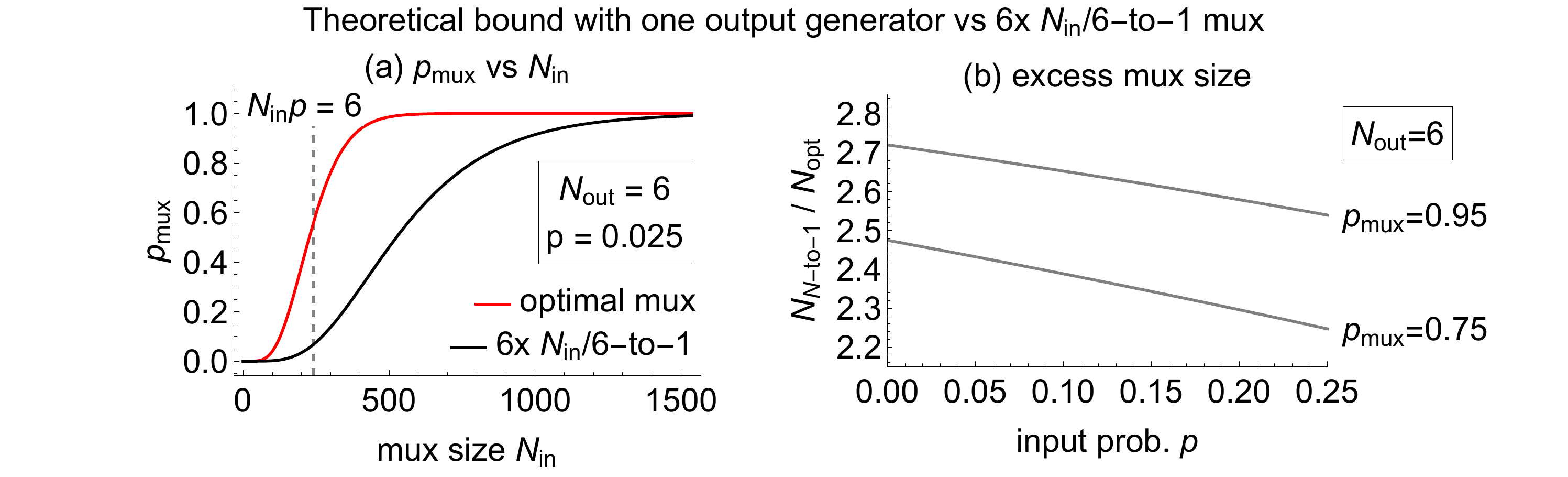}
\caption{{\it Muxing six photons for one GHZ generation circuit.} (a) Comparison of the performance of $6\times$ $N/6$-to-$1$ muxes and an optical mux with perfect $N$-to-$6$ routing capability; (b) ratio of the number of required inputs for $6\times$ $N/6$-to-$1$ versus an optimal mux, as a function of input probability $p$ for two target $p_{\rm mux}$ values.
\label{fig:muxadv6}}       
\end{center}
\end{figure*}

\begin{figure*}
\begin{center}
\includegraphics[width=0.8\textwidth]{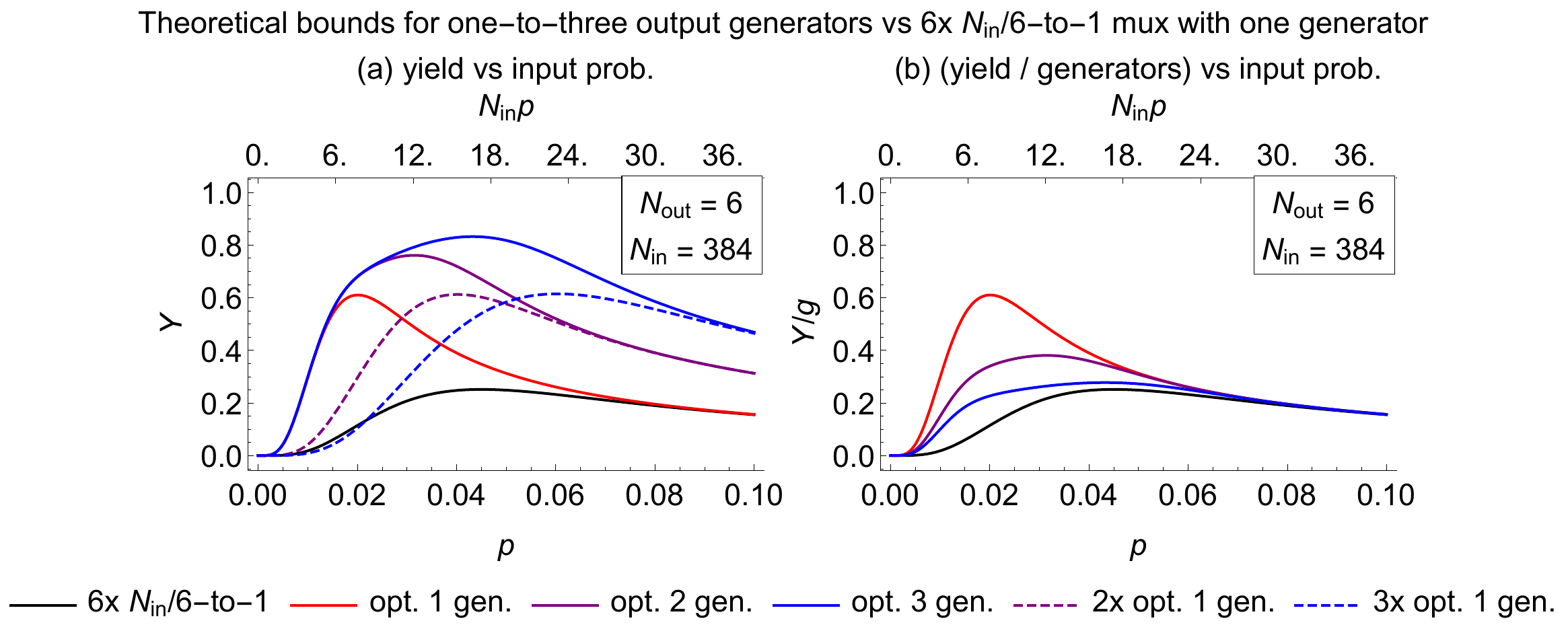}
\caption{{\it Comparison of best-achievable yield for muxing groups of six photons for different strategies with the same total number of HSPSs.}  The strategies are: $6\times$ $N/6$-to-$1$ with one generator (black), optimal muxes feeding one-three generators jointly (solid, colored), and one-three independent muxes supplying generators individually at the optimal level (dashed, colored).  (a) shows overall yield of groups of six photons, and (b) shows the overall yield divided by the number of output circuits. 
\label{fig:yieldadv6}}
\end{center}
\end{figure*}

See Fig.~{\ref{fig:muxadv6}} and Fig.~{\ref{fig:yieldadv6}} for analogs of Fig.~{\ref{fig:muxadv}} and Fig.~{\ref{fig:yieldadv}} respectively, for entanglement-generation circuits needing six photons at the input.

\section{\label{sec:reverse_hugmux_efficiency} Routing efficiency using a single-layer of MZIs to route randomly-distributed Bell-state rails into disjoint sets of modes}

The scheme referred to here is shown in Fig.~\ref{fig:rndinpschemes}(c). It uses MZIs to rearrange Bell states so that one logical rail occupies each of the groups of modes which are labeled 1, 2, 3 and 4.  Referring to MZIs with output labels (1,2)/(2,3)/(3,4)/(4,1) as type A/B/C/D respectively, Bell states can be successfully rearranged for the following cases: each rail is input to a different MZI type; two rails are input at one MZI type, and the other rails are input to different MZI types e.g. 1 type A, 2 type B, 1 type D; two rails are input at type A and two at C; two rails are input at type B and two at type D.  The Bell states cannot be rearranged if three or four rails are input at one MZI type, or if pairs of rails are input at adjacent MZI types e.g. MZI types A and B (which only cover mode sets 1, 2 and 3).  The scheme works for $45/64=70.3\%$ cases, assuming that all subsets of four modes are equally likely at the input.  In principle it is possible to do slightly better, as Bell states can be generated across some different subsets of modes in cases where $>4$ photons are heralded at the start, and the subset can be selected by the configuration of the blocking switches.

\newpage


\section{\label{sec:logic_bsg8_muxes} Routing logic for muxes in Sec.~\ref{sec:hugmux}}

\begin{table}[h]
\begin{center}
\begin{tabular}{|p{4cm}|p{5cm}|p{5cm}|} 
\hline
\textbf{Mux type} & \textbf{Routing logic} & \textbf{Hardest step and simplification} \\ 	
\hline
$4 \times $ $N/4 \! \rightarrow \! 1$ single-photon muxes for input to standard BSG, Fig.~\ref{fig:regbsgmux}(a) & Each single-photon mux determines the first input port with a photon, and uses a lookup table to output phase shifter settings.  An output herald bit indicates if all muxes succeed. & Implementing large priority encoders to determine the top heralding input port (from multiple) is nontrivial in hardware. \\
\hline
$8 \times $ $N/8 \! \rightarrow \! 1$ single-photon muxes for input to standard BSG, Fig.~\ref{fig:regbsgmux}(b) & The logic above is modified as follows: the individual muxes must output the vacuum state on demand; the muxes must work in pairs to supply one photon and one vacuum state; and the output must indicate photons in bottom rails. & Modifications are quite simple. \\
\hline
$8 \times$ $N/8$ 
($12 \times$ $N/12$) $\rightarrow \! 1$ single-photon muxes with additional MZIs for input to standard BS (GHZ) generator circuit, Fig.~\ref{fig:regbsgmux}(c) & The single-photon mux logic is as above. There is also a requirement to coordinate all the muxes and the MZIs to obtain useful patterns of output photons. & The logic has single-photon mux heralds as input bits, while output bits are for: selecting photon vs vacuum at the muxes, MZI settings, and the final arrangement of photons.  The 256 (4096) possible input patterns can be reduced to 66 (666) using wildcards (following Fig.~\ref{fig:logicreduction}). \\
\hline
\end{tabular}
\caption{\it Implementation of routing logic for muxes in Sec.~\ref{sec:hugmux}}
\label{tab:logic_bsg8_muxes}
\end{center}
\end{table}

\section{\label{sec:logic_rnd_inp_muxes} Routing logic for muxes in Sec.~\ref{sec:RandomInput}}

\begin{table}[h]
\begin{center}
\begin{tabular}{|p{4cm}|p{5cm}|p{5cm}|} 
\hline
\textbf{Mux type} & \textbf{Routing logic} & \textbf{Hardest step and simplification} \\ 	
\hline
Random-input approach with ballistic generation of photons at $N$ sources with $N_{\rm BSG}\!=\!N$, Fig.~\ref{fig:rndinpschemes}(a) & The output bits indicate: if exactly four photons are generated in different mode pairs, the pairs involved, and which photons originate in bottom rails. 
& No processing of classical signals is required before entanglement generation which relaxes demands on feedforward delay. \\
\hline
Random-input approach combined with small single-photon muxes, so that total sources $N=N_{\rm bsg}\times N_{\rm mux}$, Fig.~\ref{fig:rndinpschemes}(b)(i) & 
The muxes must work in pairs as in Table~\ref{tab:logic_bsg8_muxes} for the scheme in Fig.~\ref{fig:regbsgmux}(b). The mux pairs must be coordinated to select four photons, and the output photon pattern must be provided. & The logic takes $N_{\rm BSG}/2$ input bits from pairs of  muxes; output bits select photon vs vacuum at the muxes, and provide the final arrangement of photons.  $2^{N_{\rm BSG}/2}$ possible input patterns can be reduced to  $\binom{N_{\rm BSG}/2}{4}$ using wildcards (as per Fig.~\ref{fig:logicreduction}). \\
\hline
Random-input approach with $N$ sources and blocking switching as in Fig.~\ref{fig:rndinpschemes}(b)(i) with $N_{\rm mux}\!=\!1$ or Fig.~\ref{fig:rndinpschemes}(b)(ii). & The blocking switches act jointly to select four photons.  If this is after downcoupling then it is not possible use cases where two photons are heralded at paired modes. & The logic for selecting four photons using the blocking switches must deal with $2^{N/2}$ possible input patterns which can be reduced to $\binom{N/2}{4}$ using wildcards (as per Fig.~\ref{fig:logicreduction}). \\
\hline
Use of a single-layer of MZIs at the output of a BSG circuit with $N_{\rm BSG}>8$ to rearrange the Bell-states (in four modes from $N_{\rm BSG}/2$) into discrete mode bundles, Fig.~\ref{fig:rndinpschemes}(c) & Input bits provide the location of the Bell state.  The output bits must indicate MZI settings, whether successful rearrangement was possible, the final modes with the Bell-state, and any swaps internal to the Bell state. & 
The logic for coordinating the MZIs and providing information on the mux operation must be defined for $\binom{N_{\rm BSG}/2}{4}$ patterns of rails. \\
\hline
\end{tabular}
\caption{\it Implementation of routing logic for muxes in Sec.~\ref{sec:RandomInput}}
\label{tab:logic_rndinput_muxes}
\end{center}
\end{table}

\section{\label{sec:bn_mux_alg} Routing logic for muxes in Sec.~\ref{sec:bnmux}}

The pseudocode below is for a simple routing algorithm that can be used with the family of muxes described in Sec.~\ref{sec:bnmux} with two layers of switching and multiple output circuits.  (Note that terms such as row and column blocks are explained in the main text and Fig.~\ref{fig:bn_mux_eg}(b).) 


\begin{itemize}	
\item[$\vartriangleright$] Mark all column and row blocks as unlocked.

\item[$\vartriangleright$] {\it Loop A}: iterate over output groups.

\begin{itemize}
\item[$\vartriangleright$] {\it Loop B:} iterate over outputs (rows) within output group. 

\begin{itemize}
\item[] Find first column that can fill row either because it does, or because it contains $\ge1$ photons and is unlocked:

\item[] {\it If} success, record GMZI setting which fills the row and mark the column block as temporarily locked;	

\item[] {\it Else} mark all temporarily-locked columns as unlocked and exit loop B.

\end{itemize}
\item[$\vartriangleright$] {\it If} no failure then herald success for output group, compute row block settings to fill designated outputs, and mark corresponding row blocks as locked;

{\it Else} herald failure for output group and mark corresponding row blocks as unlocked.

\end{itemize}

\item[$\vartriangleright$] Output settings for locked rows and columns, setting defaults for unlocked rows and columns, and success/fail result for each output group.
\end{itemize}

\section{\label{sec:logic_reduction} General method for simplification of routing logic}
\begin{figure}
\begin{center}
\includegraphics[width=0.9\textwidth]{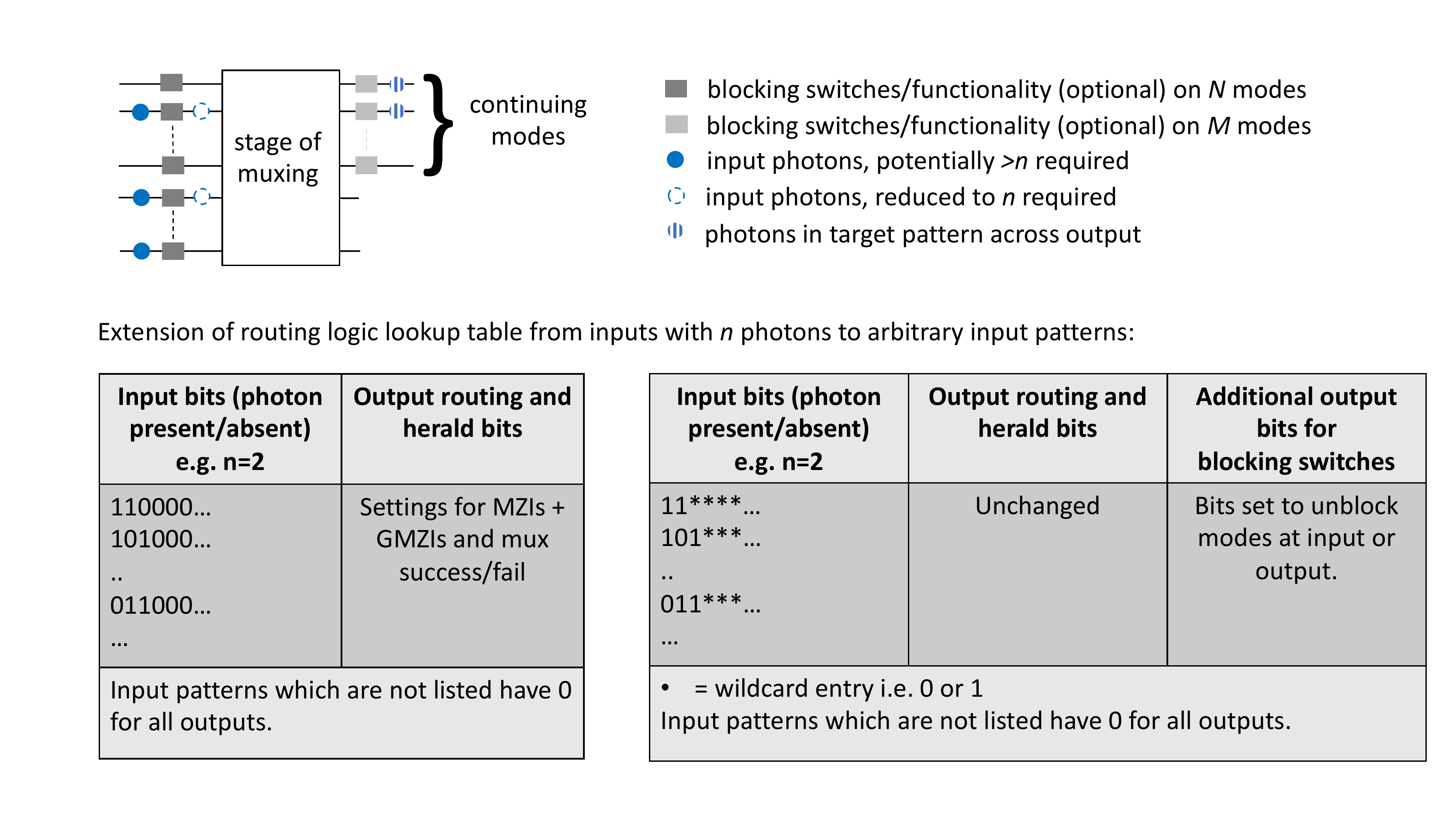}
\caption{\label{fig:logicreduction} {\it Method to extend routing logic for patterns of input bits corresponding to exactly the desired number of photons, $n$, to all possible input patterns.} Output bits are added to control blocking switches (or existing switching) so that excess photons can be discarded without affecting routing of the target photons.  Table entries for the input patterns are modified by replacing all $0$'s to the right of the final $1$ with wildcards, which avoids introducing any logical conflicts into the truth table.   Input patterns that are not useful are omitted with default values assumed for the output bits.}       
\end{center}
\end{figure}

For many switch network designs it is convenient to describe the required digital logic in terms of truth tables (or simple Boolean expressions) that relate input and output control bits for different stages of the routing algorithm.  The complexity of the logic can be captured by the numbers of input bits, output bits and entries (rows) for the truth tables that are needed (although accurate estimates of feedforward delay require simulations and optimizations that are specific to a technology node to account for the physical properties of electronics hardware \cite{EDA09}).  Often the truth tables that are needed for routing have a general structure that permits significant simplification, so that many input patterns are dealt with implicitly, and a general method to achieve this is described in Fig.~\ref{fig:logicreduction}.

\section{\label{sec:dB_sequences} De Bruijn full and reduced sequences}

\begin{figure}
\begin{center}
\includegraphics[width=0.9\textwidth]{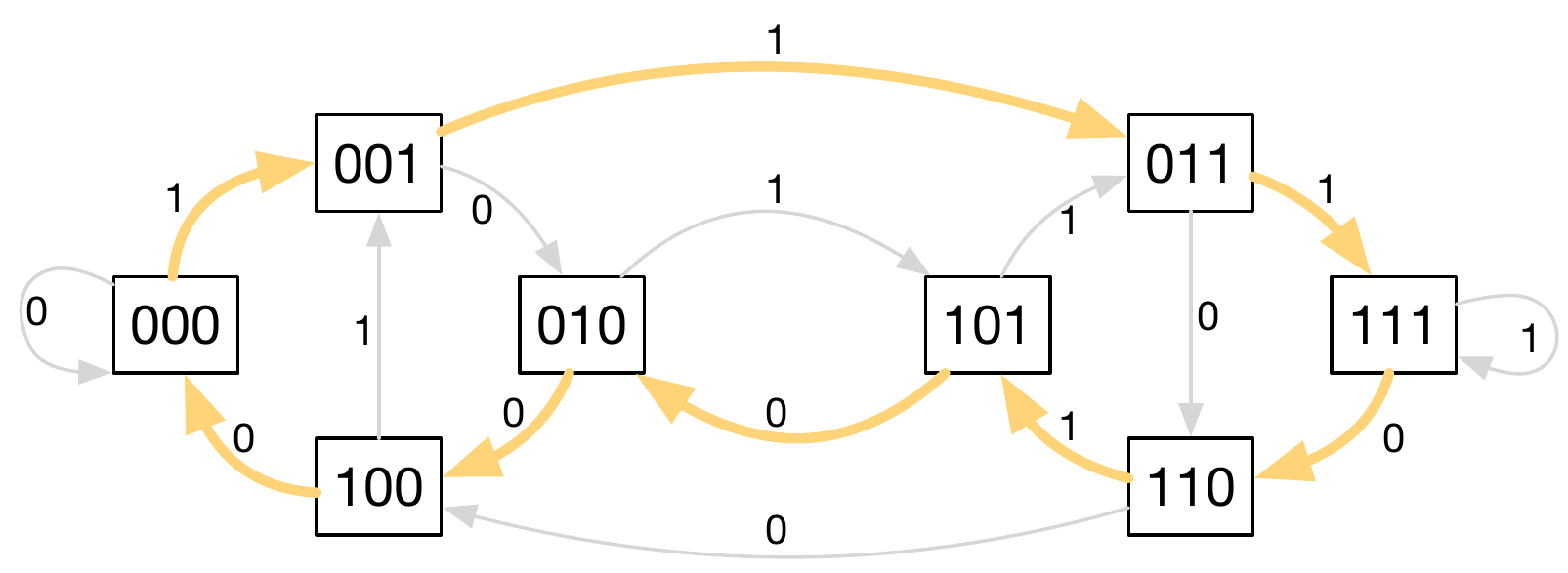}
\caption{\label{fig:db_sequence} {\it Graph showing all possible words (in square boxes) of length $L=3$ and alphabet $A=\{0,1\}$.} Arrows are labelled by one of the characters of the alphabet, an arrow connecting one word to another implies that the second word can be obtained from the first one by removing the first character and adding the characted of the arrow to the end of the word, e.g. $001 \xrightarrow{1} 011$. A valid deBruijn sequence is obtained by compiling the labels of a Hamiltonian path through the graph. A sample Hamiltonian path is marked in color with thicker lines. This would correspond to the cyclic de Bruijn sequence: ${111010001}$.}       
\end{center}
\end{figure}

Given an alphabet composed of characters $A=\{a_, a_2, ...,a_i\}$, a de Bruijn sequence with alphabet $A$ and word-length $L$ is a cyclic sequence of characters extracted from alphabet $A$ such that any possible words of length $L$ can be found exactly once within that sequence as a substring. For example:
\begin{align*}
    &A=\{0,1,2\}, L=3 \rightarrow \\ 
    &\text{de Bruijn sequence:} {0, 0, 1, 0, 2, 0, 3, 1, 1, 2, 1, 3, 2, 2, 3, 3}
\end{align*}
 De Bruijn sequences are not unique, but they are optimally short since they have lenth $|A|^L$, which is the exact number of distinct substrings of length $L$ on $A$. Finding de Bruijn sequences is straightforward. All possible ($|A|^L$) words are placed as nodes of a directed graph, where an arrow going from node $w_1 \xrightarrow{a_k} w_j$ represents that word $w_j$ can be obtained from word $w_i$ by adding character $a_k$ to the right of word $w_i$ and removing the leftmost character of $w_i$. Once such a graph is constructed, a valid de Bruijn sequence is found by composing together all the labels of arrows that lie on any Hamiltonian path through the graph. An example is shown in Fig.~\ref{fig:db_sequence}.
 
In switching schemes where the aim is to do spatio-temporal alignment of photons \textit{relative} to one another, it is possible to further reduce the size of the cyclic switches by removing all words that don't contain the character `0', since there will always be another valid word with the same alignment effect that contains that character. Obtaining the reduce de Bruijn sequence in that case is done following the same procedure outlined above, but using a graph with the reduced set of words.
 
